\documentclass[aps,prb,nofootinbib,reprint,preprintnumbers,superscriptaddress,longbibliography]{revtex4-2}
\usepackage{latexsym,epsfig,amssymb,amsfonts,amsmath,graphicx,bbm,braket,esdiff,dsfont,array,hyperref}
\pdfoutput=1
\hypersetup{
pdftitle={Parallel time-dependent variational principle algorithm for matrix product states}, pdfsubject={Time evolution of tensor networks representing quantum lattice systems with long-range interactions}, pdfauthor={Paul Secular, Nikita Gourianov, Michael Lubasch, Sergey Dolgov, Stephen R. Clark, Dieter Jaksch}, pdfkeywords={computational, strongly-correlated, quantum, physics, chemistry, tensor network, Hamiltonian, time evolution, simulation, long-range interaction, Heisenberg, Ising, spin, magnetism, TDVP, DMRG, MPS, MPI, OpenMP}, colorlinks=true, allcolors=blue}

\usepackage{xcolor}
\usepackage[shortlabels]{enumitem}
\usepackage[english]{babel}
\usepackage[subrefformat=parens,labelformat=parens,caption=false]{subfig}

\usepackage{afterpage}

\addto\captionsenglish{
  \renewcommand{\figurename}{FIG.}
  \renewcommand{\tablename}{TABLE}
  
}

\usepackage{soul}

\newcommand{\eqr}[1]{Eq.~(\ref{#1})}
\newcommand{\fir}[1]{Fig.~\ref{#1}}
\newcommand{\subfir}[1]{Fig.~\subref*{#1}}
\newcommand{\secr}[1]{Sec.~\ref{#1}}

\makeatletter
\newcommand{\colorcaption}[2][]{
  \begingroup
  \renewcommand{\@caption@fignum@sep}{ (color online). }
  \caption[#1]{#2}
  \endgroup
}
\makeatother

\allowdisplaybreaks

\begin{document}

\title{Parallel time-dependent variational principle algorithm for matrix product states}

\author{Paul Secular}
\email{paul@secular.me.uk}
\homepage{http://secular.me.uk/}
\affiliation{Department of Physics, University of Bath, Claverton Down, Bath BA2 7AY, UK}

\author{Nikita Gourianov}
\affiliation{Clarendon Laboratory, Department of Physics, University of Oxford, Oxford OX1 3PU, UK}

\author{Michael Lubasch}
\affiliation{Clarendon Laboratory, Department of Physics, University of Oxford, Oxford OX1 3PU, UK}

\author{Sergey Dolgov}
\affiliation{Department of Mathematical Sciences, University of Bath, Claverton Down, Bath BA2 7AY, UK}

\author{Stephen R. Clark}
\affiliation{H.H. Wills Physics Laboratory, University of Bristol, Bristol BS8 1TL, UK}
\affiliation{Max Planck Institute for the Structure and Dynamics of Matter, CFEL, 22761 Hamburg, Germany}

\author{Dieter Jaksch}
\affiliation{Clarendon Laboratory, Department of Physics, University of Oxford, Oxford OX1 3PU, UK}
\affiliation{Centre for Quantum Technologies, National University of Singapore, 3 Science Drive 2, Singapore 117543, Singapore}

\date{\rule[11pt]{0pt}{0pt}\today}

\begin{abstract}
Combining the time-dependent variational principle (TDVP) algorithm with the parallelization scheme introduced by Stoudenmire and White for the density matrix renormalization group (DMRG), we present the first parallel matrix product state (MPS) algorithm capable of time evolving one-dimensional (1D) quantum lattice systems with long-range interactions. We benchmark the accuracy and performance of the algorithm by simulating quenches in the long-range Ising and XY models. We show that our code scales well up to 32 processes, with parallel efficiencies as high as $86\%$. Finally, we calculate the dynamical correlation function of a 201-site Heisenberg XXX spin chain with $1/r^2$ interactions, which is challenging to compute sequentially. These results pave the way for the application of tensor networks to increasingly complex many-body systems.
\end{abstract}

\maketitle

\section{Introduction}
Although classical simulation of the quantum many-body problem is in general exponentially hard, many physically interesting, slightly entangled states can be efficiently simulated using tensor network methods \cite{vidal_efficient_2003, vidal_efficient_2004, verstraete_matrix_2006}; in particular those obeying an entanglement ``area law'' \cite{schuch_entropy_2008, eisert_colloquium:_2010}. The most common approach for one-dimensional (1D) and quasi-two-dimensional (2D) systems is the matrix product state (MPS) ansatz. This forms the basis of the modern formulation \cite{ostlund_thermodynamic_1995, dukelsky_equivalence_1998, schollwock_density-matrix_2011} of White's density matrix renormalization group (DMRG) algorithm \cite{white_density_1992} for computing ground states and low-lying excited states \cite{white_density-matrix_1993, stoudenmire_studying_2012}.

The introduction of the MPS-based time-evolving block decimation (TEBD) algorithm \cite{vidal_efficient_2003, vidal_efficient_2004, daley_time-dependent_2004} (and subsequent developments) allowed the short-time dynamics of local 1D quantum lattice models to be simulated with great success. In recent years, however, there has been a renewed interest in the dynamics of models with long-range interactions, driven by experimental advances in atomic, molecular, and optical physics. Interactions that decay as $1/r^{\alpha}$ are now realized in experiments with polar molecules ($\alpha = 3$) \cite{micheli_toolbox_2006, yan_observation_2013} and Rydberg atoms ($\alpha = 6$) \cite{schaus_observation_2012, weimer_rydberg_2010, zeiher_many-body_2016}, whilst trapped ion experiments can simulate spin models with a tunable exponent $\left( 0 \lesssim \alpha \lesssim 3 \right)$ \cite{porras_effective_2004, britton_engineered_2012, islam_emergence_2013, richerme_non-local_2014, jurcevic_quasiparticle_2014, smith_many-body_2016, zhang_observation_2017}.

In response, MPS algorithms have been developed that are able to simulate the time evolution of such models classically \cite{feiguin_time-step_2005, garcia-ripoll_time_2006, schachenmayer_dynamics_2008, pirvu_matrix_2010, stoudenmire_minimally_2010, zaletel_time-evolving_2015, haegeman_time-dependent_2011, wall_out--equilibrium_2012, lubich_dynamical_2013, lubich_time_2015, haegeman_unifying_2016, ronca_time-step_2017, paeckel_time-evolution_2019, hashizume_hybrid_2019}. One of the most promising approaches is the time-dependent variational principle (TDVP) \cite{haegeman_time-dependent_2011, lubich_dynamical_2013, lubich_time_2015, haegeman_unifying_2016}, which has found widespread use in condensed matter physics\footnote
{TDVP is applied to $1/r^\alpha$ models in Refs. \cite{koffel_entanglement_2012, hauke_spread_2013, buyskikh_entanglement_2016, haegeman_unifying_2016, halimeh_prethermalization_2017, jaschke_critical_2017, halimeh_dynamical_2017, zauner-stauber_probing_2017, zunkovic_dynamical_2018, pappalardi_scrambling_2018, lerose_impact_2019, kloss_spin_2019, lerose_quasilocalized_2019, zhou_operator_2019, piccitto_dynamical_2019}.
}, and which is beginning to find applications in quantum chemistry \cite{borrelli_simulation_2017, borrelli_theoretical_2018, kurashige_matrix_2018, baiardi_large-scale_2019, gelin_origin_2019, kloss_multiset_2019, xie_time-dependent_2019, li_numerical_2019}.

\begin{figure}[b]
  \includegraphics[width=8.6cm,keepaspectratio]{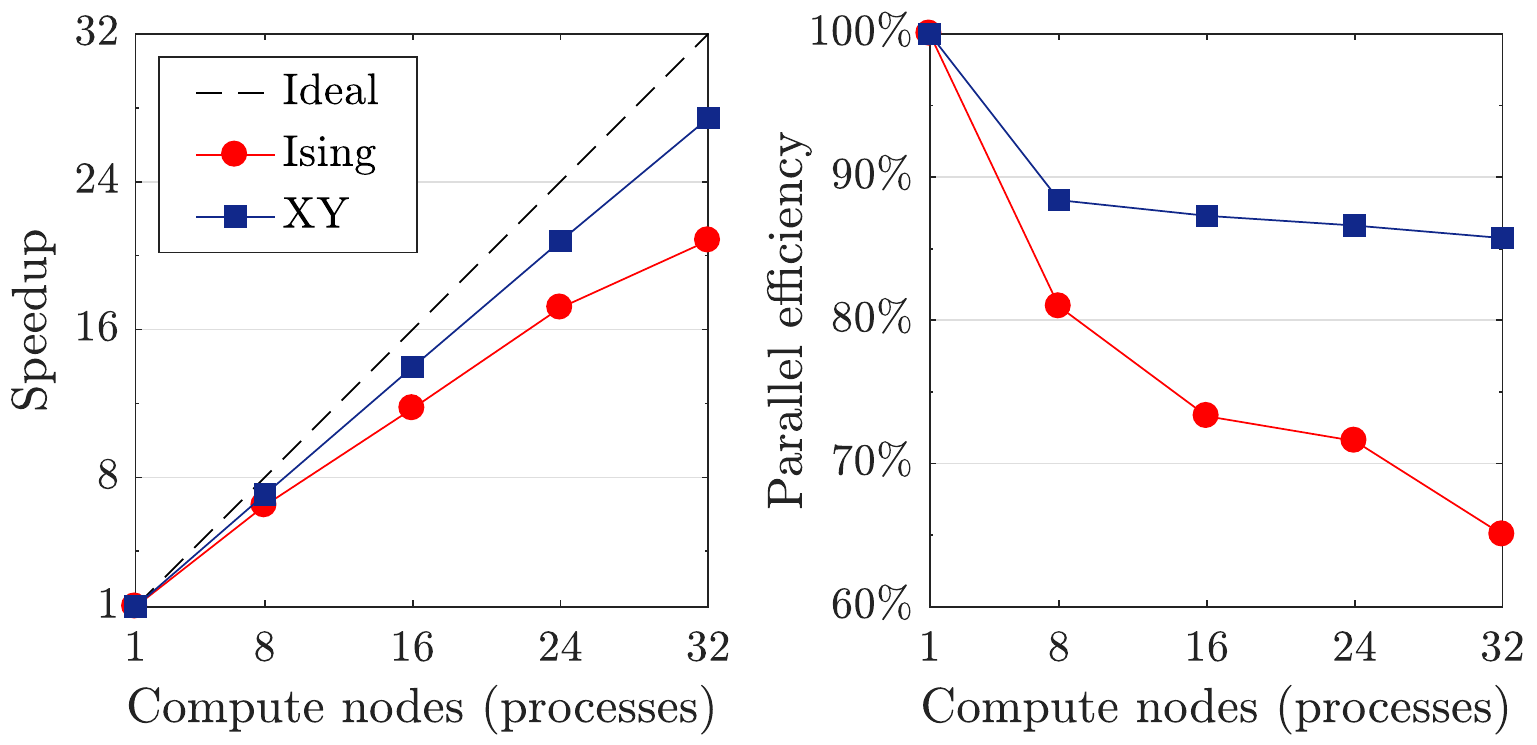}
  \caption{\label{figspeedup}Best and worse case strong scaling of the parallel time-dependent variational principle algorithm for our benchmark examples. ``Ising'' is the 129-site spin chain evolved under the $1/r^{2.3}$ transverse-field Ising Hamiltonian described in \secr{ising-benchmark}. ``XY'' is the 101-site spin chain evolved under the $1/r^{0.75}$ XY Hamiltonian described in \secr{xy-benchmark}. The benchmark described in \secr{haldane-benchmark} is excluded as it was unfeasible to run on a single process.
}
\end{figure}
TDVP is a serial algorithm that uses sequential sweeps for numerical stability. This means it cannot easily take advantage of multicore architectures or high-performance computing clusters. In fact the same is true of most MPS algorithms. Attempts to address this shortcoming include the use of parallel linear algebra operations \cite{hager_parallelization_2004, nemes_density_2014, kantian_understanding_2019, li_numerical_2019}, parallelization over quantum number blocks \cite{hager_parallelization_2004, kurashige_high-performance_2009}, and parallelization over terms in the Hamiltonian \cite{chan_algorithm_2004}. More generally, it is desirable to parallelize over different parts of an MPS, but this network-level parallelization is nontrivial. To the best of our knowledge, the only such algorithms to date are nearest-neighbor TEBD \cite{skaugen_time_2013, wall_quantum_2015, urbanek_parallel_2016, volokitin_propagating_2019}, real-space parallel DMRG \cite{stoudenmire_real-space_2013, depenbrock_tensor_2013}, and parallel infinite DMRG \cite{ueda_infinite-size_2018}, although network-level parallelization has also been proposed for projected entangled pair states (PEPS) \cite{lubasch_unifying_2014, lubasch_algorithms_2014}. A parallel time evolution method for MPS capable of handling long-range interactions has remained an open problem.

Given the close relationship between TDVP and DMRG established in Refs. \cite{lubich_time_2015, haegeman_unifying_2016}, it is natural to ask whether a parallel version of TDVP can be developed in a similar manner to real-space parallel DMRG. In this work we demonstrate that it can. The parallelization of the algorithm allows state-of-the-art calculations to be sped up by a factor of 20+ (see \fir{figspeedup}). Moreover, we show how it makes larger calculations possible that may otherwise be unfeasible. For example, in {\secr{haldane-benchmark}}, we calculate the dynamical spin-spin correlation function of a 201-site long-range Heisenberg model in a matter of days, rather than weeks.

The rest of the paper is organized as follows. We start with a background section, before moving on to introduce the parallel TDVP algorithm in \secr{sec:the-algorithm}. We provide increasingly complex benchmark examples for quantum spin-chains in \secr{sec:the-benchmarks}, and finally conclude and suggest directions for future research in \secr{sec:the-conclusion}.
\section{Background}\label{sec:background}
For completeness, we start by reviewing relevant background material, covering matrix product states, the inverse canonical gauge, matrix product operators, and the time dependent variational principle. We also establish the notation used in the rest of the paper.
\subsection{Matrix product states}
Any finite-dimensional, $N$-partite quantum state can be expressed in a given basis $\ket{\sigma_1 \sigma_2 \dots \sigma_N}$ as
\begin{equation}\label{mps-no-gauge}
  \ket{\psi} = \sum^{d_1\dots d_N}_{\sigma_1 \dots \sigma_N = 1} M_1^{\sigma_1} M_2^{\sigma_2} \dots M_N^{\sigma_N} \ket{\sigma_1 \sigma_2 \dots \sigma_N},
\end{equation}
where the $M_j^{\sigma_j}$ are matrices ($M_1^{\sigma_1}$ and $M_N^{\sigma_N}$ being row and column vectors, respectively). This decomposition is known as a tensor train \cite{oseledets_tensor-train_2011} or matrix product state (MPS) \cite{perez-garcia_matrix_2007}. To avoid index gymnastics we will often use the graphical tensor notation introduced by Penrose \cite{penrose_applications_1971} to represent MPS as tensor networks (see e.g. Ref. \cite{bridgeman_hand-waving_2017}). In this notation, tensors are displayed as nodes in a network, with their connected edges representing pairs of dummy indices in the usual Einstein notation. A contractible edge is referred to as a bond and the dimension of the corresponding dummy index is the bond dimension. Free indices (which may represent physical degrees of freedom) are shown as disconnected edges. In this notation, \eqr{mps-no-gauge} is written as
\begin{equation}
  \ket{\psi}=\raisebox{-0.6cm}{\def\svgwidth{6.6cm}
\begingroup%
  \makeatletter%
  \providecommand\color[2][]{%
    \errmessage{(Inkscape) Color is used for the text in Inkscape, but the package 'color.sty' is not loaded}%
    \renewcommand\color[2][]{}%
  }%
  \providecommand\transparent[1]{%
    \errmessage{(Inkscape) Transparency is used (non-zero) for the text in Inkscape, but the package 'transparent.sty' is not loaded}%
    \renewcommand\transparent[1]{}%
  }%
  \providecommand\rotatebox[2]{#2}%
  \newcommand*\fsize{\dimexpr\f@size pt\relax}%
  \newcommand*\lineheight[1]{\fontsize{\fsize}{#1\fsize}\selectfont}%
  \ifx\svgwidth\undefined%
    \setlength{\unitlength}{600.94487bp}%
    \ifx\svgscale\undefined%
      \relax%
    \else%
      \setlength{\unitlength}{\unitlength * \real{\svgscale}}%
    \fi%
  \else%
    \setlength{\unitlength}{\svgwidth}%
  \fi%
  \global\let\svgwidth\undefined%
  \global\let\svgscale\undefined%
  \makeatother%
  \begin{picture}(1,0.18867925)%
    \lineheight{1}%
    \setlength\tabcolsep{0pt}%
    \put(0,0){\includegraphics[width=\unitlength,page=1]{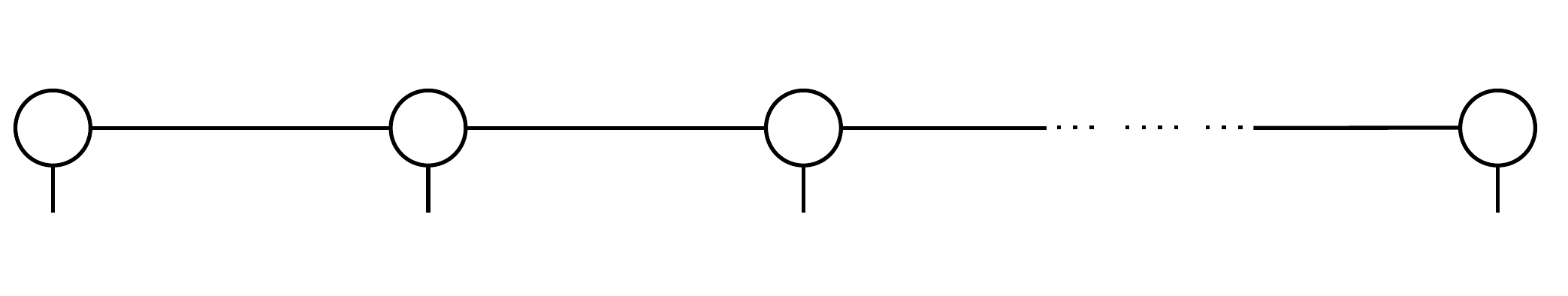}}%
    \put(0.02040817,0.042){\color[rgb]{0,0,0}\makebox(0,0)[lt]{\begin{minipage}{0.16686675\unitlength}\raggedright $\scriptstyle{d_1}$\end{minipage}}}%
    \put(0.25503867,0.042){\color[rgb]{0,0,0}\makebox(0,0)[lt]{\begin{minipage}{0.16686675\unitlength}\raggedright $\scriptstyle{d_2}$\end{minipage}}}%
    \put(0.49715738,0.042){\color[rgb]{0,0,0}\makebox(0,0)[lt]{\begin{minipage}{0.16686675\unitlength}\raggedright $\scriptstyle{d_3}$\end{minipage}}}%
    \put(0.93896162,0.042){\color[rgb]{0,0,0}\makebox(0,0)[lt]{\begin{minipage}{0.16686675\unitlength}\raggedright $\scriptstyle{d_N}$\end{minipage}}}%
    \put(0.12990664,0.145){\color[rgb]{0,0,0}\makebox(0,0)[lt]{\begin{minipage}{0.16686675\unitlength}\raggedright $\scriptstyle{\chi_1}$\end{minipage}}}%
    \put(0.36952929,0.145){\color[rgb]{0,0,0}\makebox(0,0)[lt]{\begin{minipage}{0.16686675\unitlength}\raggedright $\scriptstyle{\chi_2}$\end{minipage}}}%
    \put(0.60415979,0.145){\color[rgb]{0,0,0}\makebox(0,0)[lt]{\begin{minipage}{0.16686675\unitlength}\raggedright $\scriptstyle{\chi_3}$\end{minipage}}}%
    \put(0.00792782,0.19){\color[rgb]{0,0,0}\makebox(0,0)[lt]{\begin{minipage}{0.16686675\unitlength}\raggedright $M_1$\end{minipage}}}%
    \put(0.24755046,0.19){\color[rgb]{0,0,0}\makebox(0,0)[lt]{\begin{minipage}{0.16686675\unitlength}\raggedright $M_2$\end{minipage}}}%
    \put(0.4871731,0.19){\color[rgb]{0,0,0}\makebox(0,0)[lt]{\begin{minipage}{0.16686675\unitlength}\raggedright $M_3$\end{minipage}}}%
    \put(0.92648128,0.19){\color[rgb]{0,0,0}\makebox(0,0)[lt]{\begin{minipage}{0.16686675\unitlength}\raggedright $M_N$\end{minipage}}}%
    \put(0.80134925,0.145){\color[rgb]{0,0,0}\makebox(0,0)[lt]{\begin{minipage}{0.24489793\unitlength}\raggedright $\scriptstyle{\chi_{N-1}}$\end{minipage}}}%
  \end{picture}%
\endgroup%
},
\end{equation}
where we have explicitly labeled the dimensions of all bonds $\chi_j$, and physical edges $d_j$ (for notational clarity, dimensions are suppressed in the rest of the paper).

Here we are interested in MPS in the context of finite 1D lattice models with open boundaries. In such a model there is a one-to-one correspondence between the lattice sites and the MPS ``site tensors'' $M_j$. In quantum chemistry, the ``sites'' may instead be molecular orbitals \cite{szalay_tensor_2015}. In a lattice model, the physical dimensions $d_j$ will often be independent of $j$. For the spin-half chains considered in this paper, $d_j = d = 2$. The bonds between sites capture the entanglement present. In general, $\chi_j$ is not independent of $j$. It is thus convenient to define $\chi = \max(\chi_j)$. Exactly representing an arbitrary state as an MPS requires a $\chi = \chi_\text{exact}$ that is exponential in the number of sites. However, physical states are often well approximated by MPS of lower bond dimension. We denote by $\chi_\text{max}$ the maximum bond dimension chosen for a calculation, where typically $\chi_\textrm{max} \ll \chi_\text{exact}$.

Even with all $\chi_j$ fixed, an MPS representation is not unique, since
\begin{equation*}
    \dots M_j^{\sigma_j} M_{j+1}^{\sigma_{j+1}} \dots = \dots \tilde{M}_j^{\sigma_j} \tilde{M}_{j+1}^{\sigma_{j+1}} \dots,
\end{equation*}
where $\tilde{M}_j^{\sigma_j} = M_j^{\sigma_j} X$, and $\tilde{M}_{j+1}^{\sigma_{j+1}} = X^{-1} M_{j+1}^{\sigma_{j+1}}$ for any invertible matrix $X$. This ``gauge freedom'' can be exploited by algorithms for numerical stability. In particular, a canonical form is usually employed in which one or more tensors act as orthogonality centers \cite{schollwock_density-matrix_2011}. The parts of the MPS to the left and right of an orthogonality center form orthonormal basis states, respectively $\Ket{\Phi_{L,k}}$ and $\Ket{\Phi_{R,l}}$, where $k$ and $l$ are the indices of the states. The orthogonality center thus defines the quantum state with respect to these bases.

Bipartitioning a system into sites $[1:j]$ and $[j+1:N]$ allows us to write the Schmidt decomposition,
\begin{equation}
\Ket{\psi} = \sum_{k=1}^{\chi_j} \lambda_k \Ket{\Phi_{L,k}^{[1:j]}} \otimes \Ket{\Phi_{R,k}^{[j+1:N]}}.
\end{equation}
This can be expressed as an MPS with orthogonality center $\Lambda_j=\mathop{\mathrm{diag}}(\lambda_1 \dots \lambda_{\chi_j})$ \cite{schollwock_density-matrix_2011},
\begin{equation}\label{eq:mixed-mps1}
  \ket{\psi} = \sum_{\sigma_1 \dots \sigma_N } A_1^{\sigma_1} \dots A_j^{\sigma_j} \Lambda_j B_{j+1}
  ^{\sigma_{j+1}} \dots B_N^{\sigma_N} \ket{\sigma_1 \dots \sigma_N},
\end{equation}
where $A_p^{\sigma_p}$ $(1 \leq p \leq j)$ and $B_q^{\sigma_q}$ $(j+1 \leq q \leq N)$ satisfy
\begin{equation}
  \sum_{\sigma_p} \left(A_p^{\sigma_p}\right)^\dagger A^{\sigma_p}_p = \sum_{\sigma_q} B_q^{\sigma_q} \left(B_q^{\sigma_q}\right)^\dagger = \mathds{1}.
\end{equation}
The orthogonality center can alternatively be made into a site tensor. For example, setting $\Psi_j^{\sigma_{j}} = A_{j}^{\sigma_{j}} \Lambda_j$ gives an MPS with orthogonality center at site $j$,
\begin{equation}\label{eq:mixed-mps2}
  \ket{\psi} = \sum_{\sigma_1 \dots \sigma_N } A_1^{\sigma_1} \dots A_{j-1}^{\sigma_{j-1}} \Psi_{j}
  ^{\sigma_{j}} B_{j+1}
  ^{\sigma_{j+1}} \dots B_N^{\sigma_N} \ket{\sigma_1 \dots \sigma_N}.
\end{equation}
Eqs. (\ref{eq:mixed-mps1}) and (\ref{eq:mixed-mps2}) are known as the mixed canonical form \cite{schollwock_density-matrix_2011}. This MPS gauge is useful for serial algorithms that update tensors sequentially, as the orthogonality center can be shifted by one site whilst maintaining the orthonormality of the state. For details of this approach we refer the reader to Ref. \cite{schollwock_density-matrix_2011}.
\subsection{Inverse canonical gauge}
The key to parallelizing MPS algorithms is to use a gauge with multiple orthogonality centers. This allows different site tensors to be updated simultaneously, and to be merged back into the MPS consistently. The first such gauge was introduced by Vidal for TEBD. Any $N$-partite state can be written in Vidal's canonical form \cite{vidal_efficient_2003, vidal_efficient_2004, daley_time-dependent_2004},
\begin{equation}\label{eq:vidal-mps}
  \ket{\psi} = \sum_{\sigma_1 \dots \sigma_N } \Gamma_1^{\sigma_1} \Lambda_1 \Gamma_2^{\sigma_2} \Lambda_2 \dots \Lambda_{N-1} \Gamma_N^{\sigma_N} \ket{\sigma_1 \dots \sigma_N},
\end{equation}
where the $\Lambda_j$ are again diagonal matrices of singular values that serve as orthogonality centers. In graphical tensor notation we write this as
\begin{equation}
  \ket{\psi}=\raisebox{-0.25cm}{\def\svgwidth{6.4cm}
\begingroup%
  \makeatletter%
  \providecommand\color[2][]{%
    \errmessage{(Inkscape) Color is used for the text in Inkscape, but the package 'color.sty' is not loaded}%
    \renewcommand\color[2][]{}%
  }%
  \providecommand\transparent[1]{%
    \errmessage{(Inkscape) Transparency is used (non-zero) for the text in Inkscape, but the package 'transparent.sty' is not loaded}%
    \renewcommand\transparent[1]{}%
  }%
  \providecommand\rotatebox[2]{#2}%
  \newcommand*\fsize{\dimexpr\f@size pt\relax}%
  \newcommand*\lineheight[1]{\fontsize{\fsize}{#1\fsize}\selectfont}%
  \ifx\svgwidth\undefined%
    \setlength{\unitlength}{596bp}%
    \ifx\svgscale\undefined%
      \relax%
    \else%
      \setlength{\unitlength}{\unitlength * \real{\svgscale}}%
    \fi%
  \else%
    \setlength{\unitlength}{\svgwidth}%
  \fi%
  \global\let\svgwidth\undefined%
  \global\let\svgscale\undefined%
  \makeatother%
  \begin{picture}(1,0.1442953)%
    \lineheight{1}%
    \setlength\tabcolsep{0pt}%
    \put(0,0){\includegraphics[width=\unitlength,page=1]{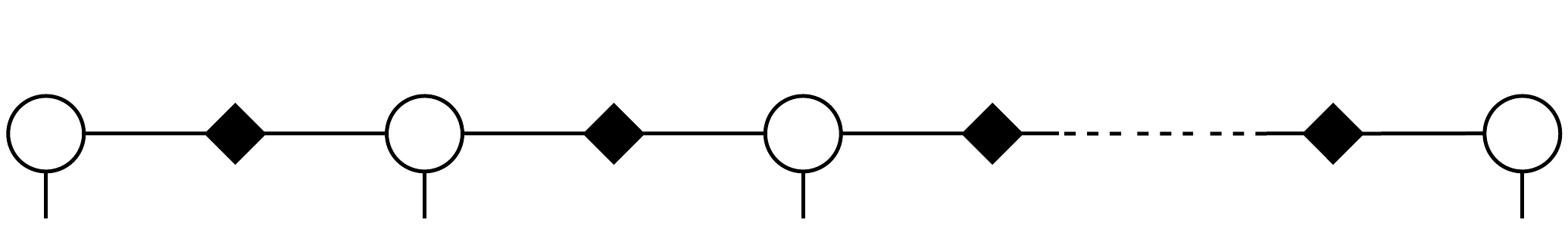}}%
    \put(0.0083,0.108){\color[rgb]{0,0,0}\makebox(0,0)[lt]{\lineheight{1.25}\smash{\begin{tabular}[t]{l}$\Gamma_1$\end{tabular}}}}%
    \put(0.247,0.108){\color[rgb]{0,0,0}\makebox(0,0)[lt]{\lineheight{1.25}\smash{\begin{tabular}[t]{l}$\Gamma_2$\end{tabular}}}}%
    \put(0.4865,0.108){\color[rgb]{0,0,0}\makebox(0,0)[lt]{\lineheight{1.25}\smash{\begin{tabular}[t]{l}$\Gamma_3$\end{tabular}}}}%
    \put(0.955,0.108){\color[rgb]{0,0,0}\makebox(0,0)[lt]{\lineheight{1.25}\smash{\begin{tabular}[t]{l}$\Gamma_N$\end{tabular}}}}%
    \put(0.13,0.108){\color[rgb]{0,0,0}\makebox(0,0)[lt]{\lineheight{1.25}\smash{\begin{tabular}[t]{l}$\Lambda_1$\end{tabular}}}}%
    \put(0.37,0.108){\color[rgb]{0,0,0}\makebox(0,0)[lt]{\lineheight{1.25}\smash{\begin{tabular}[t]{l}$\Lambda_2$\end{tabular}}}}%
    \put(0.615,0.108){\color[rgb]{0,0,0}\makebox(0,0)[lt]{\lineheight{1.25}\smash{\begin{tabular}[t]{l}$\Lambda_3$\end{tabular}}}}%
    \put(0.80,0.108){\color[rgb]{0,0,0}\makebox(0,0)[lt]{\lineheight{1.25}\smash{\begin{tabular}[t]{l}$\Lambda_{N-1}$\end{tabular}}}}%
  \end{picture}%
\endgroup%
}\ ,
\end{equation}
where the $\Gamma_j$ are site tensors. The beauty of this canonical gauge is that it simultaneously gives the Schmidt decomposition of all bipartitions.

In this work we use the inverse canonical gauge due to Stoudenmire and White \cite{stoudenmire_real-space_2013}, which is given by
\begin{equation}
  \ket{\psi}=\raisebox{-0.25cm}{\def\svgwidth{6.4cm}
\begingroup%
  \makeatletter%
  \providecommand\color[2][]{%
    \errmessage{(Inkscape) Color is used for the text in Inkscape, but the package 'color.sty' is not loaded}%
    \renewcommand\color[2][]{}%
  }%
  \providecommand\transparent[1]{%
    \errmessage{(Inkscape) Transparency is used (non-zero) for the text in Inkscape, but the package 'transparent.sty' is not loaded}%
    \renewcommand\transparent[1]{}%
  }%
  \providecommand\rotatebox[2]{#2}%
  \newcommand*\fsize{\dimexpr\f@size pt\relax}%
  \newcommand*\lineheight[1]{\fontsize{\fsize}{#1\fsize}\selectfont}%
  \ifx\svgwidth\undefined%
    \setlength{\unitlength}{596bp}%
    \ifx\svgscale\undefined%
      \relax%
    \else%
      \setlength{\unitlength}{\unitlength * \real{\svgscale}}%
    \fi%
  \else%
    \setlength{\unitlength}{\svgwidth}%
  \fi%
  \global\let\svgwidth\undefined%
  \global\let\svgscale\undefined%
  \makeatother%
  \begin{picture}(1,0.1442953)%
    \lineheight{1}%
    \setlength\tabcolsep{0pt}%
    \put(0,0){\includegraphics[width=\unitlength,page=1]{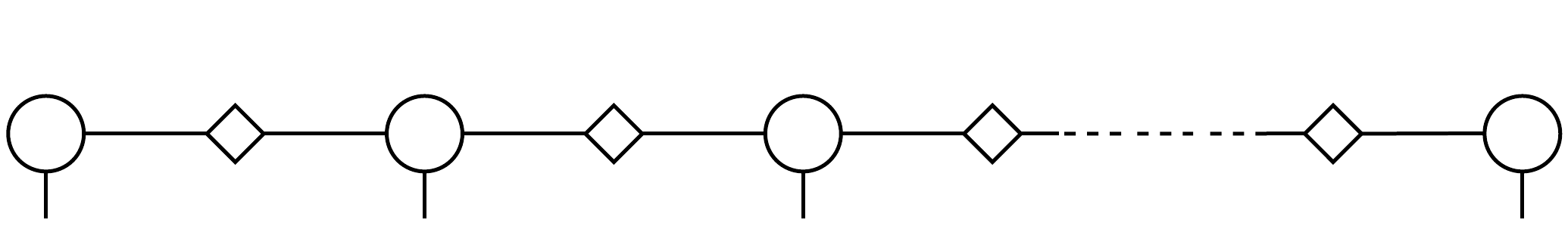}}%
    \put(0.0083,0.108){\color[rgb]{0,0,0}\makebox(0,0)[lt]{\lineheight{1.25}\smash{\begin{tabular}[t]{l}$\Psi_1$\end{tabular}}}}%
    \put(0.247,0.108){\color[rgb]{0,0,0}\makebox(0,0)[lt]{\lineheight{1.25}\smash{\begin{tabular}[t]{l}$\Psi_2$\end{tabular}}}}%
    \put(0.4865,0.108){\color[rgb]{0,0,0}\makebox(0,0)[lt]{\lineheight{1.25}\smash{\begin{tabular}[t]{l}$\Psi_3$\end{tabular}}}}%
    \put(0.95,0.108){\color[rgb]{0,0,0}\makebox(0,0)[lt]{\lineheight{1.25}\smash{\begin{tabular}[t]{l}$\Psi_N$\end{tabular}}}}%
    \put(0.13,0.108){\color[rgb]{0,0,0}\makebox(0,0)[lt]{\lineheight{1.25}\smash{\begin{tabular}[t]{l}$\textrm{V}_1$\end{tabular}}}}%
    \put(0.37,0.108){\color[rgb]{0,0,0}\makebox(0,0)[lt]{\lineheight{1.25}\smash{\begin{tabular}[t]{l}$\textrm{V}_2$\end{tabular}}}}%
    \put(0.615,0.108){\color[rgb]{0,0,0}\makebox(0,0)[lt]{\lineheight{1.25}\smash{\begin{tabular}[t]{l}$\textrm{V}_3$\end{tabular}}}}%
    \put(0.793,0.108){\color[rgb]{0,0,0}\makebox(0,0)[lt]{\lineheight{1.25}\smash{\begin{tabular}[t]{l}$\textrm{V}_{N-1}$\end{tabular}}}}%
  \end{picture}%
\endgroup%
}\ ,
\end{equation}
where $\mathrm{V}_j \equiv \Lambda_j^{-1}$. Although equivalent to the canonical gauge, it turns out to be a more natural choice for parallel TDVP and DMRG, as well as for TEBD, due to the site tensors, rather than the diagonal matrices, being orthogonality centers. An MPS in canonical form is transformed into inverse canonical form by inserting $\mathrm{V}_j \Lambda_j = \mathds{1} $ at each bond and then contracting the $\Lambda$ matrices with the $\Gamma$ site tensors, i.e.
\begin{equation*}
  \raisebox{-0.5cm}{\def\svgwidth{8.55cm}
\begingroup%
  \makeatletter%
  \providecommand\color[2][]{%
    \errmessage{(Inkscape) Color is used for the text in Inkscape, but the package 'color.sty' is not loaded}%
    \renewcommand\color[2][]{}%
  }%
  \providecommand\transparent[1]{%
    \errmessage{(Inkscape) Transparency is used (non-zero) for the text in Inkscape, but the package 'transparent.sty' is not loaded}%
    \renewcommand\transparent[1]{}%
  }%
  \providecommand\rotatebox[2]{#2}%
  \newcommand*\fsize{\dimexpr\f@size pt\relax}%
  \newcommand*\lineheight[1]{\fontsize{\fsize}{#1\fsize}\selectfont}%
  \ifx\svgwidth\undefined%
    \setlength{\unitlength}{785.83978271bp}%
    \ifx\svgscale\undefined%
      \relax%
    \else%
      \setlength{\unitlength}{\unitlength * \real{\svgscale}}%
    \fi%
  \else%
    \setlength{\unitlength}{\svgwidth}%
  \fi%
  \global\let\svgwidth\undefined%
  \global\let\svgscale\undefined%
  \makeatother%
  \begin{picture}(1,0.40821426)%
    \lineheight{1}%
    \setlength\tabcolsep{0pt}%
    \put(0,0){\includegraphics[width=\unitlength,page=1]{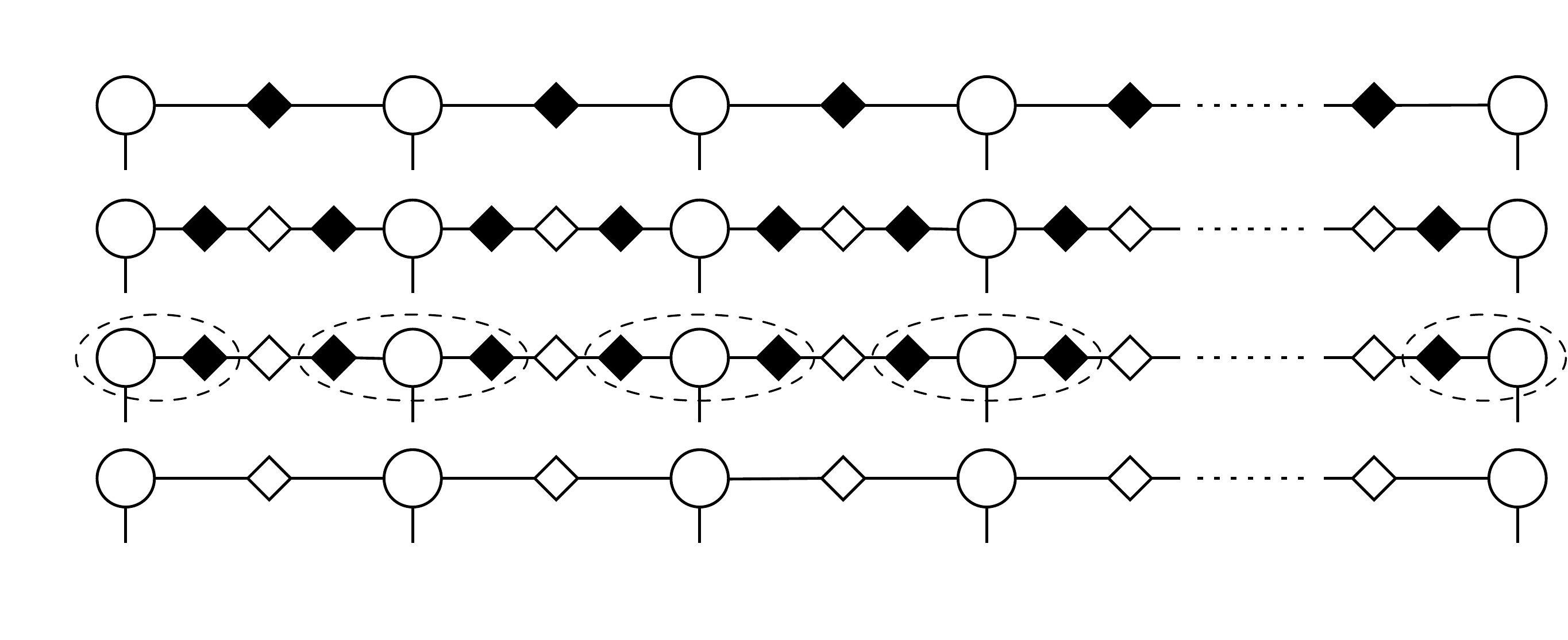}}%
    \put(0.0635,0.0125){\color[rgb]{0,0,0}\makebox(0,0)[lt]{\lineheight{1.25}\smash{\begin{tabular}[t]{l}$\Psi_1$\end{tabular}}}}%
    \put(0.247,0.0125){\color[rgb]{0,0,0}\makebox(0,0)[lt]{\lineheight{1.25}\smash{\begin{tabular}[t]{l}$\Psi_2$\end{tabular}}}}%
    \put(0.428,0.0125){\color[rgb]{0,0,0}\makebox(0,0)[lt]{\lineheight{1.25}\smash{\begin{tabular}[t]{l}$\Psi_3$\end{tabular}}}}%
    \put(0.951,0.0125){\color[rgb]{0,0,0}\makebox(0,0)[lt]{\lineheight{1.25}\smash{\begin{tabular}[t]{l}$\Psi_N$\end{tabular}}}}%
    \put(0.613,0.0125){\color[rgb]{0,0,0}\makebox(0,0)[lt]{\lineheight{1.25}\smash{\begin{tabular}[t]{l}$\Psi_4$\end{tabular}}}}%
    \put(0.153,0.0125){\color[rgb]{0,0,0}\makebox(0,0)[lt]{\lineheight{1.25}\smash{\begin{tabular}[t]{l}$\textrm{V}_1$\end{tabular}}}}%
    \put(0.335,0.0125){\color[rgb]{0,0,0}\makebox(0,0)[lt]{\lineheight{1.25}\smash{\begin{tabular}[t]{l}$\textrm{V}_2$\end{tabular}}}}%
    \put(0.52,0.0125){\color[rgb]{0,0,0}\makebox(0,0)[lt]{\lineheight{1.25}\smash{\begin{tabular}[t]{l}$\textrm{V}_3$\end{tabular}}}}%
    \put(0.70,0.0125){\color[rgb]{0,0,0}\makebox(0,0)[lt]{\lineheight{1.25}\smash{\begin{tabular}[t]{l}$\textrm{V}_4$\end{tabular}}}}%
    \put(0.838,0.0125){\color[rgb]{0,0,0}\makebox(0,0)[lt]{\lineheight{1.25}\smash{\begin{tabular}[t]{l}$\textrm{V}_{N-1}$\end{tabular}}}}%
    \put(0.0635,0.385){\color[rgb]{0,0,0}\makebox(0,0)[lt]{\lineheight{1.25}\smash{\begin{tabular}[t]{l}$\Gamma_1$\end{tabular}}}}%
    \put(0.247,0.385){\color[rgb]{0,0,0}\makebox(0,0)[lt]{\lineheight{1.25}\smash{\begin{tabular}[t]{l}$\Gamma_2$\end{tabular}}}}%
    \put(0.428,0.385){\color[rgb]{0,0,0}\makebox(0,0)[lt]{\lineheight{1.25}\smash{\begin{tabular}[t]{l}$\Gamma_3$\end{tabular}}}}%
    \put(0.951,0.385){\color[rgb]{0,0,0}\makebox(0,0)[lt]{\lineheight{1.25}\smash{\begin{tabular}[t]{l}$\Gamma_N$\end{tabular}}}}%
    \put(0.613,0.385){\color[rgb]{0,0,0}\makebox(0,0)[lt]{\lineheight{1.25}\smash{\begin{tabular}[t]{l}$\Gamma_4$\end{tabular}}}}%
    \put(0.153,0.385){\color[rgb]{0,0,0}\makebox(0,0)[lt]{\lineheight{1.25}\smash{\begin{tabular}[t]{l}$\Lambda_1$\end{tabular}}}}%
    \put(0.335,0.385){\color[rgb]{0,0,0}\makebox(0,0)[lt]{\lineheight{1.25}\smash{\begin{tabular}[t]{l}$\Lambda_2$\end{tabular}}}}%
    \put(0.52,0.385){\color[rgb]{0,0,0}\makebox(0,0)[lt]{\lineheight{1.25}\smash{\begin{tabular}[t]{l}$\Lambda_3$\end{tabular}}}}%
    \put(0.70,0.385){\color[rgb]{0,0,0}\makebox(0,0)[lt]{\lineheight{1.25}\smash{\begin{tabular}[t]{l}$\Lambda_4$\end{tabular}}}}%
    \put(0.838,0.385){\color[rgb]{0,0,0}\makebox(0,0)[lt]{\lineheight{1.25}\smash{\begin{tabular}[t]{l}$\Lambda_{N-1}$\end{tabular}}}}%
    \put(0.0,0.25){\color[rgb]{0,0,0}\makebox(0,0)[lt]{\lineheight{1.25}\smash{\begin{tabular}[t]{l}$=$\end{tabular}}}}%
    \put(0.0,0.17){\color[rgb]{0,0,0}\makebox(0,0)[lt]{\lineheight{1.25}\smash{\begin{tabular}[t]{l}$=$\end{tabular}}}}%
    \put(0.0,0.09){\color[rgb]{0,0,0}\makebox(0,0)[lt]{\lineheight{1.25}\smash{\begin{tabular}[t]{l}$=$\end{tabular}}}}%
  \end{picture}%
\endgroup%
}.
\end{equation*}
As $\mathrm{\Lambda}_j$ is diagonal, calculating $\mathrm{V}_j$ simply requires taking the reciprocal of the singular values\footnote{Taking the reciprocal of a floating-point number should be ``safe'' as long as overflows and divisions by zero are avoided.}. The contractions $\Gamma_j \mathrm{V}_j$, and $\mathrm{V}_j \Gamma_{j+1}$ are similarly cheap. We find no issue with the inversion of $\mathrm{\Lambda}_j$ as tiny singular values corresponding to numerical noise are discarded \cite{schollwock_density-matrix_2011, dongarra_singular_2018}.
\subsection{Matrix product operators}
We represent $N$-site operators using the matrix product operator (MPO) construction,
\begin{equation}
  O=\sum_{\substack{\sigma_1 \dots \sigma_N \\ \tau_1 \dots \tau_N} } O_1^{\sigma_1 \tau_1} O_2^{\sigma_2 \tau_2}\dots O_N^{\sigma_N \tau_N} \ket{\sigma_1 \dots \sigma_N}\bra{\tau_1 \dots \tau_N},
\end{equation}
where the $O_j^{\sigma_j \tau_j}$ are matrices ($O_1^{\sigma_1 \tau_1}$ and $O_N^{\sigma_N \tau_N}$ being row and column vectors, respectively). In graphical tensor notation we write this as
\begin{equation}
  \ket{\psi}=\raisebox{-0.3cm}{\def\svgwidth{6.4cm}
\begingroup%
  \makeatletter%
  \providecommand\color[2][]{%
    \errmessage{(Inkscape) Color is used for the text in Inkscape, but the package 'color.sty' is not loaded}%
    \renewcommand\color[2][]{}%
  }%
  \providecommand\transparent[1]{%
    \errmessage{(Inkscape) Transparency is used (non-zero) for the text in Inkscape, but the package 'transparent.sty' is not loaded}%
    \renewcommand\transparent[1]{}%
  }%
  \providecommand\rotatebox[2]{#2}%
  \newcommand*\fsize{\dimexpr\f@size pt\relax}%
  \newcommand*\lineheight[1]{\fontsize{\fsize}{#1\fsize}\selectfont}%
  \ifx\svgwidth\undefined%
    \setlength{\unitlength}{586bp}%
    \ifx\svgscale\undefined%
      \relax%
    \else%
      \setlength{\unitlength}{\unitlength * \real{\svgscale}}%
    \fi%
  \else%
    \setlength{\unitlength}{\svgwidth}%
  \fi%
  \global\let\svgwidth\undefined%
  \global\let\svgscale\undefined%
  \makeatother%
  \begin{picture}(1,0.16382253)%
    \lineheight{1}%
    \setlength\tabcolsep{0pt}%
    \put(0,0){\includegraphics[width=\unitlength,page=1]{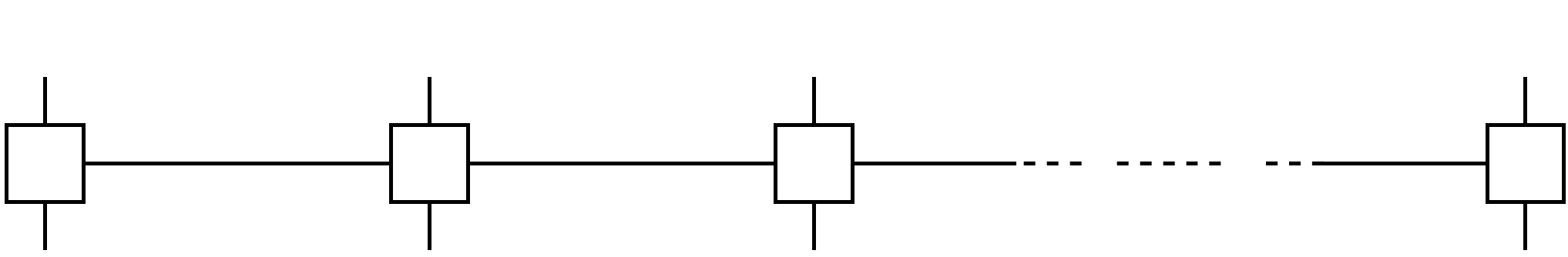}}%
    \put(0.008,0.135){\color[rgb]{0,0,0}\makebox(0,0)[lt]{\lineheight{1.25}\smash{\begin{tabular}[t]{l}$O_1$\end{tabular}}}}%
    \put(0.2516,0.135){\color[rgb]{0,0,0}\makebox(0,0)[lt]{\lineheight{1.25}\smash{\begin{tabular}[t]{l}$O_2$\end{tabular}}}}%
    \put(0.4948,0.135){\color[rgb]{0,0,0}\makebox(0,0)[lt]{\lineheight{1.25}\smash{\begin{tabular}[t]{l}$O_3$\end{tabular}}}}%
    \put(0.9478,0.135){\color[rgb]{0,0,0}\makebox(0,0)[lt]{\lineheight{1.25}\smash{\begin{tabular}[t]{l}$O_N$\end{tabular}}}}%
  \end{picture}%
\endgroup%
}\ ,
\end{equation}
where the $O_j$ are site operators (represented by squares). Site operators are analogous to MPS site tensors, but have an extra physical edge. The physical edges have the same dimensions $d_j$ as the MPS on which they act.

Local Hamiltonians are known to have a particularly compact MPO representation \cite{mcculloch_density-matrix_2007, crosswhite_finite_2008}, which means the maximum MPO bond dimension $m$ is independent of $N$. Exponentially decaying interactions can also be encoded efficiently \cite{crosswhite_applying_2008, pirvu_matrix_2010}. As the same is not true of arbitrary long-range interactions, we follow Refs. \cite{crosswhite_applying_2008, pirvu_matrix_2010, frowis_tensor_2010} in approximating power laws by sums of exponentials, giving $m = n_\text{H} \, n_\text{exps} + 2$, where $n_H$ is the number of long-range terms in the Hamiltonian, and $n_\text{exps}$ is the number of exponentials used in the approximation. In this paper we use the algorithm described in Ref. \cite{pirvu_matrix_2010}. An alternative method \cite{levenberg_method_1944,marquardt_algorithm_1963, noauthor_least-squares_nodate} is discussed in the Supplemental Material \cite{supp_material}.
\subsection{Time-dependent variational principle}
The McLachlan formulation of the Dirac-Frenkel-McLachlan time-dependent variational principle (TDVP) \cite{mclachlan_variational_1964, broeckhove_equivalence_1988} approximates the time evolution of a state $\ket{\psi}$ under the Hamiltonian $H$ by minimizing
\begin{equation}
    \left\lVert \,\mathrm{i} \diff{}{t}\Ket{\psi(t)} - H \Ket{\psi(t)} \right\rVert _2^2,
\end{equation}
with $\ket{\psi}$ kept fixed while its derivative is varied (note that we set $\hbar=1$ throughout this paper). Assuming the set of MPS with a given uniform bond dimension to be a smooth manifold (proven in Refs. \cite{holtz_manifolds_2012, uschmajew_geometry_2013, haegeman_geometry_2014}), Haegeman \textit{et al.} used this variational principle to derive a novel algorithm for real and imaginary time evolution \cite{haegeman_time-dependent_2011}.

More recently, an improved TDVP algorithm was derived for finite MPS with open boundaries, which relies on the mixed canonical gauge \cite{lubich_time_2015, haegeman_unifying_2016}. This approach leads to an effective Schr\"odinger equation for states constrained to the MPS manifold,
\begin{equation}
    \mathrm{i} \diff{}{t}\Ket{\psi(t)} = P_{T_{\ket{\psi}}} H \Ket{\psi(t)},
\end{equation}
where $P_{T_{\ket{\psi}}}$ is an orthogonal projector onto the tangent space of $\ket{\psi(t)}$. The essence of this method is that the tangent space projector can be decomposed as
\begin{eqnarray}\label{eq:proj1site}
P_{T_{\ket{\psi}}} & = & \sum_{j = 1}^{N} P_{L}^{[1:j-1]} \otimes \mathds{1}_{j} \otimes P_{R}^{[j+1:N]} \nonumber \\*
& & - \sum_{j = 1}^{N-1} P_{L}^{[1:j]} \otimes P_{R}^{[j+1:N]},
\end{eqnarray}
where
\begin{eqnarray}\label{eq:projleftandright}
 P_{L}^{[1:j-1]} & = & \sum_{k = 1}^{\chi_{j-1}} \Ket{\Phi_{L, k}^{[1:j-1]}} \Bra{\Phi_{L, k}^{[1:j-1]}}, \nonumber \\*
 P_{R}^{[j+1:N]} & = & \sum_{l = 1}^{\chi_{j+1}} \Ket{\Phi_{R, l}^{[j+1:N]}} \Bra{ \Phi_{R, l}^{[j+1:N]}},
\end{eqnarray}
meaning
\begin{equation}
    \ket{\psi(t+\delta t)} = \exp\left(-\mathrm{i} P_{T_{\ket{\psi}}} H \delta t\right) \ket{\psi(t)}
\end{equation}
can be approximated by applying a Lie-Trotter-Suzuki decomposition \cite{hatano_finding_2005} to the exponential. Consequently, one can sweep back and forth along the MPS (as in single-site DMRG \cite{white_density_2005, dolgov_alternating_2014, hubig_strictly_2015}), time evolving one site tensor at a time. This algorithm, which we refer to as 1TDVP, is symplectic, so conserves the energy and norm of a state. However, it also restricts MPS to a fixed bond dimension.

To overcome this limitation, Haegeman \textit{et al.} introduced a two-site variant (2TDVP) that similarly relies on the mixed canonical gauge \cite{haegeman_unifying_2016}. In 2TDVP, the tangent space projector of \eqr{eq:proj1site} is replaced by
\begin{eqnarray}\label{eq:proj2site}
P_{T^{[2]}_{\ket{\psi}}} & = & \sum_{j = 1}^{N-1} P_{L}^{[1:j-1]} \otimes \mathds{1}_{j} \otimes \mathds{1}_{j+1} \otimes P_{R}^{[j+2:N]} \nonumber \\*
& & - \sum_{j = 2}^{N-1} P_{L}^{[1:j-1]} \otimes \mathds{1}_{j} \otimes P_{R}^{[j+1:N]}.
\end{eqnarray}
Because MPS of different bond dimension do not belong to the same manifold, it is no longer possible to describe the time evolution of the entire MPS by a differential equation. Instead, Haegeman \textit{et al.} use a symmetric second-order Lie-Trotter-Suzuki decomposition with a discrete timestep $\delta t$ to arrive at an algorithm with the same sweeping pattern as the original two-site DMRG\footnote{Higher-order decompositions entailing more sweeps are also possible \cite{haegeman_unifying_2016}.}. Being able to dynamically vary the bond dimension makes 2TDVP particularly convenient. It has also been shown to give accurate results for a range of problems \cite{jaschke_open_2018,paeckel_time-evolution_2019}, so it is this variant we consider here.
\section{Parallel Two-site TDVP}\label{sec:the-algorithm}
In this section we introduce the parallel two-site TDVP (p2TDVP) algorithm. As a preliminary, we describe how serial 2TDVP can be carried out in the inverse canonical gauge. This is mathematically equivalent to the usual formulation given in the literature but, crucially, allows the algorithm to be parallelized. We then describe how this parallelization is carried out, and finally discuss the truncation of singular values and the algorithm's stability.
\subsection{Serial algorithm}\label{sec:serial-algorithm}
A single timestep in 2TDVP comprises a sweep from left to right followed by a sweep from right to left. During each of these two sweeps the MPS is evolved forwards in time by half a timestep. Here we only explicitly describe the left-to-right sweep, illustrated in \fir{figserial}, as the right-to-left sweep is equivalent, with just the direction reversed. Both sweeps are described formally in Appendix {\ref{appendix-serial}}.
\begin{figure}
  \includegraphics[height=6.6cm,keepaspectratio]{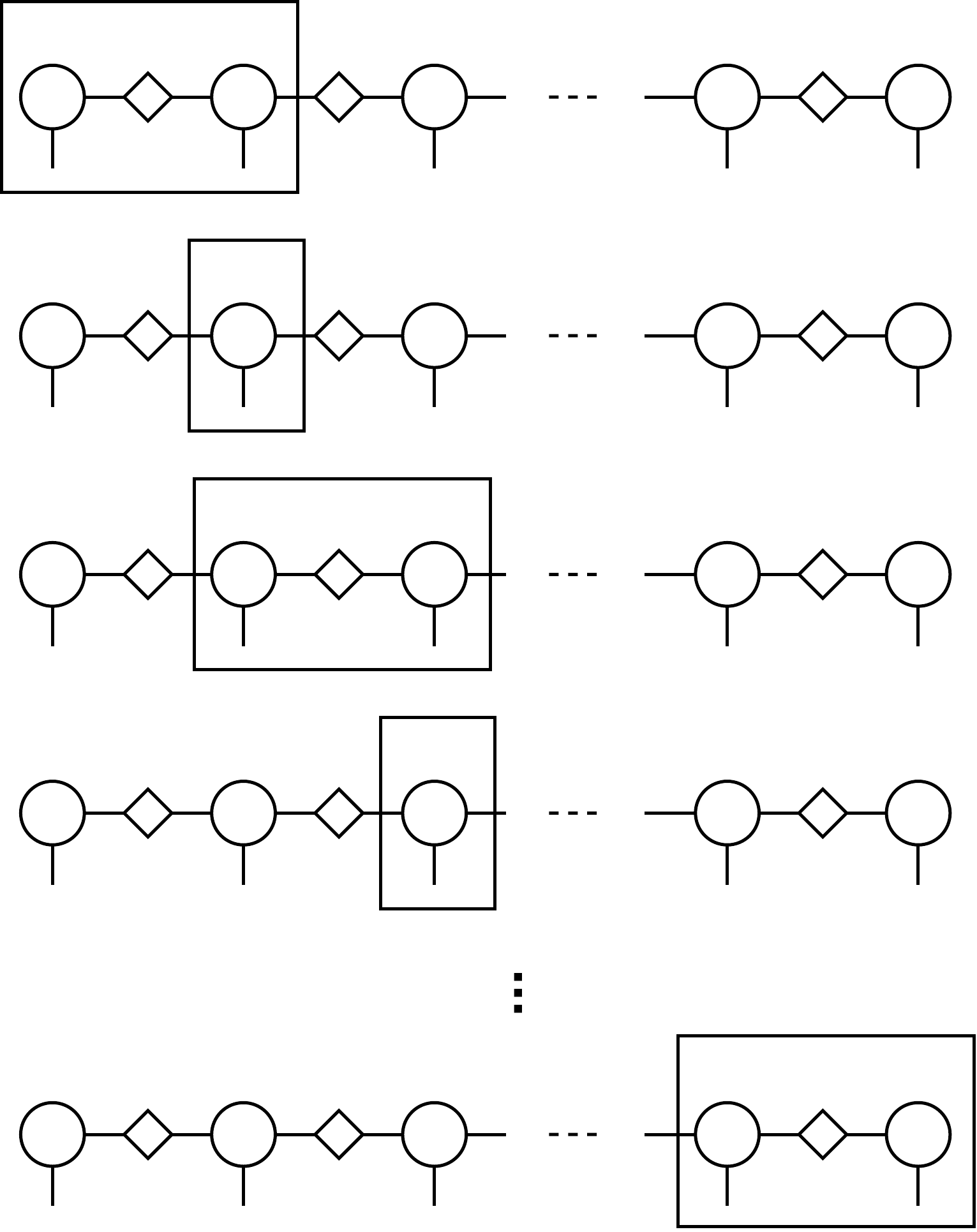}
  \caption{\label{figserial}Left-to-right sweep in serial two-site TDVP using the inverse canonical gauge. Pairs of site tensors are evolved forwards in time by half a timestep and single site tensors are evolved backwards in time by half a timestep.}
\end{figure}

\begin{figure}
	\subfloat[\label{figenvironments:left}]{
		\includegraphics[width=6.4cm,keepaspectratio]{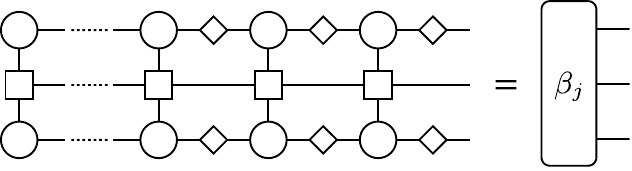}
	}\\
	\subfloat[\label{figenvironments:right}]{
		\includegraphics[width=6.4cm,keepaspectratio]{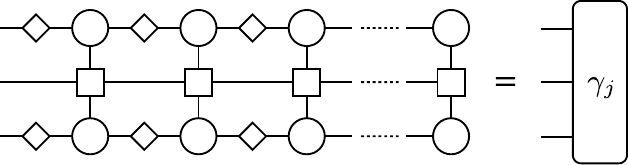}
	}
	\caption{\label{figenvironments}\protect\subref{figenvironments:left} The effective left environment $\beta_j$, and \protect\subref{figenvironments:right} the effective right environment $\gamma_j$, for an MPS site tensor $\Psi_j$. $\beta_j$ is formed from all MPS tensors to the left of $\Psi_j$, along with their Hermitian conjugates. The corresponding Hamiltonian MPO tensors are sandwiched in between. $\gamma_j$ is formed similarly, but from the tensors to the right of $\Psi_j$.}
\end{figure}

Starting from the left, the algorithm proceeds as follows: sites 1 and 2 are evolved forwards in time by half a timestep; site 2 is evolved backwards in time by half a timestep; sites 2 and 3 are evolved forwards in time by half a timestep; site 3 is evolved backwards in time by half a timestep; and so on. This pattern of local updates continues until the rightmost pair of sites is reached. The final step is to evolve this pair of sites forwards in time by half a timestep, with no backwards time evolution necessary. The whole process is then reversed and the algorithm sweeps back from right to left. As the second sweep is simply the mirror image of the first, it begins by evolving the rightmost pair of sites forwards in time by another half a timestep. One can therefore evolve the rightmost pair just once by a full timestep instead of two half timesteps (see Appendix \ref{appendix-serial}).

The one and two-site time evolution steps rely on ``effective environments'', which are the same as in DMRG. Each site tensor $\Psi_j$ has a left and a right effective environment, labeled $\beta_j$ and $\gamma_j$, respectively. These are defined in \fir{figenvironments}. The leftmost and rightmost MPS environments ($\beta_1$ and $\gamma_N$) are trivial, corresponding to $1\times1$ identity matrices. It is important to note that the effective environments need not be created from scratch at every step since previous environments can be cached and updated. At the beginning of a simulation, all righthand environments $\gamma_1 \dots \gamma_N$ are created iteratively from right to left.

\begin{figure}
  \subfloat[\label{figupdate:twosite}]{
  \includegraphics[width=6cm,keepaspectratio]{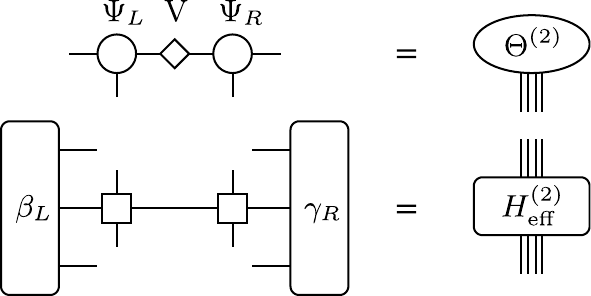}
  }\\
  \subfloat[\label{figupdate:onesite}]{
  \includegraphics[width=5cm,keepaspectratio]{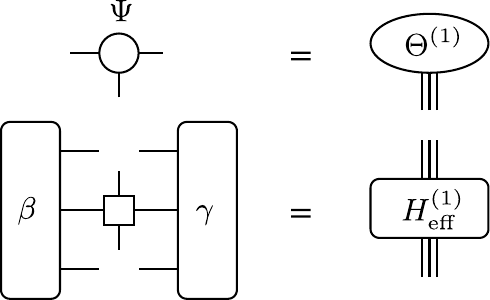}}
  \caption{\label{figupdate}Definition of the local state and its effective Hamiltonian for \protect\subref{figupdate:twosite} the two-site update, and \protect\subref{figupdate:onesite} the one-site update. The effective Hamiltonian is formed by connecting the left and right effective environments to the Hamiltonian MPO tensor(s) corresponding to the site(s) being updated. When carrying out a two-site update we use the $L$ and $R$ subscripts to refer to the left site and right site, respectively.}
\end{figure}

To time evolve two sites, $\Psi_{j}$ and $\Psi_{j+1}$, we construct an effective local state $\Theta^{(2)}$ and an effective two-site Hamiltonian $H^{(2)}_{\text{eff}}$. These are described in \subfir{figupdate:twosite}. Evolving $\Theta^{(2)}$ forwards in time means calculating
\begin{equation}
  \Theta^{(2)\prime} = \exp\left(-iH^{(2)}_{\mathrm{eff}}\delta t/2\right) \Theta^{(2)}.
\end{equation}
Using the Lanczos exponentiation \cite{park_unitary_1986} method\footnote{An alternative is the recent algorithm from Al-Mohy and Higham based on truncated Taylor series \cite{al-mohy_computing_2011}.} means that $H^{(2)}_{\mathrm{eff}}$ is not explicitly required, only $H^{(2)}_{\mathrm{eff}} \Theta^{(2)}$, so a more efficient tensor contraction pattern can be employed. \fir{figpostupdate} explains how $\Theta^{(2)\prime}$ is split back into two site tensors using the singular value decomposition (SVD). Here the smallest singular values are discarded to keep the bond dimension from growing too large (see \secr{sec:error}). After this two-site update a new lefthand environment $\beta_{j+1}$ is created from the contraction of $\beta_{j}$, $\Psi_{j}$, $\mathrm{V}_j$, $\Psi_{j}^{\dagger}$, $\mathrm{V}_j$, and the Hamiltonian MPO tensor for site $j$, i.e.
\begin{equation}
\def\svgwidth{4.2cm}
\begingroup%
  \makeatletter%
  \providecommand\color[2][]{%
    \errmessage{(Inkscape) Color is used for the text in Inkscape, but the package 'color.sty' is not loaded}%
    \renewcommand\color[2][]{}%
  }%
  \providecommand\transparent[1]{%
    \errmessage{(Inkscape) Transparency is used (non-zero) for the text in Inkscape, but the package 'transparent.sty' is not loaded}%
    \renewcommand\transparent[1]{}%
  }%
  \providecommand\rotatebox[2]{#2}%
  \newcommand*\fsize{\dimexpr\f@size pt\relax}%
  \newcommand*\lineheight[1]{\fontsize{\fsize}{#1\fsize}\selectfont}%
  \ifx\svgwidth\undefined%
    \setlength{\unitlength}{333.75bp}%
    \ifx\svgscale\undefined%
      \relax%
    \else%
      \setlength{\unitlength}{\unitlength * \real{\svgscale}}%
    \fi%
  \else%
    \setlength{\unitlength}{\svgwidth}%
  \fi%
  \global\let\svgwidth\undefined%
  \global\let\svgscale\undefined%
  \makeatother%
  \begin{picture}(1,0.41123596)%
    \lineheight{1}%
    \setlength\tabcolsep{0pt}%
    \put(0,0){\includegraphics[width=\unitlength,page=1]{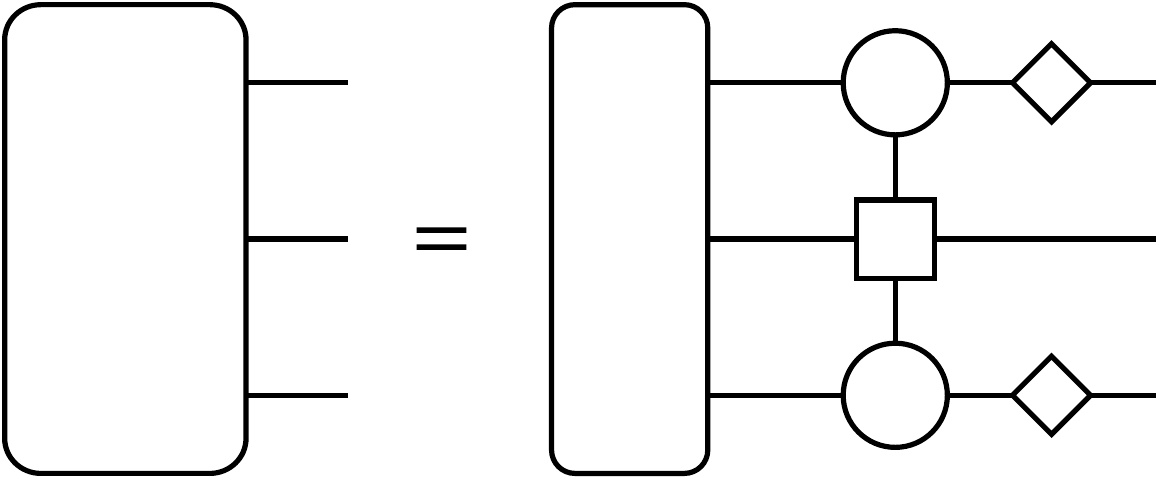}}%
    \put(0.03048596,0.18380355){\color[rgb]{0,0,0}\makebox(0,0)[lt]{\lineheight{1.25}\smash{\begin{tabular}[t]{l}$\beta_{j+1}$\end{tabular}}}}%
    \put(0.50239608,0.18380355){\color[rgb]{0,0,0}\makebox(0,0)[lt]{\lineheight{1.25}\smash{\begin{tabular}[t]{l}$\beta_j$\end{tabular}}}}%
  \end{picture}%
\endgroup%
.
\end{equation}

\begin{figure}
  \includegraphics[width=8.6cm,keepaspectratio]{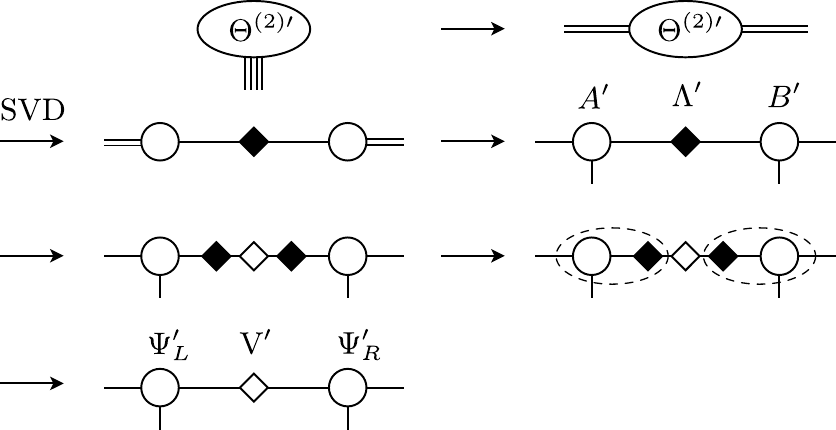}
  \caption{\label{figpostupdate}The re-splitting of $\Theta^{(2)\prime}$ following a two-site update. First $\Theta^{(2)\prime}$ is reshaped into a matrix. Then a truncated singular value decomposition (SVD) is performed, giving $A' \Lambda' B'$. $A'$ and $B'$ are reshaped into site tensors, and $\mathrm{V}'\Lambda'$ is inserted between $\Lambda'$ and $B'$. Finally, $A'\Lambda'$ is contracted to form $\Psi_L'$, and $\Lambda' B'$ is contracted to form $\Psi_R'$.}
\end{figure}

Time evolving one site $\Psi_{j+1}$ requires the construction of an effective one-site Hamiltonian $H^{(1)}_\textrm{eff}$ from $\beta_{j+1}$ and $\gamma_{j+1}$. This is described in \subfir{figupdate:onesite}. The local state $\Theta^{(1)}$ is now just the vectorization of $\Psi_{j+1}$. To time evolve $\Theta^{(1)}$ backwards in time means calculating
\begin{equation}
    \Theta^{(1)\prime} = \exp\left(+iH^{(1)}_{\mathrm{eff}}\delta t/2\right) \Theta^{(1)},
\end{equation}
but again only $H^{(1)}_{\mathrm{eff}} \Theta^{(1)}$ is actually required by the Lanczos routine. After this one-site update, $\gamma_{j+1}$ can be discarded.

A timestep in 2TDVP has the same time complexity as a sweep in two-site DMRG. The most expensive operation is the contraction of the network representing $H^{(2)}_{\mathrm{eff}} \Theta^{(2)}$, giving a bound of
\begin{equation}\label{eqnscaling}
\mathcal{O}\left( N \chi^3 m d^2 + N \chi^2 m^2 d^3 \right).
\end{equation}
This means that systems with many degrees of freedom (e.g. bosonic systems) can be very demanding. Large systems with long-range interactions are also challenging because of the size of the bond dimension required by the Hamiltonian MPO (especially if it contains multiple long-range terms). These considerations motivate the need for a parallel algorithm.
\subsection{Parallelization}\label{sec:parallel-algorithm}
The intuition behind the parallel algorithm is the fact that local updates approximately preserve the inverse canonical gauge for small $\delta t$. We thus parallelize 2TDVP by carrying out these local updates simultaneously on separate processes. More concretely, we split the $N$-site MPS into $p$ partitions, which are updated in parallel. We use a message-passing parallel programming model and assign each partition to a separate process. A partition must contain a minimum of two site tensors, so we let $p$ be an even number between 2 and $N/2$. The full sweeps of the serial algorithm are replaced by partial sweeps carried out in parallel; each process simultaneously sweeps along the tensors in its own partition following the pattern introduced by Stoudenmire and White for parallel DMRG \cite{stoudenmire_real-space_2013}. This is illustrated in \fir{figpartitioning} for two and four processes. Notice that the sweeping direction alternates for each neighboring partition. The two central processes always sweep away from the center of the MPS during the first half of a timestep.

\begin{figure}
  \subfloat[\label{figpartitioning:2procs}]{
  \includegraphics[width=8.3cm,keepaspectratio]{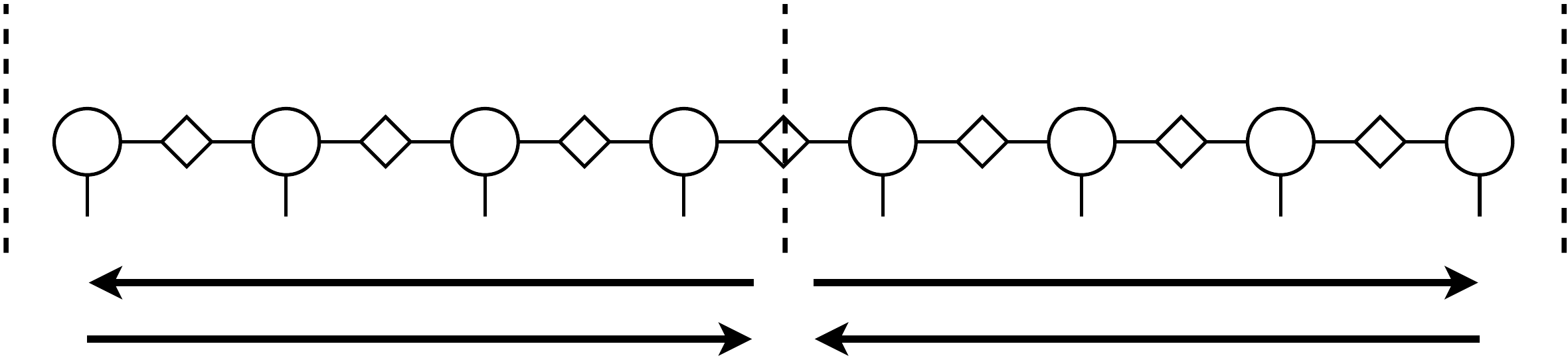}
  }\\
  \subfloat[\label{figpartitioning:4procs}]{
  \includegraphics[width=8.3cm,keepaspectratio]{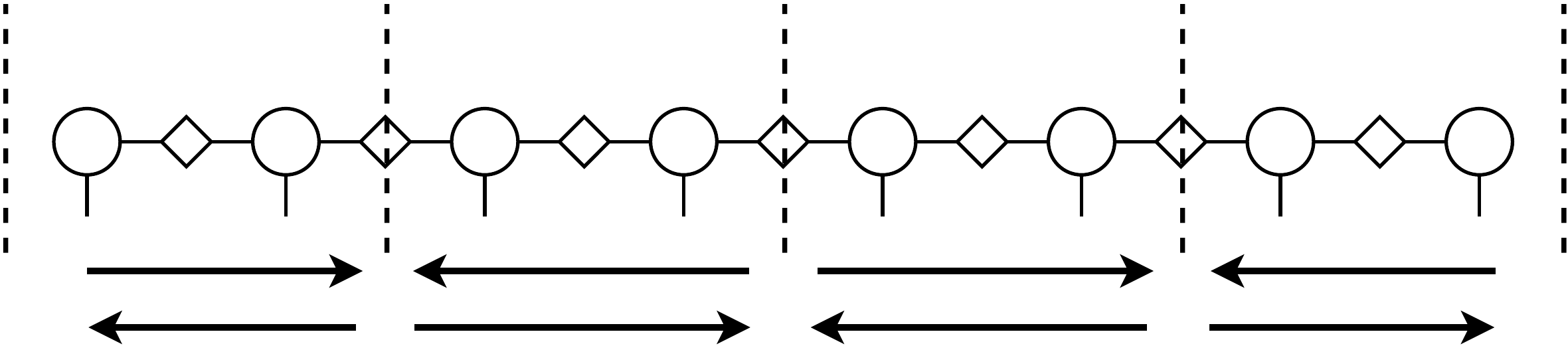}
  }
  \caption{\label{figpartitioning}Partitioning of MPS and sweeping pattern for parallel 2TDVP with \protect\subref{figpartitioning:2procs} two, and \protect\subref{figpartitioning:4procs} four processes. The dashed lines represent partition boundaries. At the start of a timestep the sweeps proceed away from the center of the MPS, with each neighboring partition sweeping in the opposite direction (top arrows). For the second half of the timestep the sweeps are reversed (bottom arrows).\\}
\end{figure}

At the start of a simulation, the necessary initial effective environments are computed sequentially, with each being assigned to the process owning the corresponding site tensor. Processes that start by sweeping right will require righthand environments and vice-versa for those that start by sweeping left.

When sweeps reach a partition boundary, it is necessary for neighboring processes to communicate. Firstly, the processes need to exchange boundary environments, and secondly, one of the processes needs the $\Psi$ tensor belonging to its neighbor in order to carry out the two-site update. This communication involves sending $\mathcal{O}\left(\chi^2 \left(m + d\right)\right)$ floating-point numbers. We let the lefthand process update the boundary sites while the righthand process waits. The lefthand process then sends the updated tensors to the righthand process and both processes update their respective effective environments. \fir{figparallel} illustrates how the local updates proceed in parallel away from partition boundaries. We describe the algorithm formally in Appendix \ref{appendix-parallel}.

\begin{figure}
  \includegraphics[height=6.5cm,keepaspectratio]{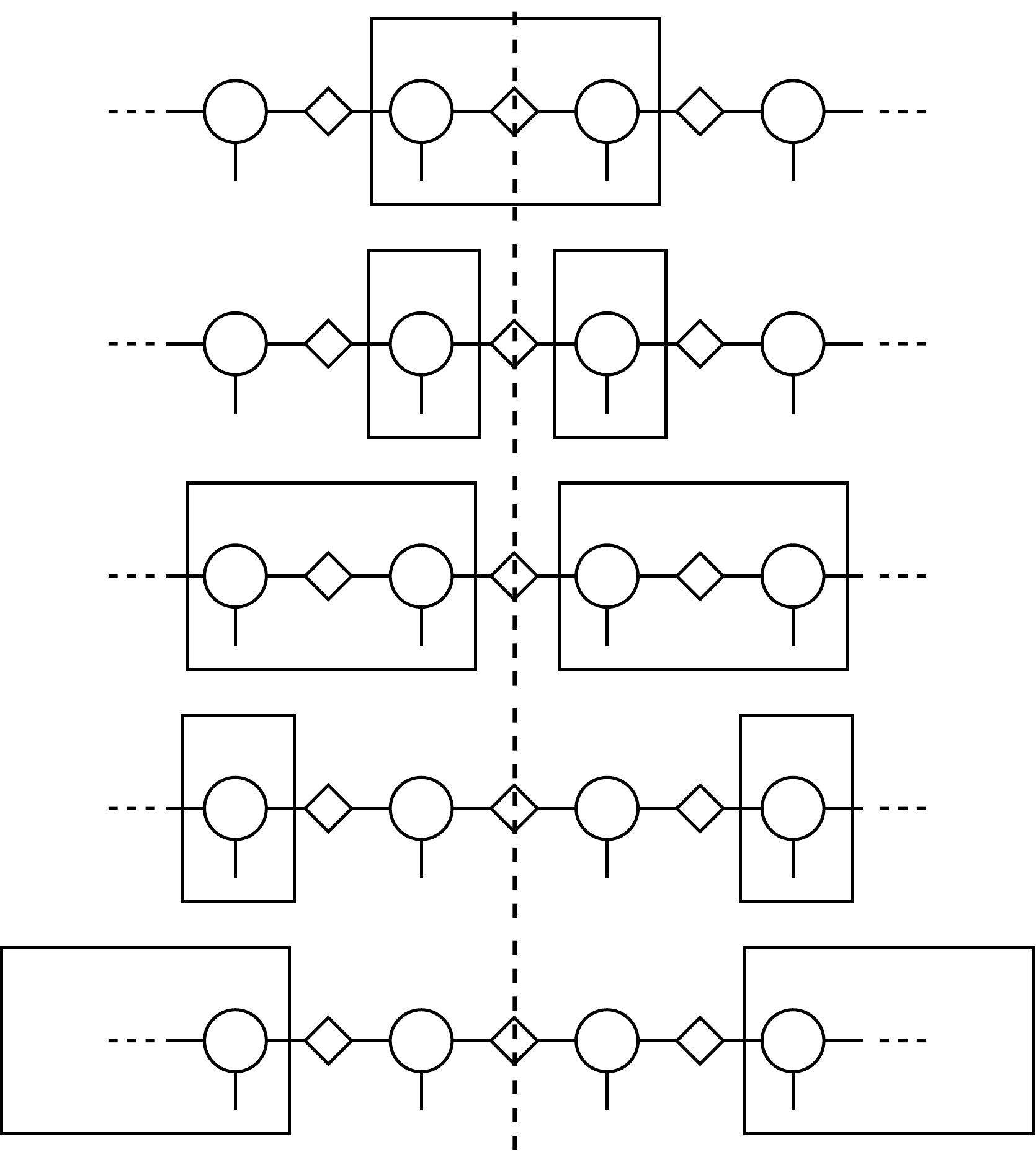}
  \caption{\label{figparallel}Time evolving local sites in parallel 2TDVP. Two neighboring processes are shown sweeping away from their shared boundary (represented by a dashed line). As in \fir{figserial}, pairs of sites are evolved forwards in time by half a timestep and single sites are evolved backwards in time by half a timestep.\\}
\end{figure}
\subsection{Truncation of singular values}\label{sec:error}
Serial 2TDVP has four sources of error: the projection, the Lie-Trotter-Suzuki decomposition, the local integration, and the truncation of Schmidt coefficients. For a detailed discussion we refer the reader to Ref. \cite{paeckel_time-evolution_2019}. The dependence of the different errors on the MPS bond dimension and the choice of $\delta t$ is rather subtle \cite{jaschke_open_2018} but, if these parameters are chosen with care, it is the truncation and projection errors that usually dominate due to the growth of bipartite entanglement \cite{goto_performance_2019}. The projection error can be prohibitively expensive to compute, especially for long-range models \cite{paeckel_time-evolution_2019, hubig_error_2018}, but the truncation error is simply calculated from the discarded singular values, as in TEBD \cite{daley_time-dependent_2004}.

In both the serial and parallel algorithms, we quantify the truncation error due to the single SVD shown in \fir{figpostupdate} using the discarded weight
\begin{equation}\label{eq:discweight}
        w = \sum_{j=\chi + 1}^{\chi'} \lambda_j^2,
\end{equation}
where $\chi'$ is the full rank of the matrix $\Theta^{(2)\prime}$, $\lambda_j$ are its singular values (sorted in order of descending magnitude), and $\chi$ is the truncated rank. We choose $\chi$ as follows. First, we define a truncation error tolerance $w_\text{max}$, which is the maximum allowed discarded weight per SVD. We then find the minimum rank $\chi_w$ such that $w \leq w_\text{max}$. Finally, we set $\chi = \min(\chi_w, \chi_\text{max})$.

The total discarded weight $w_\text{total}$ is defined as the cumulative sum of $w$ over all SVDs, over all timesteps. In the worst case, $w_\text{total}$ will grow exponentially due to a linear growth of bipartite entanglement entropy \cite{schuch_entropy_2008}. However, long-range models can exhibit a logarithmic growth of entanglement entropy, even when this growth is linear in the corresponding local model \cite{schachenmayer_entanglement_2013, buyskikh_entanglement_2016, singh_effect_2017, lerose_origin_2018, liu_confined_2019}, meaning $w_\text{total}$ will grow as a power law.
\subsection{Timestep and stability}\label{sec:stability}
The parallelization of 2TDVP introduces two further sources of error:
\begin{enumerate}[i), leftmargin=\parindent]
  \item Information about lattice sites propagates along the MPS at a finite speed meaning that each process will always be using at least one ``out-of-date'' local environment. For nonlocal 1D models, this induces an artificial locality, since instantaneous long-range interactions become effectively retarded. For $p$ parallel processes, we expect this error to be small if the characteristic velocity $v$ of the dynamics satisfies
  \begin{equation}
    v \ll (N/p) / \delta t,
  \end{equation}
  where we have assumed that each parallel partition contains $\sim (N/p)$ sites.
  \item Like all parallel MPS algorithms, the local updates in p2TDVP formally break the global gauge conditions, meaning the inverse canonical form only holds approximately. In serial TDVP the inverse canonical gauge is also technically broken, but the orthogonality of the state is preserved with respect to the last updated site (which remains an orthogonality center). In parallel TDVP this orthogonality may be lost, since different parts of the MPS are updated simultaneously.
\end{enumerate}
These errors can be controlled by reducing $\delta t$ or decreasing the number of parallel processes, meaning that it should be possible to converge a calculation, as is typically done with serial MPS algorithms. For moderate values of $\delta t$, however, the breaking of the gauge conditions at partition boundaries can cause the p2TDVP algorithm to become unstable if very small singular values are kept. To circumvent this, we define a relative SVD truncation tolerance $\varepsilon$. We discard singular values smaller than $\varepsilon \lambda_1$ (where $\lambda_1$ is the largest singular value), in addition to carrying out the truncation procedure described in \secr{sec:error}.
\begin{figure}
  \includegraphics[width=8.6cm,keepaspectratio]{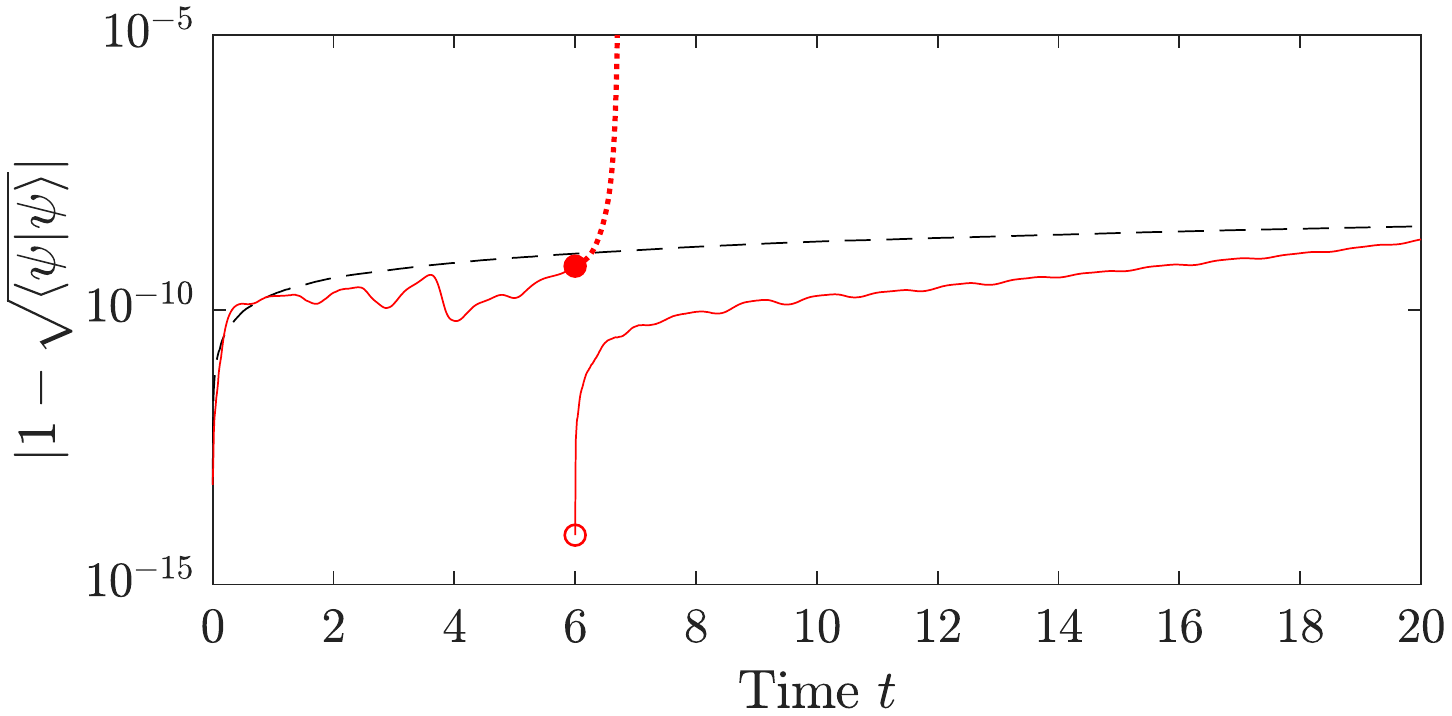}
  \caption{\label{figreorth}Error in the norm for a 641-site spin chain evolved using 32 processes under the long-range Ising Hamiltonian with $\alpha = 2.3$, following the quench described in \secr{ising-benchmark}. The solid curve shows the result for $\delta t=0.01$, $ \varepsilon = 10^{-8}$. After about 600 timesteps, the error starts to grow exponentially (dotted curve). If the MPS is reorthonormalized before this, its norm is brought back under control (righthand solid curve). At the end of the calculation $\chi=116$ and $w_\text{total}=7.0\times10^{-10}$. The dashed curve shows the result for $\delta t=0.002$, $ \varepsilon = 10^{-12}$. At the end of this calculation, $\chi=90$ and $w_\text{total}=8.8 \times 10^{-10}$. Both calculations used $m=22$, and $w_\text{max} = 10^{-16}$.
  }
\end{figure}

If the error in the norm grows unacceptably large during a simulation, the MPS can also be reorthonormalized and the effective environments recomputed. As this is a serial procedure, it should be carried out infrequently to avoid affecting the algorithm's parallel efficiency. Note that it is, however, important to ensure the initial state is orthonormal.

For the benchmark calculations described in \secr{sec:the-benchmarks}, a value of $\varepsilon=10^{-12}$ was sufficient, with no reorthonormalization necessary. However, the appropriate value of $\varepsilon$ depends on the system and choice of timestep. In \fir{figreorth}, we describe a p2TDVP simulation carried out on a 641-site spin chain using different values of $\delta t$, and $\varepsilon$. With $\delta t=0.01$, and $ \varepsilon = 10^{-8}$, the error in the norm is seen to blow up. However, reorthonormalization brings it back under control. In comparison, the calculation with $\delta t=0.002$, and $ \varepsilon = 10^{-12}$ remains stable, taking 2.4 times longer to run.
\section{Benchmarks}\label{sec:the-benchmarks}
In this section we describe the results of our numerical experiments. To test p2TDVP, we carried out benchmark calculations on spin-half models with one, two, and three long-range interaction terms. We utilized up to 32 processes, with one process assigned per compute node. Each compute node used up to 16 threads (e.g. for linear algebra operations). Full details of the test platform \cite{noauthor_balena_nodate}, software used \cite{noauthor_tntlibrary_nodate, goodyer_tnt_nodate, al-assam_tensor_2017, coulthard_engineering_2018, matlab_r2017_nodate, lehoucq_arpack_nodate, cornet_opencollab/arpack-ng_2019, ajolleyx_intel_nodate, mpi_intel_nodate, lapack, cuppen_divide_1980, gu_stable_1994, Rutter:CSD-94-799, noauthor_numerically_nodate}, and simulation parameters are provided in the Supplemental Material \cite{supp_material}.

To quantify the error introduced by the parallelization, we run our benchmark simulations on a single process and calculate the difference in the observables of interest. We also compute the infidelity,
\begin{equation}
    I \left( \ket{\psi_1}, \ket{\psi_2} \right) \equiv 1 - \frac{\left| \Braket{\psi_1|\psi_2} \right|}{\sqrt{\Braket{\psi_1|\psi_1}\Braket{\psi_2|\psi_2}}},
\end{equation}
between the serial and parallel MPS at the end of the calculations, which bounds the error in all observables.

When truncation is the dominant source of error, $I \sim w_\text{total}$ \cite{verstraete_matrix_2006, oseledets_tensor-train_2011}. Under this condition $w_\text{total}$ can be used as a proxy for $I$. In general, however, we would expect $w_\text{total}$ to provide a lower bound as it does not account for the projection error.

In the following, we define spin Hamiltonians in terms of Pauli-$X$, -$Y$, and -$Z$ operators $\sigma_i^x$, $\sigma_i^y$, and $\sigma_i^z$ (where $i$ is the index of the spin), and set the interaction strengths (and hence the energy and time scales) to unity.
\subsection{Long-range Ising model}\label{ising-benchmark}
\begin{figure}
  \includegraphics[width=8.5cm,keepaspectratio]{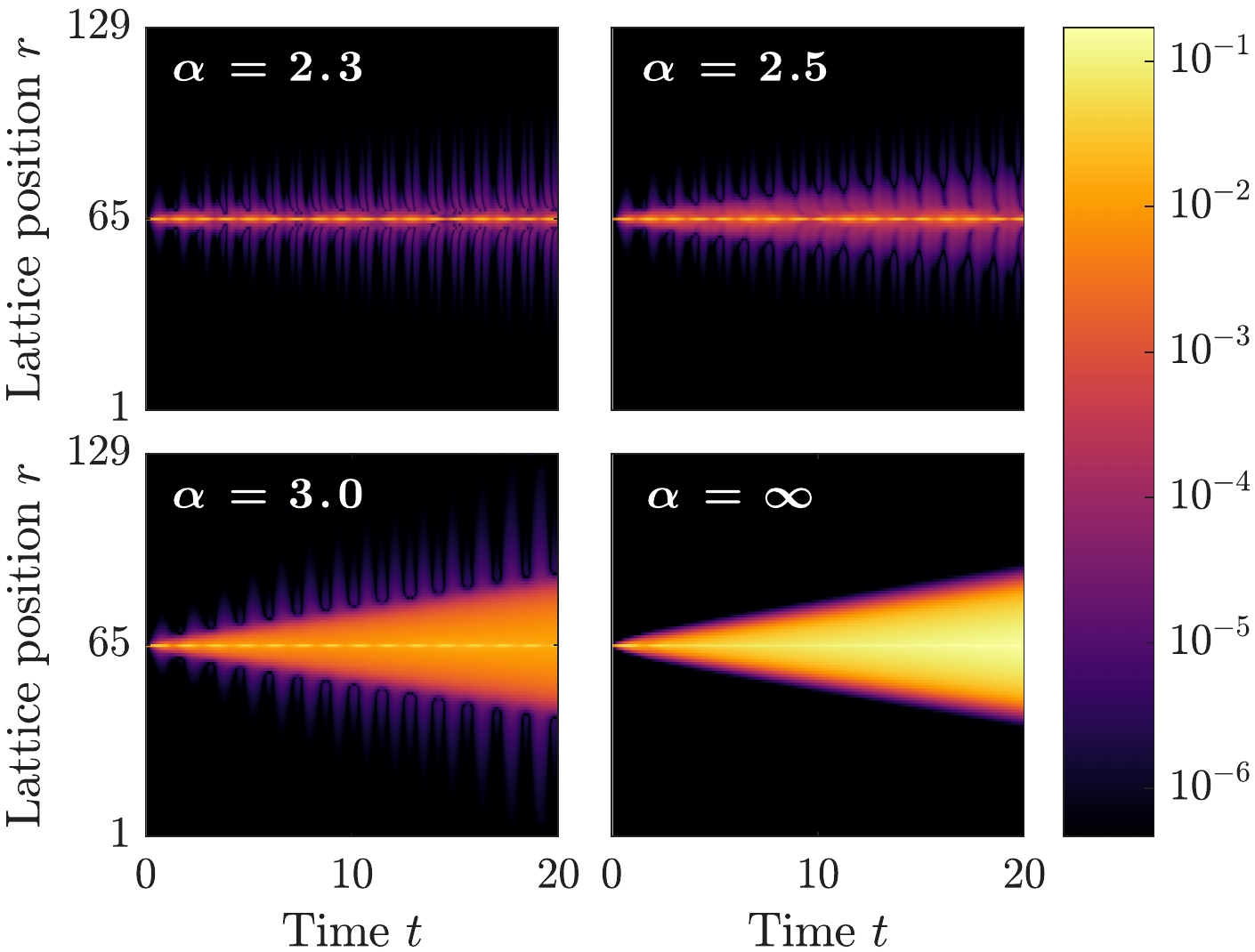}
  \caption{\label{figisinglog}Density plots of the correlation function $C_{r, 65}(t)$ for a 129-site chain evolved under the long-range Ising Hamiltonian for four values of $\alpha$. The data were calculated using p2TDVP with 32 processes. All values deviate from the serial calculations by less than $1\%$ (less than $0.2\%$ for the more accurate $\alpha=\infty$ case).
}
\end{figure}
\begin{figure*}
  \subfloat[\label{figxylog:data}]{
  \includegraphics[height=3.7cm,keepaspectratio]{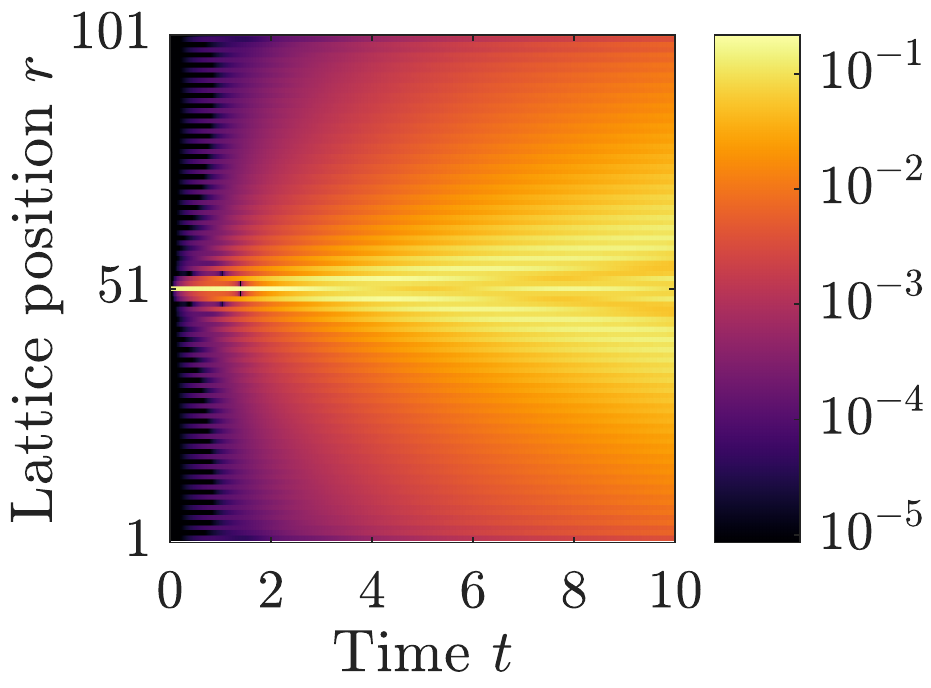}
  }
  \hspace{0.1cm}
  \subfloat[\label{figxylog:correlations}]{
  \includegraphics[height=3.7cm,keepaspectratio]{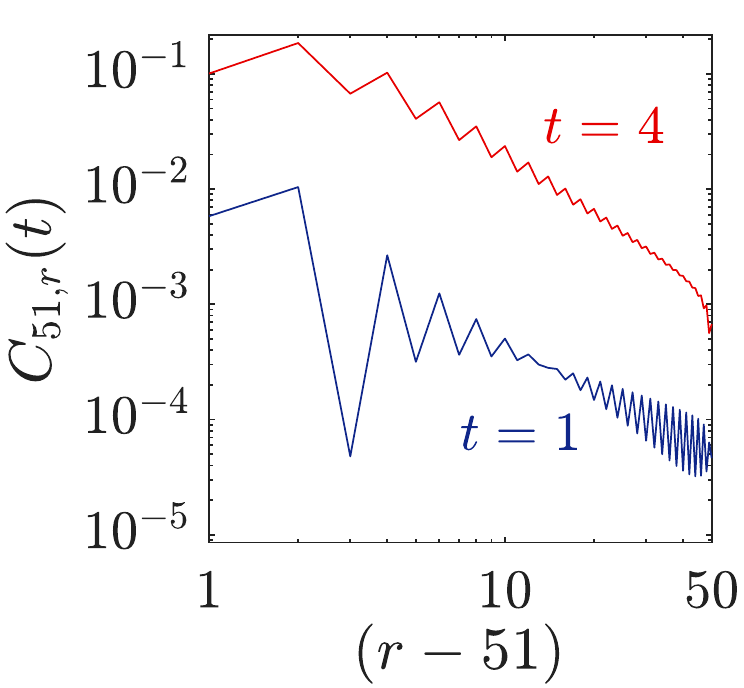}
  }
  \hspace{0.6cm}
  \subfloat[\label{figxylog:error}]{
  \includegraphics[height=3.7cm,keepaspectratio]{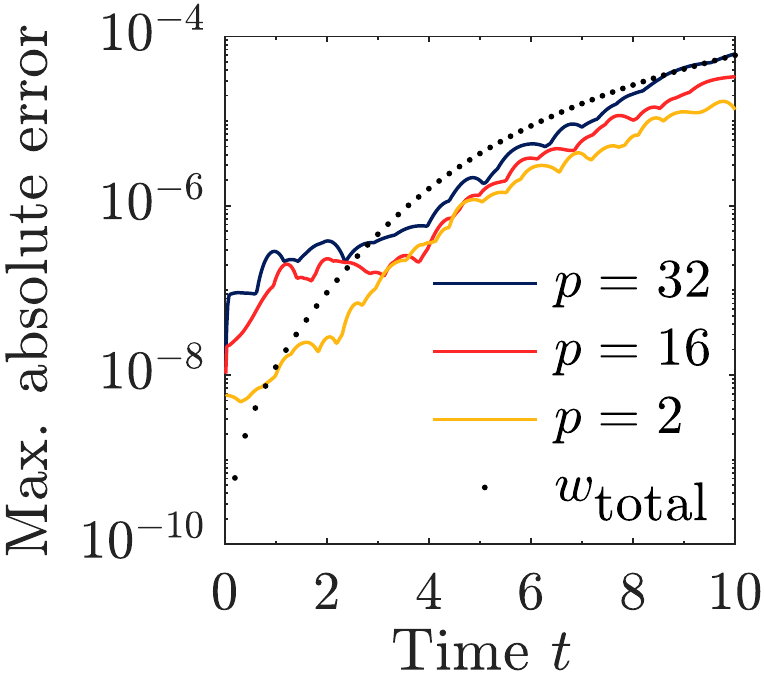}
  }
  \caption{\label{figxylog}Time evolution of the 101-site long-range XY spin chain with $\alpha = 0.75$. All calculations were carried out using $\delta t = 0.02$, $\varepsilon = 10^{-12}$, $\chi_\text{max}=256$, $m=28$, and no truncation error tolerance. \protect\subref{figxylog:data} Density plot of $C_{51,r}(t)$, calculated using 32 processes. Above the cutoff of $8.575\times10^{-6}$, all values deviate from the serial calculation by less than $1\%$ (less than $0.1\%$ after $t=2.24$). \protect\subref{figxylog:correlations} Log-log plot of $C_{51,r}(t)$ at times $t=1$ and $t=4$, calculated as in \protect\subref{figxylog:data}. On this scale, differences from the serial calculation are not visible. \protect\subref{figxylog:error} Maximum absolute deviation of $C_{51,r}(t)$ from the serial calculation for $p$ processes. $w_\textrm{total}$ from the serial calculation is shown for comparison.
  }
\end{figure*}

Our first benchmark looks at the spreading of correlations after a global quench in the ferromagnetic phase of the transverse field Ising model with short to intermediate-range interactions ($\alpha>2$). Denoting the transverse magnetic field by $B$, the Hamiltonian is
\begin{equation}\label{IsingHamiltonian}
    H = - \sum_{i<j}^N \frac{1}{\lvert i-j \rvert ^\alpha} \sigma_i^z \sigma_j^z - B \sum_{i=1}^N \sigma_i^x.
\end{equation}
We track the evolution of an equal-time spin-spin correlation function to see how well p2TDVP can capture a nonlocal observable. Here we follow Liu \textit{et al.} \cite{liu_confined_2019}, but note that this scenario was first studied using TDVP for the antiferromagnetic interaction case in Refs. \cite{hauke_spread_2013, buyskikh_entanglement_2016}.

Using a Krylov space method \cite{nauts_new_1983, luitz_ergodic_2017}, Liu \textit{et al.} accurately simulate the quench dynamics of a 19-site lattice with periodic boundaries. Starting from the ferromagnetic product state $\ket{\psi_0}=\ket{111\dots1}$, correlation confinement \cite{kormos_real-time_2017} is shown to arise due to the presence of power law interactions. This confinement is stronger the longer the range of the interactions. Stronger confinement is also shown to decrease the bipartite entanglement present, meaning the maximal bond dimension required for our simulations should decrease for smaller $\alpha$.

Liu \textit{et al.} note that correlation confinement persists when the initial state is a ferromagnetic ground state of $H$. We observe this behavior for lattices with open boundaries when applying a quench to the ground state of \eqr{IsingHamiltonian} with $\alpha=3.0$ and $B=0.1$. As in Ref. \cite{liu_confined_2019}, we quench to $B=0.27$ for various values of $\alpha$, and calculate the correlation function,
\begin{eqnarray}\label{IsingCorrelation}
    C_{r, k}(t) &=& \Braket{\psi(t) | \sigma_r^z \sigma_k^z | \psi(t)}
    \nonumber
    \\*
    && - \Braket{\psi(t) | \sigma_r^z | \psi(t)}\Braket{\psi(t) | \sigma_{k}^z | \psi(t)},
\end{eqnarray}
where $r$ is the lattice site index, $k$ is the index of the central lattice site, and $\ket{\psi(t)}$ is the state at time $t$ after the quench. As we consider chains with an odd number of spins, $k=(L+1)/2$.

\squeezetable
\begin{table}[t]
\begin{ruledtabular}
\begin{tabular}{@{}m{0.7cm}@{}m{0.7cm}@{}m{1cm}@{}m{1.31cm}@{}m{1.9cm}@{}m{1.85cm}@{}m{1.2cm}@{}}
 $\alpha$ & $m$ & $\chi_{\text{max}}$ & $w_\text{max}$ & $w_\text{total}$ & Infidelity ($I$) & Speedup
 \tabularnewline
 \\[-0.255cm]
 \hline
 \\[-0.15cm]
 2.3 & 15 & $128$ & $10^{-16}$ & $3.3 \times 10^{-11}$ & $3.8 \times 10^{-10}$ & 20.8 \tabularnewline
 2.5 & 14 & $192$ & $10^{-16}$ & $3.1 \times 10^{-11}$ & $3.4 \times 10^{-10}$ & 23.0 \tabularnewline
 3.0 & 13 & $256$ & $10^{-16}$ & $6.1 \times 10^{-11}$ & $5.6 \times 10^{-10}$ & 25.0 \tabularnewline
 $\infty$ & 3 & $512$ & $10^{-18}$ & $1.7 \times 10^{-13}$ & $4.2 \times 10^{-13}$ & 23.9
\end{tabular}
\end{ruledtabular}
\caption{Summary of results for the 129-site long-range Ising model at the end of 1000 timesteps (corresponding to $t=20$). Calculations were carried out using $\delta t = 0.02$, and $\varepsilon=10^{-12}$, on 1 and 32 processes. $\chi_{\text{max}}$ is the maximal MPS bond dimension, $m$ is the Hamiltonian MPO bond dimension, $w_\text{max}$ is the truncation error tolerance, and $w_\text{total}$ is the total discarded weight from the serial calculation. Speedups relative to the serial calculations are indicated in the last column.}
\label{tab:ising}
\end{table}

In \fir{figisinglog} we plot $C_{r, 65}(t)$ for $\alpha $ = 2.3, 2.5, 3.0, and for the nearest-neighbor case ($\alpha=\infty$). These results, computed using 32 processes, give excellent agreement with the serial calculations over many orders of magnitude. We see that the correlation confinement disappears as $\alpha \to \infty$, and a linear lightcone \cite{lieb_finite_1972, calabrese_time_2006, bravyi_lieb-robinson_2006, cevolani_universal_2018} is recovered. In fact a linear lightcone already seems to be present in the $\alpha = 3.0$ case, consistent with the bounds given in Refs. \cite{chen_finite_2019, kuwahara_strictly_2019}. Our numerical experiments lead us to conjecture that the lightcone is sublinear for $\alpha < 3$. The other striking difference between the local and nonlocal models is the existence of oscillatory long-range correlations in the latter, similar to Refs. \cite{hauke_spread_2013, buyskikh_entanglement_2016}. The relatively large system size used here allows us to conjecture that these correlations decay as a power-law with an exponent approximately equal to $\alpha$.

At the end of the simulations, we compute the infidelities $I$ between the serial and parallel calculations. In Table \ref{tab:ising} we show these for 32 processes, along with the total discarded weights $w_\textrm{total}$ from the end of the serial calculations. We find that the ratio $I/w_\textrm{total}$ grows with decreasing $\alpha$, from 2.5 for the nearest-neighbor case, to 11.5 for $\alpha=2.3$.

A strong scaling analysis for the $\alpha=2.3$ case is shown in \fir{figspeedup}. A speedup of 20.8 was achieved using 32 processes, although by this point the parallel efficiency drops below $70\%$. We find greater speedups for the simulations with larger bond dimensions (summarized in Table \ref{tab:ising} for 32 processes). This is to be expected as the computational complexity of the linear algebra operations asymptotically dominates over the parallel overheads.
\subsection{Long-range XY model}\label{xy-benchmark}
We next simulate a local quench in the antiferromagnetic XY model,
\begin{equation}\label{xyHamiltonian}
    H = \frac{1}{2} \sum_{i<j}^N \frac{1}{\lvert i-j \rvert ^\alpha} \left(\sigma_i^x \sigma_j^x + \sigma_i^y \sigma_j^y\right),
\end{equation}
with very long-range interactions ($\alpha < 1$). In this regime, information can spread through the system almost instantaneously \cite{hauke_spread_2013, eisert_breakdown_2013, gong_persistence_2014, storch_interplay_2015, luitz_emergent_2019}. This is an important test case as it is not \textit{a priori} clear how accurately p2TDVP will capture the nonlocal {\cite{hauke_spread_2013}} propagation of information from a single site.

Following Haegeman \textit{et al.} \cite{haegeman_unifying_2016, haegeman_notitle_nodate}, we calculate the ground state $\ket{\psi_0}$ of $H$ for a 101-site spin chain and apply a U(1) symmetry-breaking perturbation,
\begin{equation}
U = \exp\left(\mathrm{i} \pi \sigma_{51}^y / 4\right),
\end{equation}
to the central spin. We then examine the evolution of the single-site observable \cite{gong_persistence_2014, haegeman_unifying_2016}
\begin{equation}\label{xyCorrelation}
    C_{51,r}(t) = \left| \Braket{\psi(t) | \sigma_r^x | \psi(t)} - \Braket{\psi_0 | \sigma_r^x | \psi_0} \right|,
\end{equation}
where the perturbed state at time $t$ is given by
\begin{equation}
\ket{\psi(t)} = \mathrm{e}^{-\mathrm{i}Ht} U \ket{\psi_0}.
\end{equation}
Using p2TDVP we reproduce the results of Ref. \cite{haegeman_unifying_2016} for $\alpha = 0.75$, which is the most interesting case as it illustrates the breakdown of lightcone dynamics. It is also the most numerically challenging due to the large MPO bond dimension.

\squeezetable
\begin{table}[t]
\begin{ruledtabular}
\begin{tabular}{@{}m{2cm}m{4.3cm}@{}m{2.1cm}@{}}
 Processes ($p$) & Total discarded weight ($w_\textrm{total}$) & Infidelity ($I$)
 \tabularnewline
 \\[-0.255cm]
 \hline
 \\[-0.15cm]
 1 & $6.0\times10^{-5}$ & N/A \tabularnewline
 2 & $6.0\times10^{-5}$ & $5.1\times10^{-4}$ \tabularnewline
 8 & $6.5\times10^{-5}$ & $5.3\times10^{-4}$ \tabularnewline
 16 & $6.9\times10^{-5}$ & $5.6\times10^{-4}$ \tabularnewline
 32 & $7.9\times10^{-5}$ & $6.1\times10^{-4}$
\end{tabular}
\end{ruledtabular}
\caption{Total discarded weights and infidelities for the 101-site long-range XY model at the end of 500 timesteps.}
\label{tab:xy}
\end{table}

As shown in {\fir{figspeedup}}, this calculation scales well up to 32 processes with an efficiency $\geq 86\%$. For 32 processes this corresponds to a speedup of 27.4. The scaling is near optimal because the MPS saturates our chosen $\chi_\text{max}$ after just one timestep. We have excluded the time taken to compute expectation values, but note here that we were also able to compute these in parallel, as discussed in Ref. \cite{stoudenmire_real-space_2013}, since $C_{51,r}$ depends on $\sigma_r^x$, which is a single-site observable.

The time evolution of $C_{51,r}(t)$, calculated using 32 processes, is shown in \subfir{figxylog:data}. The results deviate from the serial calculation by less than $1\%$, except for small values at the start of the simulation ($t < 0.8$). The spatial profile of $C_{51,r}$ is seen most clearly in \subfir{figxylog:correlations}. Its value oscillates, but appears to decay algebraically with $r$, in excellent agreement with Fig. 5 of Ref. \cite{haegeman_unifying_2016}.

In \subfir{figxylog:error} we show the maximum absolute deviation in $C_{51,r}(t)$ from the serial calculation for 2, 16, and 32 processes, along with the discarded weight from the serial calculation. The deviation has a weak dependence on the number of processes $p$, but appears to be approximately bounded by $w_\textrm{total}$ (except at the beginning of the simulation where the parallelization error seems to dominate). The infidelities $I$ and total discarded weights $w_\text{total}$ from the end of the calculations are shown in Table {\ref{tab:xy}}. These also depend weakly on $p$, with $I$ being less than an order of magnitude larger than $w_\textrm{total}$.
\subsection{Long-range XXX model}\label{haldane-benchmark}

In our final benchmark, we test p2TDVP with a U(1) symmetric MPS \cite{singh_tensor_2011, al-assam_tensor_2017} by simulating the long-range isotropic Heisenberg (XXX) Hamiltonian with $\alpha = 2$,
\begin{equation}\label{HaldaneHamiltonian}
    H = \frac{1}{4}\sum_{i<j}^N \frac{1}{\lvert i-j \rvert ^2} \left( \sigma_i^x \sigma_j^x + \sigma_i^y \sigma_j^y + \sigma_i^z \sigma_j^z \right).
\end{equation}
In the thermodynamic limit this is equivalent to the exactly solvable spin-half Haldane-Shastry model \cite{haldane_exact_1988, shastry_exact_1988}, which was argued in Ref. \cite{zaletel_time-evolving_2015} to provide a stringent test case as it is both long-ranged and critical. Here we instead use p2TDVP to time evolve a 201-site spin chain with open boundaries in order to calculate the dynamical spin-spin correlation function,
\begin{equation}\label{HaldaneCorrelation}
    C(r-k,t) = \Braket{\psi_0|\sigma^z_r(t) \sigma^z_k(0)|\psi_0},
\end{equation}
where $\ket{\psi_0}$ is the ground state of \eqr{HaldaneHamiltonian}, and $k$ is the central lattice site (i.e. $k = 101$). As a $\sigma^z$ perturbation does not break the U(1) symmetry of $\ket{\psi_0}$, the $Z$-component of spin is conserved. This allows us to take advantage of symmetric block-sparse tensors \cite{singh_tensor_2011, al-assam_tensor_2017}, and hence use a relatively large bond dimension of $\chi_\textrm{max} = 1024$.

\begin{figure}
	\includegraphics[width=8.6cm]{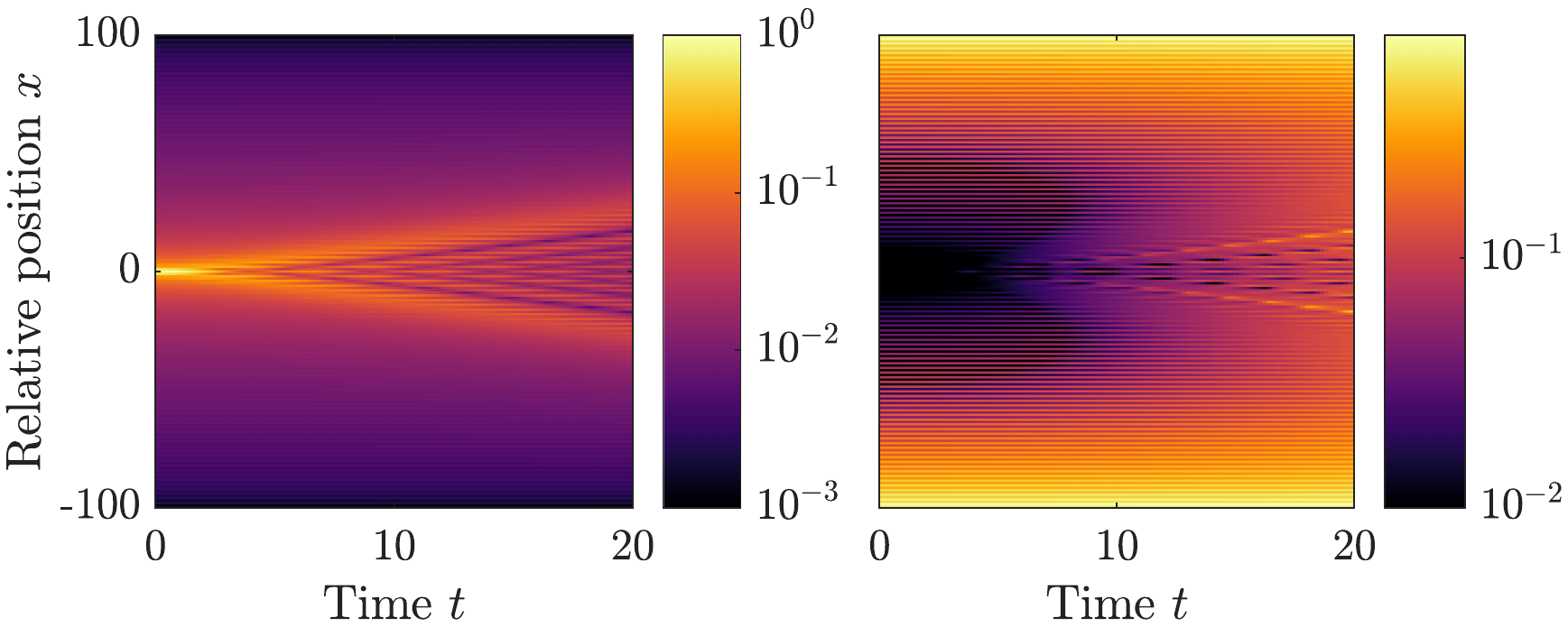}
	\caption{\label{figXXX}(Left) Density plot of $\left|C(x, t)\right|$ in the $t$-$x$ plane for the 201-site long-range ($1/r^2$) XXX spin chain. At the edges, $\left|C(x,t)\right|$ drops to $1.2\times10^{-3}$. (Right) Relative difference $\eta_\infty$ from the exact thermodynamic limit result. Towards the edges, $\eta_\infty$ grows to $0.78$. The calculation was carried out using p2TDVP on 32 processes with $m = 38$, $\delta t = 0.025$, $\varepsilon = 10^{-12}$, and $w_\textrm{max} = 10^{-16}$. $w_\text{total}$ at the end of the calculation was $7.4\times10^{-11}$.}
\end{figure}

\begin{figure}
	\includegraphics[width=8.45cm]{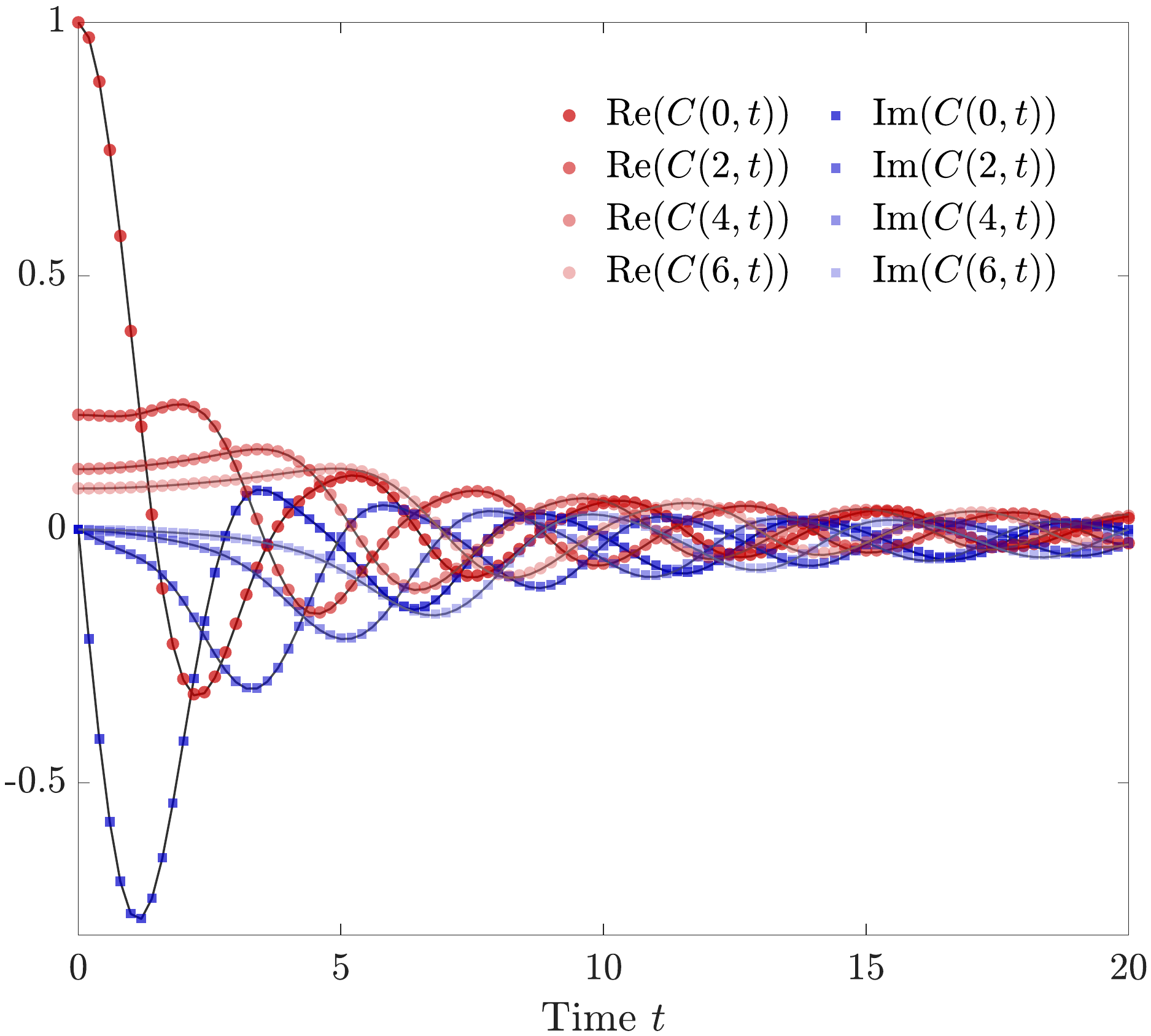}
	\caption{\label{fighaldane}Real and imaginary parts of $C(x, t)$ for $x= 0, 2, 4, 6$. The markers are the results of the p2TDVP calculation described in \fir{figXXX}. The solid lines show the exact thermodynamic limit results for comparison.}
\end{figure}

In the thermodynamic limit, the dynamical spin-spin correlation function is given by \cite{haldane_exact_1993}
\begin{eqnarray}\label{HaldaneFunction}
    C_\infty(x, t) &=& \frac{(-1)^x}{4} \int_{-1}^{1} d\lambda_1 \int_{-1}^{1} d\lambda_2 \ e^{\mathrm{i} ( Qx - Et )},
    \nonumber
    \\*
    Q &=& \pi \lambda_1 \lambda_2,
    \nonumber
    \\*
    E &=& \frac{\pi^2}{4} ( \lambda_1^2 + \lambda_2^2 - 2 \lambda_1^2 \lambda_2^2 ).
\end{eqnarray}
In \fir{figXXX} we show the magnitude of $C(x,t)$ calculated using p2TDVP on 32 processes. We also show the relative difference $\eta_\infty$ from $C_\infty(x, t)$, where
\begin{equation}
  \eta_\infty = \frac{\left|C(x, t) - C_\infty(x, t) \right|}{\left|C_\infty(x,t)\right|}.
\end{equation}
There is quantitative difference between the calculations, but they agree well qualitatively (except towards the edges where $\left|C(x,t)\right|$ drops off exponentially in the finite system due to the open boundaries). In \fir{fighaldane} we plot the real and imaginary parts of $C(x,t)$ for $x=0,2,4,6$. Again, we see good qualitative agreement with the analytic result, and with Fig. 4 of Ref. \cite{zaletel_time-evolving_2015}.

The simulation took 3.4 days to run, excluding the calculation of $C(x,t$), although a different partitioning of the MPS should give a slight speedup (see Supplemental Material \cite{supp_material}). The same computation would likely take weeks to run on a single compute node with serial 2TDVP, making a full scaling analysis impractical.

As we cannot easily calculate the error introduced by the parallel splitting for this system, we repeat the simulation on a smaller lattice of 65 sites with $\chi_\textrm{max} = 512$. Denoting the results calculated on 1 and 32 processes by $C_s$ and $C_p$, respectively, we find a maximum relative difference of $\max(\eta_p) = 3.1 \times10^{-5}$, where
\begin{equation}
  \eta_p = \frac{\left|C_\textrm{p}(x, t) - C_\textrm{s}(x, t) \right|}{\left|C_\textrm{s}(x,t)\right|}.
\end{equation}
In contrast, $\min(\eta_\infty) = 5.3\times10^{-4}$ for both the serial and parallel simulations. At least for this smaller system then, the parallelization error is negligible compared to the deviation from the thermodynamic limit (see \fir{figxxx65sites}).

\begin{figure}
  \includegraphics[width=8.6cm]{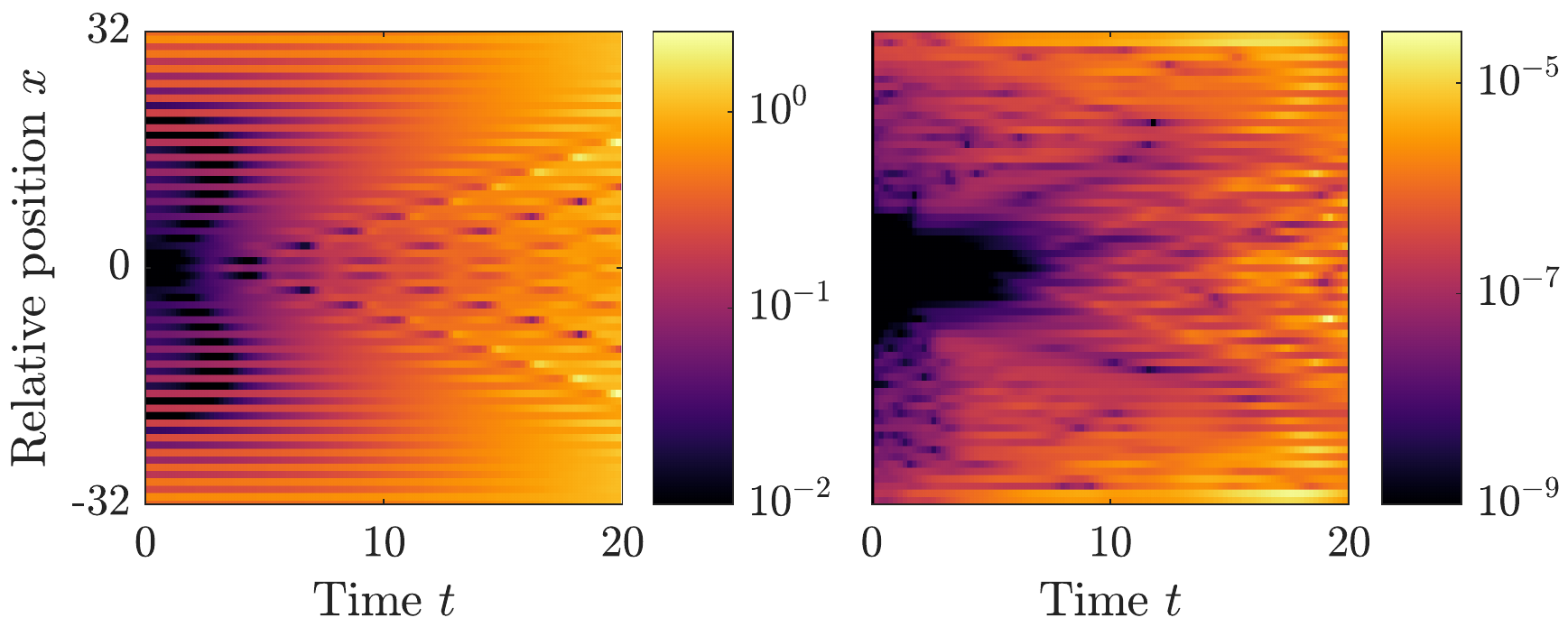}
  \caption{\label{figxxx65sites}Error in $C(x,t)$ for the 65-site long-range XXX model calculation carried out with p2TDVP on 32 processes. (Left) Relative difference $\eta_\infty$ from the thermodynamic limit result. (Right) Relative difference $\eta_p$ from the serial 2TDVP calculation. The ratio $\eta_p / \eta_\infty < 1.3 \times 10^{-4}$ for all $x$ and $t$. At the end of the calculation, $w_\text{total}$ was $5.8\times10^{-11}$ ($4.5\times10^{-11}$ for the serial calculation).}
\end{figure}

\section{Discussion}\label{sec:the-conclusion}
We have introduced a parallel version of the two-site TDVP algorithm (p2TDVP) and applied it to quenches in paradigmatic spin-half models with power law decaying interactions. To assess our algorithm's accuracy we calculated onsite expectation values, equal-time two-point correlation functions, and a dynamical spin-spin correlation function. Remarkably, we have shown that demanding calculations can be accelerated at the cost of very little additional error. Though the parallel splitting can potentially lead to instability, we have explained how this can be worked around. Speedups are system dependent, but we have demonstrated parallel efficiencies of 65--86\% with 32 processes. This suggests that it should be possible to use our algorithm to simulate systems in a week that would otherwise take many months. The use of a dynamical load balancer may further improve this efficiency.

As a next step, p2TDVP could be applied to fermionic models. It is not yet clear how accurately p2TDVP can simulate 2D systems, but fermionic models in two dimensions can be especially challenging for all numerical methods \cite{stoudenmire_studying_2012, simons_collaboration_on_the_many-electron_problem_solutions_2015}, underlining the need for a parallel algorithm. Targeting larger 1D systems should be more straightforward. Large system sizes are important for the study of open quantum dynamics \cite{schroder_simulating_2016} and transport properties \cite{kloss_spin_2019}, and for distinguishing between many-body localized and thermal phases \cite{abanin_colloquium:_2019, safavi-naini_quantum_2019}.

In Ref. \cite{depenbrock_tensor_2013} it was established that single-site DMRG can be parallelized. We therefore expect that a parallel variant of one-site TDVP (p1TDVP) could be developed using the same approach. This would enable a parallel version of the hybrid method discussed in Refs. \cite{paeckel_time-evolution_2019, goto_performance_2019, chanda_time_2019}, whereby a simulation starts with 2TDVP and switches to 1TDVP when $\chi_\textrm{max}$ is saturated. 1TDVP is faster, and can give more accurate results for some observables \cite{goto_performance_2019}. A parallel version should scale well as the fixed bond dimension would allow for optimal load balancing. It may similarly be possible to apply the parallelization scheme presented here to other related MPS-local time-evolution methods \cite{garcia-ripoll_time_2006, paeckel_time-evolution_2019, li_numerical_2019}.

We have focused on real time evolution, but p2TDVP can also be used for imaginary time evolution. This might prove beneficial for cases where parallel DMRG fails to converge. As this evolution is non-unitary, however, one has to pay particularly careful attention to the orthonormality of the MPS.

An obvious extension to this work would be the combination of parallel TDVP with established MPS techniques such as non-Abelian symmetries \cite{mcculloch_density-matrix_2007}, different local bases \cite{brockt_matrix-product-state_2015, motruk_density_2016, pastori_disentangling_2019, rams_breaking_2019, krumnow_towards_2019}, and infinite boundary conditions \cite{phien_infinite_2012}. A code combining these features would be of great benefit to the community. Generalizing to other tensor network types is a further avenue to explore. TDVP can be extended to tree tensor network states \cite{haegeman_unifying_2016, bauernfeind_time_2019}, so it would be worthwhile to see if our algorithm can be modified to work with these or other networks that admit a canonical form \cite{evenbly_gauge_2018, zaletel_isometric_2019}.

Finally, another promising application is the solution of general partial differential equations (PDEs). MPS-based PDE solvers, such as the multigrid renormalization method \cite{lubasch_multigrid_2018}, can be exponentially faster than standard PDE solvers. It is exciting to anticipate an additional speedup for such methods via the parallelization technique reported here.

\begin{figure*}
  \subfloat[\label{sfigUML:step1}]{
  \includegraphics[width=7cm]{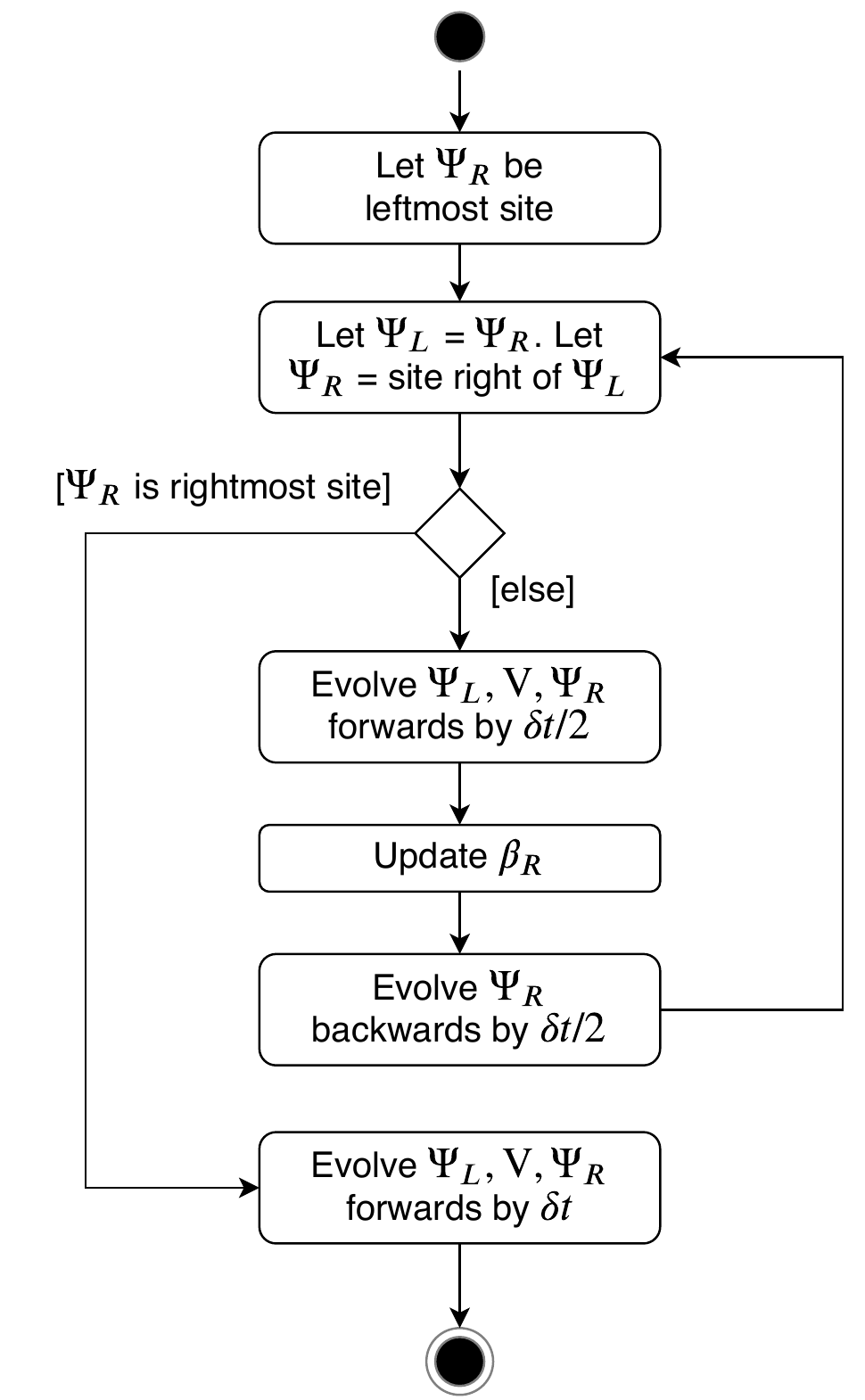}
  }
  \hspace{2cm}
  \subfloat[\label{sfigUML:step2}]{
  \includegraphics[width=6.2cm]{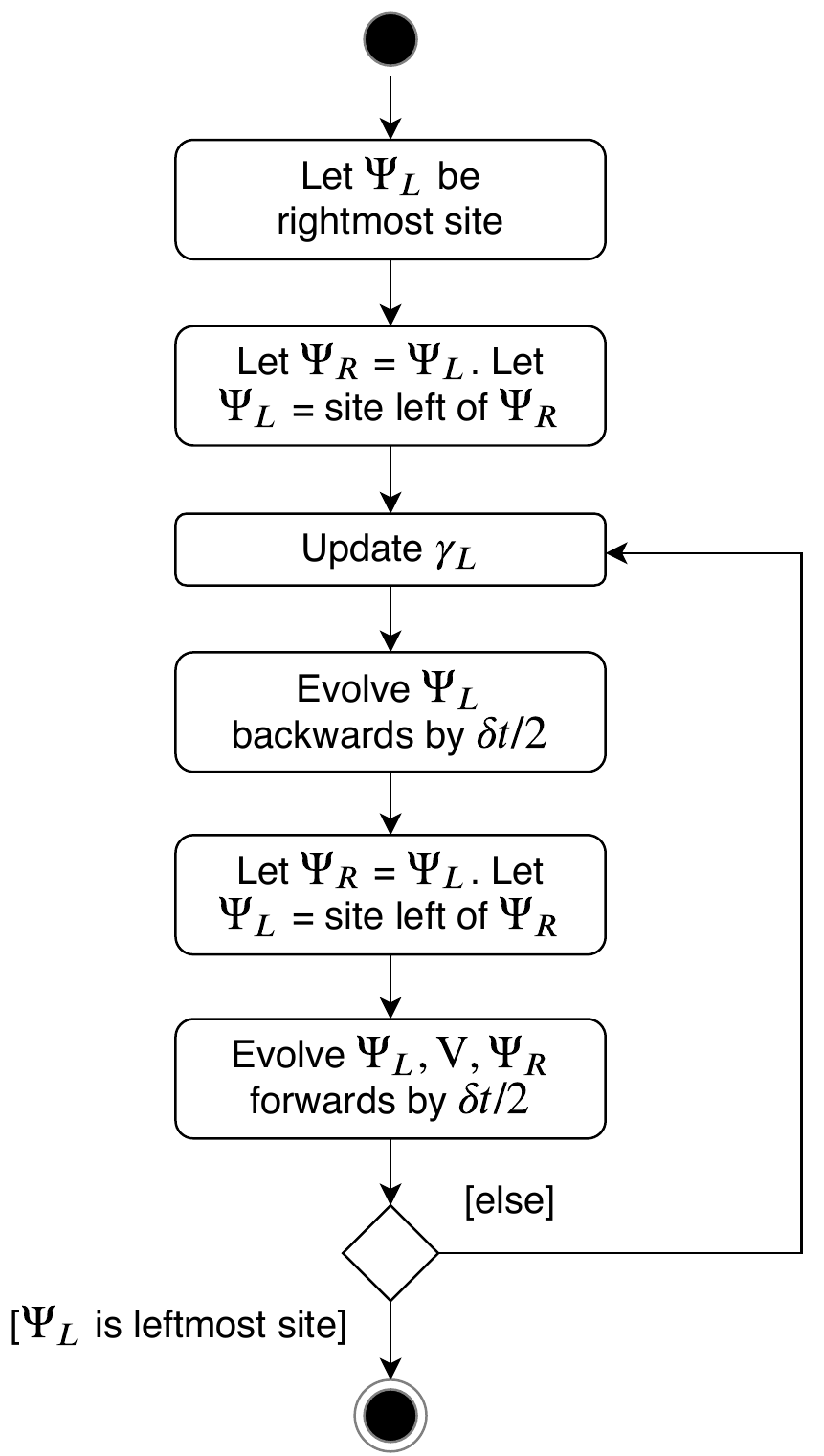}
  }
  \caption{\label{sfigUML2}UML activity diagrams illustrating the \protect\subref{sfigUML:step1} left-to-right, and \protect\subref{sfigUML:step2} right-to-left, sweeps constituting a single timestep in the serial 2TDVP algorithm, when carried out in the inverse canonical gauge. Note that ``site tensor'' is abbreviated as ``site''. For simplicity, we assume $N>2$ (the trivial $N=2$ case only requires the first sweep).\\\\}
\end{figure*}

\begin{acknowledgments}
	P.S. thanks Jonathan Coulthard, James Grant, \hbox{Fangli} Liu, Leon Schoonderwoerd, and Frank Pollmann for helpful discussions. This research made use of the Balena High Performance Computing Service at the \hbox{University} of Bath, and used Tensor Network \hbox{Theory} \hbox{Library} code written by Sarah Al-Assam, Chris Goodyer, and Jonathan Coulthard. The density plots in this paper use the `Inferno' color map by Nathaniel J. Smith and \hbox{Stefan} van der Walt. P.S. is supported by \hbox{ClusterVision} and the University of Bath. M.L. and D.J. are grateful for funding from the UK's Engineering and \hbox{Physical} Sciences Research Council (EPSRC) for grant Nos. EP/M013243/1 and EP/K038311/1. D.J. also acknowledges EPSRC grant No. EP/P01058X/1. S.R.C. gratefully acknowledges support from the EPSRC under grant No. EP/P025110/2.\\
\end{acknowledgments}

{\em Author contributions.---}S.R.C and D.J. proposed the research direction; N.G. and M.L. conceptualized the project; N.G., M.L., and P.S. designed the algorithm; N.G. developed the serial TDVP code with supervision from M.L.; P.S. developed the parallel TDVP code, conducted the numerical experiments, and wrote the manuscript, with supervision from S.D. and S.R.C.; all authors discussed the results and edited the manuscript.
\appendix
\section{\texorpdfstring{\MakeUppercase{Serial Algorithm}}{Serial Algorithm}}\label{appendix-serial}
We use unified modelling language (UML) activity diagrams to describe a single timestep in the serial 2TDVP algorithm for an MPS in the inverse canonical gauge. Figs.~\subref*{sfigUML:step1} and ~\subref*{sfigUML:step2} describe the left-to-right sweep (illustrated schematically in {\fir{figserial}}) and right-to-left sweep, respectively. For each two-site update, the left and right site tensors are labeled $\Psi_L$ and $\Psi_R$, with $\textrm{V}$ being the diagonal matrix sandwiched between them.

As noted in Section \ref{sec:serial-algorithm}, one can evolve the rightmost pair of sites at the end of the first sweep by a single full timestep. This means that the second sweep does not need to carry out a forwards time evolution step for the rightmost two sites. This same approach can be used in the parallel version of the algorithm (see \fir{pfigUML1}).

\section{\texorpdfstring{\MakeUppercase{Parallel Algorithm}}{Parallel Algorithm}} \label{appendix-parallel}
\begin{figure*}
  \includegraphics[width=17.94cm,keepaspectratio]{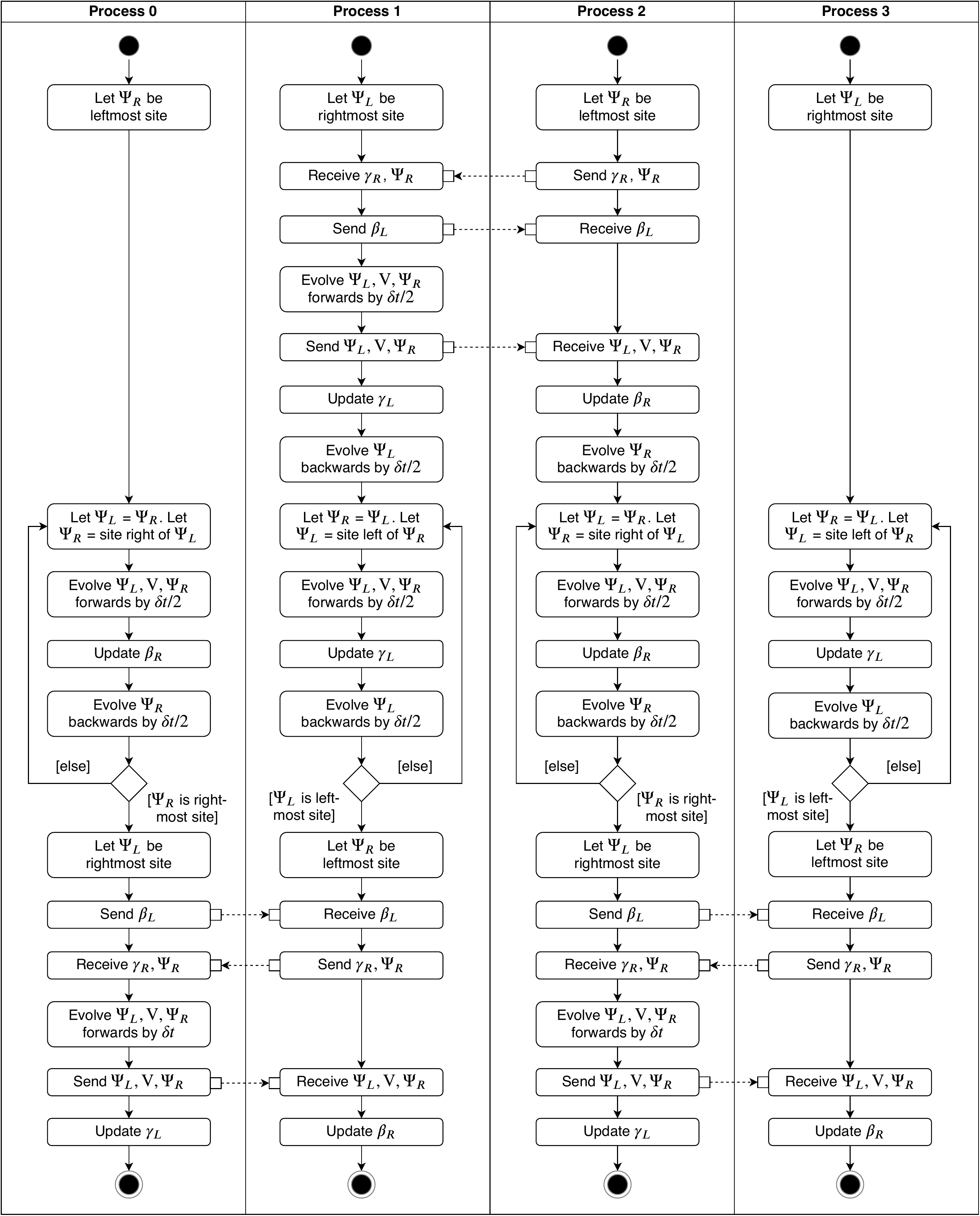}
  \caption{\label{pfigUML1}UML activity diagram describing the first half of a p2TDVP timestep on four processes. Dashed lines represent message-passing communication, and ``site'' is shorthand for ``site tensor''.
  }
\end{figure*}

\begin{figure*}
  \includegraphics[width=17.94cm,keepaspectratio]{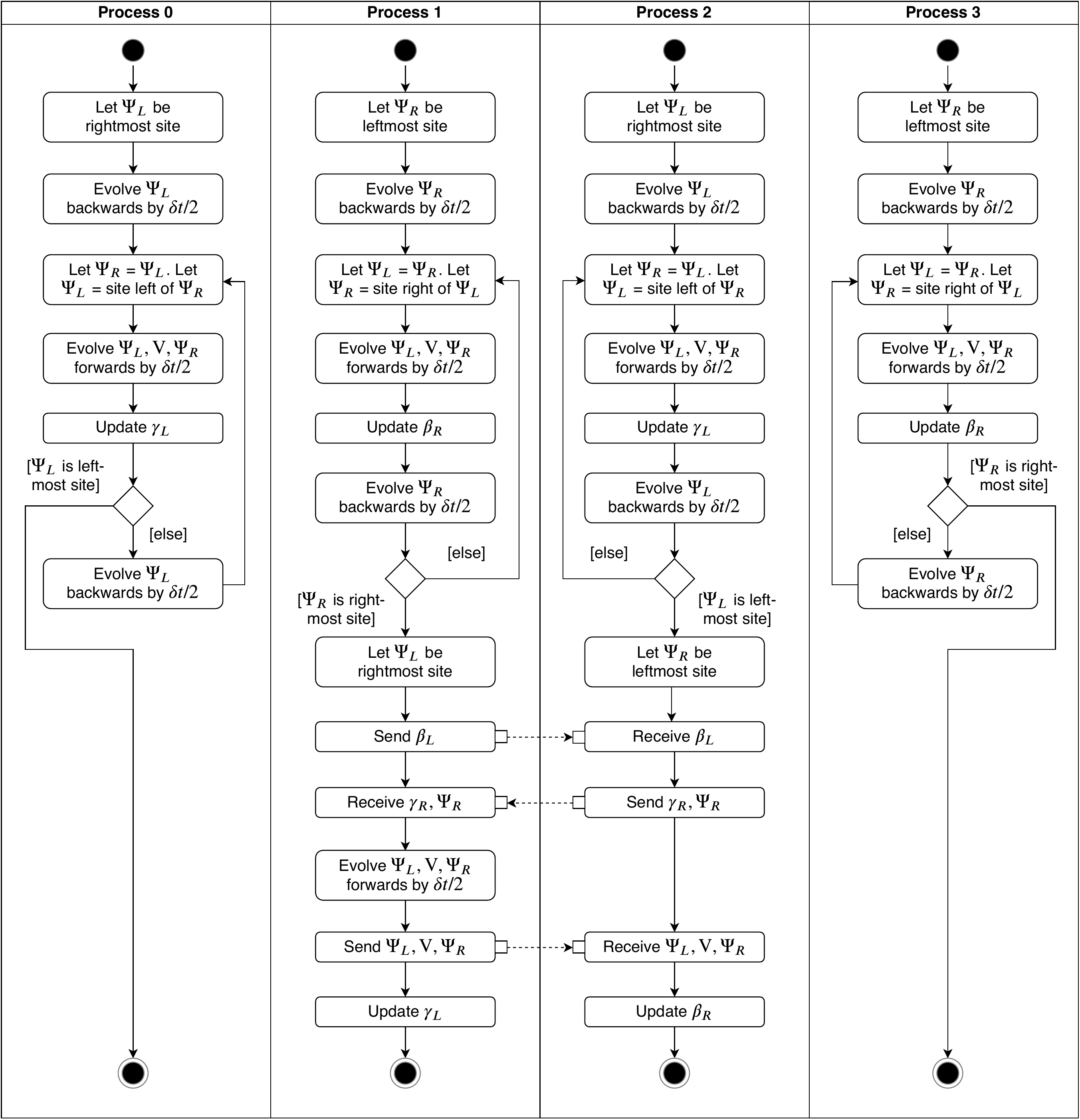}
  \caption{\label{pfigUML2}UML activity diagram describing the second half of a p2TDVP timestep on four processes. The sweep on each partition occurs in the opposite direction to the corresponding sweep in \fir{pfigUML1}.\\}
\end{figure*}

Here we use UML activity diagrams to describe a single timestep in the p2TDVP algorithm. Figs. {\ref{pfigUML1}} and {\ref{pfigUML2}} describe the first and second sweeps, respectively. For concreteness we show four parallel processes (illustrated schematically in Fig.~\subref*{figpartitioning:4procs}). This case contains all the logic necessary to generalize to $p$ processes.

\interlinepenalty=10000


\begin{thebibliography}{153}%
\makeatletter
\providecommand \@ifxundefined [1]{%
 \@ifx{#1\undefined}
}%
\providecommand \@ifnum [1]{%
 \ifnum #1\expandafter \@firstoftwo
 \else \expandafter \@secondoftwo
 \fi
}%
\providecommand \@ifx [1]{%
 \ifx #1\expandafter \@firstoftwo
 \else \expandafter \@secondoftwo
 \fi
}%
\providecommand \natexlab [1]{#1}%
\providecommand \enquote  [1]{``#1''}%
\providecommand \bibnamefont  [1]{#1}%
\providecommand \bibfnamefont [1]{#1}%
\providecommand \citenamefont [1]{#1}%
\providecommand \href@noop [0]{\@secondoftwo}%
\providecommand \href [0]{\begingroup \@sanitize@url \@href}%
\providecommand \@href[1]{\@@startlink{#1}\@@href}%
\providecommand \@@href[1]{\endgroup#1\@@endlink}%
\providecommand \@sanitize@url [0]{\catcode `\\12\catcode `\$12\catcode
  `\&12\catcode `\#12\catcode `\^12\catcode `\_12\catcode `\%12\relax}%
\providecommand \@@startlink[1]{}%
\providecommand \@@endlink[0]{}%
\providecommand \url  [0]{\begingroup\@sanitize@url \@url }%
\providecommand \@url [1]{\endgroup\@href {#1}{\urlprefix }}%
\providecommand \urlprefix  [0]{URL }%
\providecommand \Eprint [0]{\href }%
\providecommand \doibase [0]{https://doi.org/}%
\providecommand \selectlanguage [0]{\@gobble}%
\providecommand \bibinfo  [0]{\@secondoftwo}%
\providecommand \bibfield  [0]{\@secondoftwo}%
\providecommand \translation [1]{[#1]}%
\providecommand \BibitemOpen [0]{}%
\providecommand \bibitemStop [0]{}%
\providecommand \bibitemNoStop [0]{.\EOS\space}%
\providecommand \EOS [0]{\spacefactor3000\relax}%
\providecommand \BibitemShut  [1]{\csname bibitem#1\endcsname}%
\let\auto@bib@innerbib\@empty
\bibitem [{\citenamefont {Vidal}(2003)}]{vidal_efficient_2003}%
  \BibitemOpen
  \bibfield  {author} {\bibinfo {author} {\bibfnamefont {G.}~\bibnamefont
  {Vidal}},\ }\bibfield  {title} {\bibinfo {title} {Efficient {Classical}
  {Simulation} of {Slightly} {Entangled} {Quantum} {Computations}},\ }\href
  {https://doi.org/10.1103/PhysRevLett.91.147902} {\bibfield  {journal}
  {\bibinfo  {journal} {Phys. Rev. Lett.}\ }\textbf {\bibinfo {volume} {91}},\
  \bibinfo {pages} {147902} (\bibinfo {year} {2003})}\BibitemShut {NoStop}%
\bibitem [{\citenamefont {Vidal}(2004)}]{vidal_efficient_2004}%
  \BibitemOpen
  \bibfield  {author} {\bibinfo {author} {\bibfnamefont {G.}~\bibnamefont
  {Vidal}},\ }\bibfield  {title} {\bibinfo {title} {Efficient {Simulation} of
  {One}-{Dimensional} {Quantum} {Many}-{Body} {Systems}},\ }\href
  {https://doi.org/10.1103/PhysRevLett.93.040502} {\bibfield  {journal}
  {\bibinfo  {journal} {Phys. Rev. Lett.}\ }\textbf {\bibinfo {volume} {93}},\
  \bibinfo {pages} {040502} (\bibinfo {year} {2004})}\BibitemShut {NoStop}%
\bibitem [{\citenamefont {Verstraete}\ and\ \citenamefont
  {Cirac}(2006)}]{verstraete_matrix_2006}%
  \BibitemOpen
  \bibfield  {author} {\bibinfo {author} {\bibfnamefont {F.}~\bibnamefont
  {Verstraete}}\ and\ \bibinfo {author} {\bibfnamefont {J.~I.}\ \bibnamefont
  {Cirac}},\ }\bibfield  {title} {\bibinfo {title} {Matrix product states
  represent ground states faithfully},\ }\href
  {https://doi.org/10.1103/PhysRevB.73.094423} {\bibfield  {journal} {\bibinfo
  {journal} {Phys. Rev. B}\ }\textbf {\bibinfo {volume} {73}},\ \bibinfo
  {pages} {094423} (\bibinfo {year} {2006})}\BibitemShut {NoStop}%
\bibitem [{\citenamefont {Schuch}\ \emph {et~al.}(2008)\citenamefont {Schuch},
  \citenamefont {Wolf}, \citenamefont {Verstraete},\ and\ \citenamefont
  {Cirac}}]{schuch_entropy_2008}%
  \BibitemOpen
  \bibfield  {author} {\bibinfo {author} {\bibfnamefont {N.}~\bibnamefont
  {Schuch}}, \bibinfo {author} {\bibfnamefont {M.~M.}\ \bibnamefont {Wolf}},
  \bibinfo {author} {\bibfnamefont {F.}~\bibnamefont {Verstraete}},\ and\
  \bibinfo {author} {\bibfnamefont {J.~I.}\ \bibnamefont {Cirac}},\ }\bibfield
  {title} {\bibinfo {title} {Entropy {Scaling} and {Simulability} by {Matrix}
  {Product} {States}},\ }\href {https://doi.org/10.1103/PhysRevLett.100.030504}
  {\bibfield  {journal} {\bibinfo  {journal} {Phys. Rev. Lett.}\ }\textbf
  {\bibinfo {volume} {100}},\ \bibinfo {pages} {030504} (\bibinfo {year}
  {2008})}\BibitemShut {NoStop}%
\bibitem [{\citenamefont {Eisert}\ \emph {et~al.}(2010)\citenamefont {Eisert},
  \citenamefont {Cramer},\ and\ \citenamefont
  {Plenio}}]{eisert_colloquium:_2010}%
  \BibitemOpen
  \bibfield  {author} {\bibinfo {author} {\bibfnamefont {J.}~\bibnamefont
  {Eisert}}, \bibinfo {author} {\bibfnamefont {M.}~\bibnamefont {Cramer}},\
  and\ \bibinfo {author} {\bibfnamefont {M.~B.}\ \bibnamefont {Plenio}},\
  }\bibfield  {title} {\bibinfo {title} {Colloquium: {Area} laws for the
  entanglement entropy},\ }\href {https://doi.org/10.1103/RevModPhys.82.277}
  {\bibfield  {journal} {\bibinfo  {journal} {Rev. Mod. Phys.}\ }\textbf
  {\bibinfo {volume} {82}},\ \bibinfo {pages} {277} (\bibinfo {year}
  {2010})}\BibitemShut {NoStop}%
\bibitem [{\citenamefont {{\"O}stlund}\ and\ \citenamefont
  {Rommer}(1995)}]{ostlund_thermodynamic_1995}%
  \BibitemOpen
  \bibfield  {author} {\bibinfo {author} {\bibfnamefont {S.}~\bibnamefont
  {{\"O}stlund}}\ and\ \bibinfo {author} {\bibfnamefont {S.}~\bibnamefont
  {Rommer}},\ }\bibfield  {title} {\bibinfo {title} {Thermodynamic {Limit} of
  {Density} {Matrix} {Renormalization}},\ }\href
  {https://doi.org/10.1103/PhysRevLett.75.3537} {\bibfield  {journal} {\bibinfo
   {journal} {Phys. Rev. Lett.}\ }\textbf {\bibinfo {volume} {75}},\ \bibinfo
  {pages} {3537} (\bibinfo {year} {1995})}\BibitemShut {NoStop}%
\bibitem [{\citenamefont {Dukelsky}\ \emph {et~al.}(1998)\citenamefont
  {Dukelsky}, \citenamefont {Mart{\'i}n-Delgado}, \citenamefont {Nishino},\
  and\ \citenamefont {Sierra}}]{dukelsky_equivalence_1998}%
  \BibitemOpen
  \bibfield  {author} {\bibinfo {author} {\bibfnamefont {J.}~\bibnamefont
  {Dukelsky}}, \bibinfo {author} {\bibfnamefont {M.~A.}\ \bibnamefont
  {Mart{\'i}n-Delgado}}, \bibinfo {author} {\bibfnamefont {T.}~\bibnamefont
  {Nishino}},\ and\ \bibinfo {author} {\bibfnamefont {G.}~\bibnamefont
  {Sierra}},\ }\bibfield  {title} {\bibinfo {title} {Equivalence of the
  variational matrix product method and the density matrix renormalization
  group applied to spin chains},\ }\href
  {https://doi.org/10.1209/epl/i1998-00381-x} {\bibfield  {journal} {\bibinfo
  {journal} {EPL}\ }\textbf {\bibinfo {volume} {43}},\ \bibinfo {pages} {457}
  (\bibinfo {year} {1998})}\BibitemShut {NoStop}%
\bibitem [{\citenamefont
  {Schollw{\"o}ck}(2011)}]{schollwock_density-matrix_2011}%
  \BibitemOpen
  \bibfield  {author} {\bibinfo {author} {\bibfnamefont {U.}~\bibnamefont
  {Schollw{\"o}ck}},\ }\bibfield  {title} {\bibinfo {title} {The density-matrix
  renormalization group in the age of matrix product states},\ }\href
  {https://doi.org/10.1016/j.aop.2010.09.012} {\bibfield  {journal} {\bibinfo
  {journal} {Ann. Phys. (N. Y.)}\ }\bibinfo {series} {January 2011 {Special}
  {Issue}},\ \textbf {\bibinfo {volume} {326}},\ \bibinfo {pages} {96}
  (\bibinfo {year} {2011})}\BibitemShut {NoStop}%
\bibitem [{\citenamefont {White}(1992)}]{white_density_1992}%
  \BibitemOpen
  \bibfield  {author} {\bibinfo {author} {\bibfnamefont {S.~R.}\ \bibnamefont
  {White}},\ }\bibfield  {title} {\bibinfo {title} {Density matrix formulation
  for quantum renormalization groups},\ }\href
  {https://doi.org/10.1103/PhysRevLett.69.2863} {\bibfield  {journal} {\bibinfo
   {journal} {Phys. Rev. Lett.}\ }\textbf {\bibinfo {volume} {69}},\ \bibinfo
  {pages} {2863} (\bibinfo {year} {1992})}\BibitemShut {NoStop}%
\bibitem [{\citenamefont {White}(1993)}]{white_density-matrix_1993}%
  \BibitemOpen
  \bibfield  {author} {\bibinfo {author} {\bibfnamefont {S.~R.}\ \bibnamefont
  {White}},\ }\bibfield  {title} {\bibinfo {title} {Density-matrix algorithms
  for quantum renormalization groups},\ }\href
  {https://doi.org/10.1103/PhysRevB.48.10345} {\bibfield  {journal} {\bibinfo
  {journal} {Phys. Rev. B}\ }\textbf {\bibinfo {volume} {48}},\ \bibinfo
  {pages} {10345} (\bibinfo {year} {1993})}\BibitemShut {NoStop}%
\bibitem [{\citenamefont {Stoudenmire}\ and\ \citenamefont
  {White}(2012)}]{stoudenmire_studying_2012}%
  \BibitemOpen
  \bibfield  {author} {\bibinfo {author} {\bibfnamefont {E.}~\bibnamefont
  {Stoudenmire}}\ and\ \bibinfo {author} {\bibfnamefont {S.~R.}\ \bibnamefont
  {White}},\ }\bibfield  {title} {\bibinfo {title} {Studying
  {Two}-{Dimensional} {Systems} with the {Density} {Matrix} {Renormalization}
  {Group}},\ }\href {https://doi.org/10.1146/annurev-conmatphys-020911-125018}
  {\bibfield  {journal} {\bibinfo  {journal} {Annu. Rev. Condens. Matter
  Phys.}\ }\textbf {\bibinfo {volume} {3}},\ \bibinfo {pages} {111} (\bibinfo
  {year} {2012})}\BibitemShut {NoStop}%
\bibitem [{\citenamefont {Daley}\ \emph {et~al.}(2004)\citenamefont {Daley},
  \citenamefont {Kollath}, \citenamefont {Schollw{\"o}ck},\ and\ \citenamefont
  {Vidal}}]{daley_time-dependent_2004}%
  \BibitemOpen
  \bibfield  {author} {\bibinfo {author} {\bibfnamefont {A.~J.}\ \bibnamefont
  {Daley}}, \bibinfo {author} {\bibfnamefont {C.}~\bibnamefont {Kollath}},
  \bibinfo {author} {\bibfnamefont {U.}~\bibnamefont {Schollw{\"o}ck}},\ and\
  \bibinfo {author} {\bibfnamefont {G.}~\bibnamefont {Vidal}},\ }\bibfield
  {title} {\bibinfo {title} {Time-dependent density-matrix
  renormalization-group using adaptive effective {Hilbert} spaces},\ }\href
  {https://doi.org/10.1088/1742-5468/2004/04/P04005} {\bibfield  {journal}
  {\bibinfo  {journal} {J. Stat. Mech.}\ }\textbf {\bibinfo {volume} {2004}},\
  \bibinfo {pages} {P04005} (\bibinfo {year} {2004})}\BibitemShut {NoStop}%
\bibitem [{\citenamefont {Micheli}\ \emph {et~al.}(2006)\citenamefont
  {Micheli}, \citenamefont {Brennen},\ and\ \citenamefont
  {Zoller}}]{micheli_toolbox_2006}%
  \BibitemOpen
  \bibfield  {author} {\bibinfo {author} {\bibfnamefont {A.}~\bibnamefont
  {Micheli}}, \bibinfo {author} {\bibfnamefont {G.~K.}\ \bibnamefont
  {Brennen}},\ and\ \bibinfo {author} {\bibfnamefont {P.}~\bibnamefont
  {Zoller}},\ }\bibfield  {title} {\bibinfo {title} {A toolbox for lattice-spin
  models with polar molecules},\ }\href {https://doi.org/10.1038/nphys287}
  {\bibfield  {journal} {\bibinfo  {journal} {Nat. Phys.}\ }\textbf {\bibinfo
  {volume} {2}},\ \bibinfo {pages} {341} (\bibinfo {year} {2006})}\BibitemShut
  {NoStop}%
\bibitem [{\citenamefont {Yan}\ \emph {et~al.}(2013)\citenamefont {Yan},
  \citenamefont {Moses}, \citenamefont {Gadway}, \citenamefont {Covey},
  \citenamefont {Hazzard}, \citenamefont {Rey}, \citenamefont {Jin},\ and\
  \citenamefont {Ye}}]{yan_observation_2013}%
  \BibitemOpen
  \bibfield  {author} {\bibinfo {author} {\bibfnamefont {B.}~\bibnamefont
  {Yan}}, \bibinfo {author} {\bibfnamefont {S.~A.}\ \bibnamefont {Moses}},
  \bibinfo {author} {\bibfnamefont {B.}~\bibnamefont {Gadway}}, \bibinfo
  {author} {\bibfnamefont {J.~P.}\ \bibnamefont {Covey}}, \bibinfo {author}
  {\bibfnamefont {K.~R.~A.}\ \bibnamefont {Hazzard}}, \bibinfo {author}
  {\bibfnamefont {A.~M.}\ \bibnamefont {Rey}}, \bibinfo {author} {\bibfnamefont
  {D.~S.}\ \bibnamefont {Jin}},\ and\ \bibinfo {author} {\bibfnamefont
  {J.}~\bibnamefont {Ye}},\ }\bibfield  {title} {\bibinfo {title} {Observation
  of dipolar spin-exchange interactions with lattice-confined polar
  molecules},\ }\href {https://doi.org/10.1038/nature12483} {\bibfield
  {journal} {\bibinfo  {journal} {Nature}\ }\textbf {\bibinfo {volume} {501}},\
  \bibinfo {pages} {521} (\bibinfo {year} {2013})}\BibitemShut {NoStop}%
\bibitem [{\citenamefont {Schau{\ss}}\ \emph {et~al.}(2012)\citenamefont
  {Schau{\ss}}, \citenamefont {Cheneau}, \citenamefont {Endres}, \citenamefont
  {Fukuhara}, \citenamefont {Hild}, \citenamefont {Omran}, \citenamefont
  {Pohl}, \citenamefont {Gross}, \citenamefont {Kuhr},\ and\ \citenamefont
  {Bloch}}]{schaus_observation_2012}%
  \BibitemOpen
  \bibfield  {author} {\bibinfo {author} {\bibfnamefont {P.}~\bibnamefont
  {Schau{\ss}}}, \bibinfo {author} {\bibfnamefont {M.}~\bibnamefont {Cheneau}},
  \bibinfo {author} {\bibfnamefont {M.}~\bibnamefont {Endres}}, \bibinfo
  {author} {\bibfnamefont {T.}~\bibnamefont {Fukuhara}}, \bibinfo {author}
  {\bibfnamefont {S.}~\bibnamefont {Hild}}, \bibinfo {author} {\bibfnamefont
  {A.}~\bibnamefont {Omran}}, \bibinfo {author} {\bibfnamefont
  {T.}~\bibnamefont {Pohl}}, \bibinfo {author} {\bibfnamefont {C.}~\bibnamefont
  {Gross}}, \bibinfo {author} {\bibfnamefont {S.}~\bibnamefont {Kuhr}},\ and\
  \bibinfo {author} {\bibfnamefont {I.}~\bibnamefont {Bloch}},\ }\bibfield
  {title} {\bibinfo {title} {Observation of spatially ordered structures in a
  two-dimensional {Rydberg} gas},\ }\href {https://doi.org/10.1038/nature11596}
  {\bibfield  {journal} {\bibinfo  {journal} {Nature}\ }\textbf {\bibinfo
  {volume} {491}},\ \bibinfo {pages} {87} (\bibinfo {year} {2012})}\BibitemShut
  {NoStop}%
\bibitem [{\citenamefont {Weimer}\ \emph {et~al.}(2010)\citenamefont {Weimer},
  \citenamefont {M{\"u}ller}, \citenamefont {Lesanovsky}, \citenamefont
  {Zoller},\ and\ \citenamefont {B{\"u}chler}}]{weimer_rydberg_2010}%
  \BibitemOpen
  \bibfield  {author} {\bibinfo {author} {\bibfnamefont {H.}~\bibnamefont
  {Weimer}}, \bibinfo {author} {\bibfnamefont {M.}~\bibnamefont {M{\"u}ller}},
  \bibinfo {author} {\bibfnamefont {I.}~\bibnamefont {Lesanovsky}}, \bibinfo
  {author} {\bibfnamefont {P.}~\bibnamefont {Zoller}},\ and\ \bibinfo {author}
  {\bibfnamefont {H.~P.}\ \bibnamefont {B{\"u}chler}},\ }\bibfield  {title}
  {\bibinfo {title} {A {Rydberg} quantum simulator},\ }\href
  {https://doi.org/10.1038/nphys1614} {\bibfield  {journal} {\bibinfo
  {journal} {Nat. Phys.}\ }\textbf {\bibinfo {volume} {6}},\ \bibinfo {pages}
  {382} (\bibinfo {year} {2010})}\BibitemShut {NoStop}%
\bibitem [{\citenamefont {Zeiher}\ \emph {et~al.}(2016)\citenamefont {Zeiher},
  \citenamefont {van Bijnen}, \citenamefont {Schau{\ss}}, \citenamefont {Hild},
  \citenamefont {Choi}, \citenamefont {Pohl}, \citenamefont {Bloch},\ and\
  \citenamefont {Gross}}]{zeiher_many-body_2016}%
  \BibitemOpen
  \bibfield  {author} {\bibinfo {author} {\bibfnamefont {J.}~\bibnamefont
  {Zeiher}}, \bibinfo {author} {\bibfnamefont {R.}~\bibnamefont {van Bijnen}},
  \bibinfo {author} {\bibfnamefont {P.}~\bibnamefont {Schau{\ss}}}, \bibinfo
  {author} {\bibfnamefont {S.}~\bibnamefont {Hild}}, \bibinfo {author}
  {\bibfnamefont {J.-y.}\ \bibnamefont {Choi}}, \bibinfo {author}
  {\bibfnamefont {T.}~\bibnamefont {Pohl}}, \bibinfo {author} {\bibfnamefont
  {I.}~\bibnamefont {Bloch}},\ and\ \bibinfo {author} {\bibfnamefont
  {C.}~\bibnamefont {Gross}},\ }\bibfield  {title} {\bibinfo {title} {Many-body
  interferometry of a {Rydberg}-dressed spin lattice},\ }\href
  {https://doi.org/10.1038/nphys3835} {\bibfield  {journal} {\bibinfo
  {journal} {Nat. Phys.}\ }\textbf {\bibinfo {volume} {12}},\ \bibinfo {pages}
  {1095} (\bibinfo {year} {2016})}\BibitemShut {NoStop}%
\bibitem [{\citenamefont {Porras}\ and\ \citenamefont
  {Cirac}(2004)}]{porras_effective_2004}%
  \BibitemOpen
  \bibfield  {author} {\bibinfo {author} {\bibfnamefont {D.}~\bibnamefont
  {Porras}}\ and\ \bibinfo {author} {\bibfnamefont {J.~I.}\ \bibnamefont
  {Cirac}},\ }\bibfield  {title} {\bibinfo {title} {Effective {Quantum} {Spin}
  {Systems} with {Trapped} {Ions}},\ }\href
  {https://doi.org/10.1103/PhysRevLett.92.207901} {\bibfield  {journal}
  {\bibinfo  {journal} {Phys. Rev. Lett.}\ }\textbf {\bibinfo {volume} {92}},\
  \bibinfo {pages} {207901} (\bibinfo {year} {2004})}\BibitemShut {NoStop}%
\bibitem [{\citenamefont {Britton}\ \emph {et~al.}(2012)\citenamefont
  {Britton}, \citenamefont {Sawyer}, \citenamefont {Keith}, \citenamefont
  {Wang}, \citenamefont {Freericks}, \citenamefont {Uys}, \citenamefont
  {Biercuk},\ and\ \citenamefont {Bollinger}}]{britton_engineered_2012}%
  \BibitemOpen
  \bibfield  {author} {\bibinfo {author} {\bibfnamefont {J.~W.}\ \bibnamefont
  {Britton}}, \bibinfo {author} {\bibfnamefont {B.~C.}\ \bibnamefont {Sawyer}},
  \bibinfo {author} {\bibfnamefont {A.~C.}\ \bibnamefont {Keith}}, \bibinfo
  {author} {\bibfnamefont {C.-C.~J.}\ \bibnamefont {Wang}}, \bibinfo {author}
  {\bibfnamefont {J.~K.}\ \bibnamefont {Freericks}}, \bibinfo {author}
  {\bibfnamefont {H.}~\bibnamefont {Uys}}, \bibinfo {author} {\bibfnamefont
  {M.~J.}\ \bibnamefont {Biercuk}},\ and\ \bibinfo {author} {\bibfnamefont
  {J.~J.}\ \bibnamefont {Bollinger}},\ }\bibfield  {title} {\bibinfo {title}
  {Engineered two-dimensional {Ising} interactions in a trapped-ion quantum
  simulator with hundreds of spins},\ }\href
  {https://doi.org/10.1038/nature10981} {\bibfield  {journal} {\bibinfo
  {journal} {Nature}\ }\textbf {\bibinfo {volume} {484}},\ \bibinfo {pages}
  {489} (\bibinfo {year} {2012})}\BibitemShut {NoStop}%
\bibitem [{\citenamefont {Islam}\ \emph {et~al.}(2013)\citenamefont {Islam},
  \citenamefont {Senko}, \citenamefont {Campbell}, \citenamefont {Korenblit},
  \citenamefont {Smith}, \citenamefont {Lee}, \citenamefont {Edwards},
  \citenamefont {Wang}, \citenamefont {Freericks},\ and\ \citenamefont
  {Monroe}}]{islam_emergence_2013}%
  \BibitemOpen
  \bibfield  {author} {\bibinfo {author} {\bibfnamefont {R.}~\bibnamefont
  {Islam}}, \bibinfo {author} {\bibfnamefont {C.}~\bibnamefont {Senko}},
  \bibinfo {author} {\bibfnamefont {W.~C.}\ \bibnamefont {Campbell}}, \bibinfo
  {author} {\bibfnamefont {S.}~\bibnamefont {Korenblit}}, \bibinfo {author}
  {\bibfnamefont {J.}~\bibnamefont {Smith}}, \bibinfo {author} {\bibfnamefont
  {A.}~\bibnamefont {Lee}}, \bibinfo {author} {\bibfnamefont {E.~E.}\
  \bibnamefont {Edwards}}, \bibinfo {author} {\bibfnamefont {C.-C.~J.}\
  \bibnamefont {Wang}}, \bibinfo {author} {\bibfnamefont {J.~K.}\ \bibnamefont
  {Freericks}},\ and\ \bibinfo {author} {\bibfnamefont {C.}~\bibnamefont
  {Monroe}},\ }\bibfield  {title} {\bibinfo {title} {Emergence and
  {Frustration} of {Magnetism} with {Variable}-{Range} {Interactions} in a
  {Quantum} {Simulator}},\ }\href {https://doi.org/10.1126/science.1232296}
  {\bibfield  {journal} {\bibinfo  {journal} {Science}\ }\textbf {\bibinfo
  {volume} {340}},\ \bibinfo {pages} {583} (\bibinfo {year}
  {2013})}\BibitemShut {NoStop}%
\bibitem [{\citenamefont {Richerme}\ \emph {et~al.}(2014)\citenamefont
  {Richerme}, \citenamefont {Gong}, \citenamefont {Lee}, \citenamefont {Senko},
  \citenamefont {Smith}, \citenamefont {Foss-Feig}, \citenamefont {Michalakis},
  \citenamefont {Gorshkov},\ and\ \citenamefont
  {Monroe}}]{richerme_non-local_2014}%
  \BibitemOpen
  \bibfield  {author} {\bibinfo {author} {\bibfnamefont {P.}~\bibnamefont
  {Richerme}}, \bibinfo {author} {\bibfnamefont {Z.-X.}\ \bibnamefont {Gong}},
  \bibinfo {author} {\bibfnamefont {A.}~\bibnamefont {Lee}}, \bibinfo {author}
  {\bibfnamefont {C.}~\bibnamefont {Senko}}, \bibinfo {author} {\bibfnamefont
  {J.}~\bibnamefont {Smith}}, \bibinfo {author} {\bibfnamefont
  {M.}~\bibnamefont {Foss-Feig}}, \bibinfo {author} {\bibfnamefont
  {S.}~\bibnamefont {Michalakis}}, \bibinfo {author} {\bibfnamefont {A.~V.}\
  \bibnamefont {Gorshkov}},\ and\ \bibinfo {author} {\bibfnamefont
  {C.}~\bibnamefont {Monroe}},\ }\bibfield  {title} {\bibinfo {title}
  {Non-local propagation of correlations in quantum systems with long-range
  interactions},\ }\href {https://doi.org/10.1038/nature13450} {\bibfield
  {journal} {\bibinfo  {journal} {Nature}\ }\textbf {\bibinfo {volume} {511}},\
  \bibinfo {pages} {198} (\bibinfo {year} {2014})}\BibitemShut {NoStop}%
\bibitem [{\citenamefont {Jurcevic}\ \emph {et~al.}(2014)\citenamefont
  {Jurcevic}, \citenamefont {Lanyon}, \citenamefont {Hauke}, \citenamefont
  {Hempel}, \citenamefont {Zoller}, \citenamefont {Blatt},\ and\ \citenamefont
  {Roos}}]{jurcevic_quasiparticle_2014}%
  \BibitemOpen
  \bibfield  {author} {\bibinfo {author} {\bibfnamefont {P.}~\bibnamefont
  {Jurcevic}}, \bibinfo {author} {\bibfnamefont {B.~P.}\ \bibnamefont
  {Lanyon}}, \bibinfo {author} {\bibfnamefont {P.}~\bibnamefont {Hauke}},
  \bibinfo {author} {\bibfnamefont {C.}~\bibnamefont {Hempel}}, \bibinfo
  {author} {\bibfnamefont {P.}~\bibnamefont {Zoller}}, \bibinfo {author}
  {\bibfnamefont {R.}~\bibnamefont {Blatt}},\ and\ \bibinfo {author}
  {\bibfnamefont {C.~F.}\ \bibnamefont {Roos}},\ }\bibfield  {title} {\bibinfo
  {title} {Quasiparticle engineering and entanglement propagation in a quantum
  many-body system},\ }\href {https://doi.org/10.1038/nature13461} {\bibfield
  {journal} {\bibinfo  {journal} {Nature}\ }\textbf {\bibinfo {volume} {511}},\
  \bibinfo {pages} {202} (\bibinfo {year} {2014})}\BibitemShut {NoStop}%
\bibitem [{\citenamefont {Smith}\ \emph {et~al.}(2016)\citenamefont {Smith},
  \citenamefont {Lee}, \citenamefont {Richerme}, \citenamefont {Neyenhuis},
  \citenamefont {Hess}, \citenamefont {Hauke}, \citenamefont {Heyl},
  \citenamefont {Huse},\ and\ \citenamefont {Monroe}}]{smith_many-body_2016}%
  \BibitemOpen
  \bibfield  {author} {\bibinfo {author} {\bibfnamefont {J.}~\bibnamefont
  {Smith}}, \bibinfo {author} {\bibfnamefont {A.}~\bibnamefont {Lee}}, \bibinfo
  {author} {\bibfnamefont {P.}~\bibnamefont {Richerme}}, \bibinfo {author}
  {\bibfnamefont {B.}~\bibnamefont {Neyenhuis}}, \bibinfo {author}
  {\bibfnamefont {P.~W.}\ \bibnamefont {Hess}}, \bibinfo {author}
  {\bibfnamefont {P.}~\bibnamefont {Hauke}}, \bibinfo {author} {\bibfnamefont
  {M.}~\bibnamefont {Heyl}}, \bibinfo {author} {\bibfnamefont {D.~A.}\
  \bibnamefont {Huse}},\ and\ \bibinfo {author} {\bibfnamefont
  {C.}~\bibnamefont {\hbox{Monroe}}},\ }\bibfield  {title} {\bibinfo {title}
  {Many-body localization in a quantum simulator with programmable random
  disorder},\ }\href {https://doi.org/10.1038/nphys3783} {\bibfield  {journal}
  {\bibinfo  {journal} {Nat. Phys.}\ }\textbf {\bibinfo {volume} {12}},\
  \bibinfo {pages} {907} (\bibinfo {year} {2016})}\BibitemShut {NoStop}%
\bibitem [{\citenamefont {Zhang}\ \emph {et~al.}(2017)\citenamefont {Zhang},
  \citenamefont {Pagano}, \citenamefont {Hess}, \citenamefont {Kyprianidis},
  \citenamefont {Becker}, \citenamefont {Kaplan}, \citenamefont {Gorshkov},
  \citenamefont {Gong},\ and\ \citenamefont {Monroe}}]{zhang_observation_2017}%
  \BibitemOpen
  \bibfield  {author} {\bibinfo {author} {\bibfnamefont {J.}~\bibnamefont
  {Zhang}}, \bibinfo {author} {\bibfnamefont {G.}~\bibnamefont {Pagano}},
  \bibinfo {author} {\bibfnamefont {P.~W.}\ \bibnamefont {Hess}}, \bibinfo
  {author} {\bibfnamefont {A.}~\bibnamefont {Kyprianidis}}, \bibinfo {author}
  {\bibfnamefont {P.}~\bibnamefont {Becker}}, \bibinfo {author} {\bibfnamefont
  {H.}~\bibnamefont {Kaplan}}, \bibinfo {author} {\bibfnamefont {A.~V.}\
  \bibnamefont {Gorshkov}}, \bibinfo {author} {\bibfnamefont {Z.-X.}\
  \bibnamefont {Gong}},\ and\ \bibinfo {author} {\bibfnamefont
  {C.}~\bibnamefont {Monroe}},\ }\bibfield  {title} {\bibinfo {title}
  {Observation of a many-body dynamical phase transition with a 53-qubit
  quantum simulator},\ }\href {https://doi.org/10.1038/nature24654} {\bibfield
  {journal} {\bibinfo  {journal} {Nature}\ }\textbf {\bibinfo {volume} {551}},\
  \bibinfo {pages} {601} (\bibinfo {year} {2017})}\BibitemShut {NoStop}%
\bibitem [{\citenamefont {Feiguin}\ and\ \citenamefont
  {White}(2005)}]{feiguin_time-step_2005}%
  \BibitemOpen
  \bibfield  {author} {\bibinfo {author} {\bibfnamefont {A.~E.}\ \bibnamefont
  {Feiguin}}\ and\ \bibinfo {author} {\bibfnamefont {S.~R.}\ \bibnamefont
  {White}},\ }\bibfield  {title} {\bibinfo {title} {Time-step targeting methods
  for real-time dynamics using the density matrix renormalization group},\
  }\href {https://doi.org/10.1103/PhysRevB.72.020404} {\bibfield  {journal}
  {\bibinfo  {journal} {Phys. Rev. B}\ }\textbf {\bibinfo {volume} {72}},\
  \bibinfo {pages} {020404} (\bibinfo {year} {2005})}\BibitemShut {NoStop}%
\bibitem [{\citenamefont {Garc{\'i}a-Ripoll}(2006)}]{garcia-ripoll_time_2006}%
  \BibitemOpen
  \bibfield  {author} {\bibinfo {author} {\bibfnamefont {J.~J.}\ \bibnamefont
  {Garc{\'i}a-Ripoll}},\ }\bibfield  {title} {\bibinfo {title} {Time evolution
  of {Matrix} {Product} {States}},\ }\href
  {https://doi.org/10.1088/1367-2630/8/12/305} {\bibfield  {journal} {\bibinfo
  {journal} {New J. Phys.}\ }\textbf {\bibinfo {volume} {8}},\ \bibinfo {pages}
  {305} (\bibinfo {year} {2006})}\BibitemShut {NoStop}%
\bibitem [{\citenamefont {Schachenmayer}(2008)}]{schachenmayer_dynamics_2008}%
  \BibitemOpen
  \bibfield  {author} {\bibinfo {author} {\bibfnamefont {J.}~\bibnamefont
  {Schachenmayer}},\ }\emph {\bibinfo {title} {Dynamics and {Long}-{Range}
  {Interactions} in 1d {Quantum} {Systems}}},\ \href@noop {} {\bibinfo {type}
  {Diploma {Thesis}}},\ \bibinfo  {school} {Technische Universit{\"a}t
  M{\"u}nchen}, \bibinfo{year}{2008}, \url{http://qoqms.phys.strath.ac.uk/downloads/schachenmayer_diploma_thesis.pdf}\BibitemShut {NoStop}%
\bibitem [{\citenamefont {Pirvu}\ \emph {et~al.}(2010)\citenamefont {Pirvu},
  \citenamefont {Murg}, \citenamefont {Cirac},\ and\ \citenamefont
  {Verstraete}}]{pirvu_matrix_2010}%
  \BibitemOpen
  \bibfield  {author} {\bibinfo {author} {\bibfnamefont {B.}~\bibnamefont
  {Pirvu}}, \bibinfo {author} {\bibfnamefont {V.}~\bibnamefont {Murg}},
  \bibinfo {author} {\bibfnamefont {J.~I.}\ \bibnamefont {Cirac}},\ and\
  \bibinfo {author} {\bibfnamefont {F.}~\bibnamefont {Verstraete}},\ }\bibfield
   {title} {\bibinfo {title} {Matrix product operator representations},\ }\href
  {https://doi.org/10.1088/1367-2630/12/2/025012} {\bibfield  {journal}
  {\bibinfo  {journal} {New J. Phys.}\ }\textbf {\bibinfo {volume} {12}},\
  \bibinfo {pages} {025012} (\bibinfo {year} {2010})}\BibitemShut {NoStop}%
\bibitem [{\citenamefont {Stoudenmire}\ and\ \citenamefont
  {White}(2010)}]{stoudenmire_minimally_2010}%
  \BibitemOpen
  \bibfield  {author} {\bibinfo {author} {\bibfnamefont {E.~M.}\ \bibnamefont
  {Stoudenmire}}\ and\ \bibinfo {author} {\bibfnamefont {S.~R.}\ \bibnamefont
  {White}},\ }\bibfield  {title} {\bibinfo {title} {Minimally entangled typical
  thermal state algorithms},\ }\href
  {https://doi.org/10.1088/1367-2630/12/5/055026} {\bibfield  {journal}
  {\bibinfo  {journal} {New J. Phys.}\ }\textbf {\bibinfo {volume} {12}},\
  \bibinfo {pages} {055026} (\bibinfo {year} {2010})}\BibitemShut {NoStop}%
\bibitem [{\citenamefont {Zaletel}\ \emph {et~al.}(2015)\citenamefont
  {Zaletel}, \citenamefont {Mong}, \citenamefont {Karrasch}, \citenamefont
  {Moore},\ and\ \citenamefont {Pollmann}}]{zaletel_time-evolving_2015}%
  \BibitemOpen
  \bibfield  {author} {\bibinfo {author} {\bibfnamefont {M.~P.}\ \bibnamefont
  {Zaletel}}, \bibinfo {author} {\bibfnamefont {R.~S.~K.}\ \bibnamefont
  {Mong}}, \bibinfo {author} {\bibfnamefont {C.}~\bibnamefont {Karrasch}},
  \bibinfo {author} {\bibfnamefont {J.~E.}\ \bibnamefont {Moore}},\ and\
  \bibinfo {author} {\bibfnamefont {F.}~\bibnamefont {Pollmann}},\ }\bibfield
  {title} {\bibinfo {title} {Time-evolving a matrix product state with
  long-ranged interactions},\ }\href
  {https://doi.org/10.1103/PhysRevB.91.165112} {\bibfield  {journal} {\bibinfo
  {journal} {Phys. Rev. B}\ }\textbf {\bibinfo {volume} {91}},\ \bibinfo
  {pages} {165112} (\bibinfo {year} {2015})}\BibitemShut {NoStop}%
\bibitem [{\citenamefont {Haegeman}\ \emph {et~al.}(2011)\citenamefont
  {Haegeman}, \citenamefont {Cirac}, \citenamefont {Osborne}, \citenamefont
  {Pi{\v z}orn}, \citenamefont {Verschelde},\ and\ \citenamefont
  {Verstraete}}]{haegeman_time-dependent_2011}%
  \BibitemOpen
  \bibfield  {author} {\bibinfo {author} {\bibfnamefont {J.}~\bibnamefont
  {Haegeman}}, \bibinfo {author} {\bibfnamefont {J.~I.}\ \bibnamefont {Cirac}},
  \bibinfo {author} {\bibfnamefont {T.~J.}\ \bibnamefont {Osborne}}, \bibinfo
  {author} {\bibfnamefont {I.}~\bibnamefont {Pi{\v z}orn}}, \bibinfo {author}
  {\bibfnamefont {H.}~\bibnamefont {Verschelde}},\ and\ \bibinfo {author}
  {\bibfnamefont {F.}~\bibnamefont {Verstraete}},\ }\bibfield  {title}
  {\bibinfo {title} {Time-{Dependent} {Variational} {Principle} for {Quantum}
  {Lattices}},\ }\href {https://doi.org/10.1103/PhysRevLett.107.070601}
  {\bibfield  {journal} {\bibinfo  {journal} {Phys. Rev. Lett.}\ }\textbf
  {\bibinfo {volume} {107}},\ \bibinfo {pages} {070601} (\bibinfo {year}
  {2011})}\BibitemShut {NoStop}%
\bibitem [{\citenamefont {Wall}\ and\ \citenamefont
  {Carr}(2012)}]{wall_out--equilibrium_2012}%
  \BibitemOpen
  \bibfield  {author} {\bibinfo {author} {\bibfnamefont {M.~L.}\ \bibnamefont
  {Wall}}\ and\ \bibinfo {author} {\bibfnamefont {L.~D.}\ \bibnamefont
  {Carr}},\ }\bibfield  {title} {\bibinfo {title} {Out-of-equilibrium dynamics
  with matrix product states},\ }\href
  {https://doi.org/10.1088/1367-2630/14/12/125015} {\bibfield  {journal}
  {\bibinfo  {journal} {New J. Phys.}\ }\textbf {\bibinfo {volume} {14}},\
  \bibinfo {pages} {125015} (\bibinfo {year} {2012})}\BibitemShut {NoStop}%
\bibitem [{\citenamefont {Lubich}\ \emph {et~al.}(2013)\citenamefont {Lubich},
  \citenamefont {Rohwedder}, \citenamefont {Schneider},\ and\ \citenamefont
  {Vandereycken}}]{lubich_dynamical_2013}%
  \BibitemOpen
  \bibfield  {author} {\bibinfo {author} {\bibfnamefont {C.}~\bibnamefont
  {Lubich}}, \bibinfo {author} {\bibfnamefont {T.}~\bibnamefont {Rohwedder}},
  \bibinfo {author} {\bibfnamefont {R.}~\bibnamefont {Schneider}},\ and\
  \bibinfo {author} {\bibfnamefont {B.}~\bibnamefont {\hbox{Vandereycken}}},\
  }\bibfield  {title} {\bibinfo {title} {Dynamical {Approximation} by
  {Hierarchical} {Tucker} and {Tensor}-{Train} {Tensors}},\ }\href
  {https://doi.org/10.1137/120885723} {\bibfield  {journal} {\bibinfo
  {journal} {SIAM J. \hbox{Matrix} Anal. Appl.}\ }\textbf {\bibinfo {volume} {34}},\
  \bibinfo {pages} {470} (\bibinfo {year} {2013})}\BibitemShut {NoStop}%
\bibitem [{\citenamefont {Lubich}\ \emph {et~al.}(2015)\citenamefont {Lubich},
  \citenamefont {Oseledets},\ and\ \citenamefont
  {Vandereycken}}]{lubich_time_2015}%
  \BibitemOpen
  \bibfield  {author} {\bibinfo {author} {\bibfnamefont {C.}~\bibnamefont
  {Lubich}}, \bibinfo {author} {\bibfnamefont {I.}~\bibnamefont {Oseledets}},\
  and\ \bibinfo {author} {\bibfnamefont {B.}~\bibnamefont {Vandereycken}},\
  }\bibfield  {title} {\bibinfo {title} {Time {Integration} of {Tensor}
  {Trains}},\ }\href {https://doi.org/10.1137/140976546} {\bibfield  {journal}
  {\bibinfo  {journal} {SIAM J. Numer. Anal.}\ }\textbf {\bibinfo {volume}
  {53}},\ \bibinfo {pages} {917} (\bibinfo {year} {2015})}\BibitemShut
  {NoStop}%
\bibitem [{\citenamefont {Haegeman}\ \emph {et~al.}(2016)\citenamefont
  {Haegeman}, \citenamefont {Lubich}, \citenamefont {Oseledets}, \citenamefont
  {Vandereycken},\ and\ \citenamefont {Verstraete}}]{haegeman_unifying_2016}%
  \BibitemOpen
  \bibfield  {author} {\bibinfo {author} {\bibfnamefont {J.}~\bibnamefont
  {Haegeman}}, \bibinfo {author} {\bibfnamefont {C.}~\bibnamefont {Lubich}},
  \bibinfo {author} {\bibfnamefont {I.}~\bibnamefont {Oseledets}}, \bibinfo
  {author} {\bibfnamefont {B.}~\bibnamefont {Vandereycken}},\ and\ \bibinfo
  {author} {\bibfnamefont {F.}~\bibnamefont {Verstraete}},\ }\bibfield  {title}
  {\bibinfo {title} {Unifying time evolution and optimization with matrix
  product states},\ }\href {https://doi.org/10.1103/PhysRevB.94.165116}
  {\bibfield  {journal} {\bibinfo  {journal} {Phys. Rev. B}\ }\textbf {\bibinfo
  {volume} {94}},\ \bibinfo {pages} {165116} (\bibinfo {year}
  {2016})}\BibitemShut {NoStop}%
\bibitem [{\citenamefont {Ronca}\ \emph {et~al.}(2017)\citenamefont {Ronca},
  \citenamefont {Li}, \citenamefont {Jimenez-Hoyos},\ and\ \citenamefont
  {Chan}}]{ronca_time-step_2017}%
  \BibitemOpen
  \bibfield  {author} {\bibinfo {author} {\bibfnamefont {E.}~\bibnamefont
  {Ronca}}, \bibinfo {author} {\bibfnamefont {Z.}~\bibnamefont {Li}}, \bibinfo
  {author} {\bibfnamefont {C.~A.}\ \bibnamefont {Jimenez-Hoyos}},\ and\
  \bibinfo {author} {\bibfnamefont {G.~K.-L.}\ \bibnamefont {Chan}},\
  }\bibfield  {title} {\bibinfo {title} {Time-{Step} {Targeting}
  {Time}-{Dependent} and {Dynamical} {Density} {Matrix} {Renormalization}
  {Group} {Algorithms} with ab {Initio} {Hamiltonians}},\ }\href
  {https://doi.org/10.1021/acs.jctc.7b00682} {\bibfield  {journal} {\bibinfo
  {journal} {J. Chem. Theory Comput.}\ }\textbf {\bibinfo {volume} {13}},\
  \bibinfo {pages} {5560} (\bibinfo {year} {2017})}\BibitemShut {NoStop}%
\bibitem [{\citenamefont {Paeckel}\ \emph {et~al.}(2019)\citenamefont
  {Paeckel}, \citenamefont {K{\"o}hler}, \citenamefont {Swoboda}, \citenamefont
  {Manmana}, \citenamefont {Schollw{\"o}ck},\ and\ \citenamefont
  {Hubig}}]{paeckel_time-evolution_2019}%
  \BibitemOpen
  \bibfield  {author} {\bibinfo {author} {\bibfnamefont {S.}~\bibnamefont
  {Paeckel}}, \bibinfo {author} {\bibfnamefont {T.}~\bibnamefont {K{\"o}hler}},
  \bibinfo {author} {\bibfnamefont {A.}~\bibnamefont {Swoboda}}, \bibinfo
  {author} {\bibfnamefont {S.~R.}\ \bibnamefont {Manmana}}, \bibinfo {author}
  {\bibfnamefont {U.}~\bibnamefont {Schollw{\"o}ck}},\ and\ \bibinfo {author}
  {\bibfnamefont {C.}~\bibnamefont {Hubig}},\ }\bibfield  {title} {\bibinfo
  {title} {Time-evolution methods for matrix-product states},\ }\href
  {https://doi.org/10.1016/j.aop.2019.167998} {\bibfield  {journal} {\bibinfo
  {journal} {Ann. Phys. (N. Y.)}\ }\textbf {\bibinfo {volume} {411}},\ \bibinfo
  {pages} {167998} (\bibinfo {year} {2019})}\BibitemShut {NoStop}%
\bibitem [{\citenamefont {Hashizume}\ \emph {et~al.}(2019)\citenamefont
  {Hashizume}, \citenamefont {Halimeh},\ and\ \citenamefont
  {McCulloch}}]{hashizume_hybrid_2019}%
  \BibitemOpen
  \bibfield  {author} {\bibinfo {author} {\bibfnamefont {T.}~\bibnamefont
  {Hashizume}}, \bibinfo {author} {\bibfnamefont {J.~C.}\ \bibnamefont
  {Halimeh}},\ and\ \bibinfo {author} {\bibfnamefont {I.~P.}\ \bibnamefont
  {McCulloch}},\ }\bibfield  {title} {\bibinfo {title} {Hybrid infinite
  time-evolving block decimation algorithm for long-range multi-dimensional
  quantum many-body systems},\ }\Eprint {https://arxiv.org/abs/1910.10726v1}
  {arXiv:1910.10726v1} [cond-mat.str-el]\BibitemShut
  {NoStop}%
\bibitem [{\citenamefont {Koffel}\ \emph {et~al.}(2012)\citenamefont {Koffel},
  \citenamefont {Lewenstein},\ and\ \citenamefont
  {Tagliacozzo}}]{koffel_entanglement_2012}%
  \BibitemOpen
  \bibfield  {author} {\bibinfo {author} {\bibfnamefont {T.}~\bibnamefont
  {Koffel}}, \bibinfo {author} {\bibfnamefont {M.}~\bibnamefont {Lewenstein}},\
  and\ \bibinfo {author} {\bibfnamefont {L.}~\bibnamefont {Tagliacozzo}},\
  }\bibfield  {title} {\bibinfo {title} {Entanglement {Entropy} for the
  {Long}-{Range} {Ising} {Chain} in a {Transverse} {Field}},\ }\href
  {https://doi.org/10.1103/PhysRevLett.109.267203} {\bibfield  {journal}
  {\bibinfo  {journal} {Phys. Rev. Lett.}\ }\textbf {\bibinfo {volume} {109}},\
  \bibinfo {pages} {267203} (\bibinfo {year} {2012})}\BibitemShut {NoStop}%
\bibitem [{\citenamefont {Hauke}\ and\ \citenamefont
  {Tagliacozzo}(2013)}]{hauke_spread_2013}%
  \BibitemOpen
  \bibfield  {author} {\bibinfo {author} {\bibfnamefont {P.}~\bibnamefont
  {Hauke}}\ and\ \bibinfo {author} {\bibfnamefont {L.}~\bibnamefont
  {Tagliacozzo}},\ }\bibfield  {title} {\bibinfo {title} {Spread of
  {Correlations} in {Long}-{Range} {Interacting} {Quantum} {Systems}},\ }\href
  {https://doi.org/10.1103/PhysRevLett.111.207202} {\bibfield  {journal}
  {\bibinfo  {journal} {Phys. Rev. Lett.}\ }\textbf {\bibinfo {volume} {111}},\
  \bibinfo {pages} {207202} (\bibinfo {year} {2013})}\BibitemShut {NoStop}%
\bibitem [{\citenamefont {Buyskikh}\ \emph {et~al.}(2016)\citenamefont
  {Buyskikh}, \citenamefont {Fagotti}, \citenamefont {Schachenmayer},
  \citenamefont {Essler},\ and\ \citenamefont
  {Daley}}]{buyskikh_entanglement_2016}%
  \BibitemOpen
  \bibfield  {author} {\bibinfo {author} {\bibfnamefont {A.~S.}\ \bibnamefont
  {Buyskikh}}, \bibinfo {author} {\bibfnamefont {M.}~\bibnamefont {Fagotti}},
  \bibinfo {author} {\bibfnamefont {J.}~\bibnamefont {Schachenmayer}}, \bibinfo
  {author} {\bibfnamefont {F.}~\bibnamefont {Essler}},\ and\ \bibinfo {author}
  {\bibfnamefont {A.~J.}\ \bibnamefont {Daley}},\ }\bibfield  {title} {\bibinfo
  {title} {Entanglement growth and correlation spreading with variable-range
  interactions in spin and fermionic tunneling models},\ }\href
  {https://doi.org/10.1103/PhysRevA.93.053620} {\bibfield  {journal} {\bibinfo
  {journal} {Phys. Rev. A}\ }\textbf {\bibinfo {volume} {93}},\ \bibinfo
  {pages} {053620} (\bibinfo {year} {2016})}\BibitemShut {NoStop}%
\bibitem [{\citenamefont {Halimeh}\ \emph {et~al.}(2017)\citenamefont
  {Halimeh}, \citenamefont {Zauner-Stauber}, \citenamefont {McCulloch},
  \citenamefont {de~Vega}, \citenamefont {Schollw{\"o}ck},\ and\ \citenamefont
  {Kastner}}]{halimeh_prethermalization_2017}%
  \BibitemOpen
  \bibfield  {author} {\bibinfo {author} {\bibfnamefont {J.~C.}\ \bibnamefont
  {Halimeh}}, \bibinfo {author} {\bibfnamefont {V.}~\bibnamefont
  {Zauner-Stauber}}, \bibinfo {author} {\bibfnamefont {I.~P.}\ \bibnamefont
  {McCulloch}}, \bibinfo {author} {\bibfnamefont {I.}~\bibnamefont {de~Vega}},
  \bibinfo {author} {\bibfnamefont {U.}~\bibnamefont {Schollw{\"o}ck}},\ and\
  \bibinfo {author} {\bibfnamefont {M.}~\bibnamefont {Kastner}},\ }\bibfield
  {title} {\bibinfo {title} {Prethermalization and persistent order in the
  absence of a thermal phase transition},\ }\href
  {https://doi.org/10.1103/PhysRevB.95.024302} {\bibfield  {journal} {\bibinfo
  {journal} {Phys. Rev. B}\ }\textbf {\bibinfo {volume} {95}},\ \bibinfo
  {pages} {024302} (\bibinfo {year} {2017})}\BibitemShut {NoStop}%
\bibitem [{\citenamefont {Jaschke}\ \emph {et~al.}(2017)\citenamefont
  {Jaschke}, \citenamefont {Maeda}, \citenamefont {Whalen}, \citenamefont
  {Wall},\ and\ \citenamefont {Carr}}]{jaschke_critical_2017}%
  \BibitemOpen
  \bibfield  {author} {\bibinfo {author} {\bibfnamefont {D.}~\bibnamefont
  {Jaschke}}, \bibinfo {author} {\bibfnamefont {K.}~\bibnamefont {Maeda}},
  \bibinfo {author} {\bibfnamefont {J.~D.}\ \bibnamefont {Whalen}}, \bibinfo
  {author} {\bibfnamefont {M.~L.}\ \bibnamefont {Wall}},\ and\ \bibinfo
  {author} {\bibfnamefont {L.~D.}\ \bibnamefont {Carr}},\ }\bibfield  {title}
  {\bibinfo {title} {Critical phenomena and {Kibble}{\textendash}{Zurek}
  scaling in the long-range quantum {Ising} chain},\ }\href
  {https://doi.org/10.1088/1367-2630/aa65bc} {\bibfield  {journal} {\bibinfo
  {journal} {New J. Phys.}\ }\textbf {\bibinfo {volume} {19}},\ \bibinfo
  {pages} {033032} (\bibinfo {year} {2017})}\BibitemShut {NoStop}%
\bibitem [{\citenamefont {Halimeh}\ and\ \citenamefont
  {Zauner-Stauber}(2017)}]{halimeh_dynamical_2017}%
  \BibitemOpen
  \bibfield  {author} {\bibinfo {author} {\bibfnamefont {J.~C.}\ \bibnamefont
  {Halimeh}}\ and\ \bibinfo {author} {\bibfnamefont {V.}~\bibnamefont
  {Zauner-Stauber}},\ }\bibfield  {title} {\bibinfo {title} {Dynamical phase
  diagram of quantum spin chains with long-range interactions},\ }\href
  {https://doi.org/10.1103/PhysRevB.96.134427} {\bibfield  {journal} {\bibinfo
  {journal} {Phys. Rev. B}\ }\textbf {\bibinfo {volume} {96}},\ \bibinfo
  {pages} {134427} (\bibinfo {year} {2017})}\BibitemShut {NoStop}%
\bibitem [{\citenamefont {Zauner-Stauber}\ and\ \citenamefont
  {Halimeh}(2017)}]{zauner-stauber_probing_2017}%
  \BibitemOpen
  \bibfield  {author} {\bibinfo {author} {\bibfnamefont {V.}~\bibnamefont
  {Zauner-Stauber}}\ and\ \bibinfo {author} {\bibfnamefont {J.~C.}\
  \bibnamefont {Halimeh}},\ }\bibfield  {title} {\bibinfo {title} {Probing the
  anomalous dynamical phase in long-range quantum spin chains through
  {Fisher}-zero lines},\ }\href {https://doi.org/10.1103/PhysRevE.96.062118}
  {\bibfield  {journal} {\bibinfo  {journal} {Phys. Rev. E}\ }\textbf {\bibinfo
  {volume} {96}},\ \bibinfo {pages} {062118} (\bibinfo {year}
  {2017})}\BibitemShut {NoStop}%
\bibitem [{\citenamefont {{\v Z}unkovi{\v c}}\ \emph
  {et~al.}(2018)\citenamefont {{\v Z}unkovi{\v c}}, \citenamefont {Heyl},
  \citenamefont {Knap},\ and\ \citenamefont {Silva}}]{zunkovic_dynamical_2018}%
  \BibitemOpen
  \bibfield  {author} {\bibinfo {author} {\bibfnamefont {B.}~\bibnamefont {{\v
  Z}unkovi{\v c}}}, \bibinfo {author} {\bibfnamefont {M.}~\bibnamefont {Heyl}},
  \bibinfo {author} {\bibfnamefont {M.}~\bibnamefont {Knap}},\ and\ \bibinfo
  {author} {\bibfnamefont {A.}~\bibnamefont {Silva}},\ }\bibfield  {title}
  {\bibinfo {title} {Dynamical {Quantum} {Phase} {Transitions} in {Spin}
  {Chains} with {Long}-{Range} {Interactions}: {Merging} {Different} {Concepts}
  of {Nonequilibrium} {Criticality}},\ }\href
  {https://doi.org/10.1103/PhysRevLett.120.130601} {\bibfield  {journal}
  {\bibinfo  {journal} {Phys. Rev. Lett.}\ }\textbf {\bibinfo {volume} {120}},\
  \bibinfo {pages} {130601} (\bibinfo {year} {2018})}\BibitemShut {NoStop}%
\bibitem [{\citenamefont {Pappalardi}\ \emph {et~al.}(2018)\citenamefont
  {Pappalardi}, \citenamefont {Russomanno}, \citenamefont {{\v Z}unkovi{\v c}},
  \citenamefont {Iemini}, \citenamefont {Silva},\ and\ \citenamefont
  {Fazio}}]{pappalardi_scrambling_2018}%
  \BibitemOpen
  \bibfield  {author} {\bibinfo {author} {\bibfnamefont {S.}~\bibnamefont
  {Pappalardi}}, \bibinfo {author} {\bibfnamefont {A.}~\bibnamefont
  {Russomanno}}, \bibinfo {author} {\bibfnamefont {B.}~\bibnamefont {{\v
  Z}unkovi{\v c}}}, \bibinfo {author} {\bibfnamefont {F.}~\bibnamefont
  {Iemini}}, \bibinfo {author} {\bibfnamefont {A.}~\bibnamefont {Silva}},\ and\
  \bibinfo {author} {\bibfnamefont {R.}~\bibnamefont {Fazio}},\ }\bibfield
  {title} {\bibinfo {title} {Scrambling and entanglement spreading in
  long-range spin chains},\ }\href {https://doi.org/10.1103/PhysRevB.98.134303}
  {\bibfield  {journal} {\bibinfo  {journal} {Phys. Rev. B}\ }\textbf {\bibinfo
  {volume} {98}},\ \bibinfo {pages} {134303} (\bibinfo {year}
  {2018})}\BibitemShut {NoStop}%
\bibitem [{\citenamefont {Lerose}\ \emph
  {et~al.}(2019{\natexlab{a}})\citenamefont {Lerose}, \citenamefont {{\v
  Z}unkovi{\v c}}, \citenamefont {Marino}, \citenamefont {Gambassi},\ and\
  \citenamefont {Silva}}]{lerose_impact_2019}%
  \BibitemOpen
  \bibfield  {author} {\bibinfo {author} {\bibfnamefont {A.}~\bibnamefont
  {Lerose}}, \bibinfo {author} {\bibfnamefont {B.}~\bibnamefont {{\v
  Z}unkovi{\v c}}}, \bibinfo {author} {\bibfnamefont {J.}~\bibnamefont
  {Marino}}, \bibinfo {author} {\bibfnamefont {A.}~\bibnamefont {Gambassi}},\
  and\ \bibinfo {author} {\bibfnamefont {A.}~\bibnamefont {Silva}},\ }\bibfield
   {title} {\bibinfo {title} {Impact of nonequilibrium fluctuations on
  prethermal dynamical phase transitions in long-range interacting spin
  chains},\ }\href {https://doi.org/10.1103/PhysRevB.99.045128} {\bibfield
  {journal} {\bibinfo  {journal} {Phys. Rev. B}\ }\textbf {\bibinfo {volume}
  {99}},\ \bibinfo {pages} {045128} (\bibinfo {year}
  {2019}{\natexlab{a}})}\BibitemShut {NoStop}%
\bibitem [{\citenamefont {Kloss}\ and\ \citenamefont
  {Bar~Lev}(2019)}]{kloss_spin_2019}%
  \BibitemOpen
  \bibfield  {author} {\bibinfo {author} {\bibfnamefont {B.}~\bibnamefont
  {Kloss}}\ and\ \bibinfo {author} {\bibfnamefont {Y.}~\bibnamefont
  {Bar~Lev}},\ }\bibfield  {title} {\bibinfo {title} {Spin transport in a
  long-range-interacting spin chain},\ }\href
  {https://doi.org/10.1103/PhysRevA.99.032114} {\bibfield  {journal} {\bibinfo
  {journal} {Phys. Rev. A}\ }\textbf {\bibinfo {volume} {99}},\ \bibinfo
  {pages} {032114} (\bibinfo {year} {2019})}\BibitemShut {NoStop}%
\bibitem [{\citenamefont {Lerose}\ \emph
  {et~al.}(2019{\natexlab{b}})\citenamefont {Lerose}, \citenamefont {{\v
  Z}unkovi{\v c}}, \citenamefont {Silva},\ and\ \citenamefont
  {Gambassi}}]{lerose_quasilocalized_2019}%
  \BibitemOpen
  \bibfield  {author} {\bibinfo {author} {\bibfnamefont {A.}~\bibnamefont
  {Lerose}}, \bibinfo {author} {\bibfnamefont {B.}~\bibnamefont {{\v
  Z}unkovi{\v c}}}, \bibinfo {author} {\bibfnamefont {A.}~\bibnamefont
  {Silva}},\ and\ \bibinfo {author} {\bibfnamefont {A.}~\bibnamefont
  {Gambassi}},\ }\bibfield  {title} {\bibinfo {title} {Quasilocalized
  excitations induced by long-range interactions in translationally invariant
  quantum spin chains},\ }\href {https://doi.org/10.1103/PhysRevB.99.121112}
  {\bibfield  {journal} {\bibinfo  {journal} {Phys. Rev. B}\ }\textbf {\bibinfo
  {volume} {99}},\ \bibinfo {pages} {121112} (\bibinfo {year}
  {2019}{\natexlab{b}})}\BibitemShut {NoStop}%
\bibitem [{\citenamefont {Zhou}\ \emph {et~al.}(2019)\citenamefont {Zhou},
  \citenamefont {Xu}, \citenamefont {Chen}, \citenamefont {Guo},\ and\
  \citenamefont {Swingle}}]{zhou_operator_2019}%
  \BibitemOpen
  \bibfield  {author} {\bibinfo {author} {\bibfnamefont {T.}~\bibnamefont
  {Zhou}}, \bibinfo {author} {\bibfnamefont {S.}~\bibnamefont {Xu}}, \bibinfo
  {author} {\bibfnamefont {X.}~\bibnamefont {Chen}}, \bibinfo {author}
  {\bibfnamefont {A.}~\bibnamefont {Guo}},\ and\ \bibinfo {author}
  {\bibfnamefont {B.}~\bibnamefont {Swingle}},\ }\bibfield  {title} {\bibinfo
  {title} {Operator {L\'evy} Flight: Light Cones in Chaotic Long-Range
  Interacting Systems},\ }\href{https://doi.org/10.1103/PhysRevLett.124.180601}
  {\bibfield {journal} {\bibinfo {journal} {Phys. Rev. Lett.}\ }\textbf {\bibinfo {volume} {124}},\ \bibinfo {pages} {180601}\ (\bibinfo {year} {2020})}\BibitemShut {NoStop}%
\bibitem [{\citenamefont {Piccitto}\ \emph {et~al.}(2019)\citenamefont
  {Piccitto}, \citenamefont {{\v Z}unkovi{\v c}},\ and\ \citenamefont
  {Silva}}]{piccitto_dynamical_2019}%
  \BibitemOpen
  \bibfield  {author} {\bibinfo {author} {\bibfnamefont {G.}~\bibnamefont
  {Piccitto}}, \bibinfo {author} {\bibfnamefont {B.}~\bibnamefont {{\v
  Z}unkovi{\v c}}},\ and\ \bibinfo {author} {\bibfnamefont {A.}~\bibnamefont
  {Silva}},\ }\bibfield  {title} {\bibinfo {title} {Dynamical phase diagram of
  a quantum {Ising} chain with long-range interactions},\ }\href
  {https://doi.org/10.1103/PhysRevB.100.180402} {\bibfield  {journal} {\bibinfo
   {journal} {Phys. Rev. B}\ }\textbf {\bibinfo {volume} {100}},\ \bibinfo
  {pages} {180402} (\bibinfo {year} {2019})}\BibitemShut {NoStop}%
\bibitem [{\citenamefont {Borrelli}\ and\ \citenamefont
  {Gelin}(2017)}]{borrelli_simulation_2017}%
  \BibitemOpen
  \bibfield  {author} {\bibinfo {author} {\bibfnamefont {R.}~\bibnamefont
  {Borrelli}}\ and\ \bibinfo {author} {\bibfnamefont {M.~F.}\ \bibnamefont
  {Gelin}},\ }\bibfield  {title} {\bibinfo {title} {Simulation of {Quantum}
  {Dynamics} of {Excitonic} {Systems} at {Finite} {Temperature}: an efficient
  method based on {Thermo} {Field} {Dynamics}},\ }\href
  {https://doi.org/10.1038/s41598-017-08901-2} {\bibfield  {journal} {\bibinfo
  {journal} {Sci. Rep.}\ }\textbf {\bibinfo {volume} {7}},\ \bibinfo {pages}
  {1} (\bibinfo {year} {2017})}\BibitemShut {NoStop}%
\bibitem [{\citenamefont {Borrelli}(2018)}]{borrelli_theoretical_2018}%
  \BibitemOpen
  \bibfield  {author} {\bibinfo {author} {\bibfnamefont {R.}~\bibnamefont
  {Borrelli}},\ }\bibfield  {title} {\bibinfo {title} {Theoretical study of
  charge-transfer processes at finite temperature using a novel thermal
  {Schr{\"o}dinger} equation},\ }\href
  {https://doi.org/10.1016/j.chemphys.2018.06.005} {\bibfield  {journal}
  {\bibinfo  {journal} {Chem. Phys.}\ }\textbf {\bibinfo {volume} {515}},\
  \bibinfo {pages} {236} (\bibinfo {year} {2018})}\BibitemShut {NoStop}%
\bibitem [{\citenamefont {Kurashige}(2018)}]{kurashige_matrix_2018}%
  \BibitemOpen
  \bibfield  {author} {\bibinfo {author} {\bibfnamefont {Y.}~\bibnamefont
  {Kurashige}},\ }\bibfield  {title} {\bibinfo {title} {Matrix product state
  formulation of the multiconfiguration time-dependent {Hartree} theory},\
  }\href {https://doi.org/10.1063/1.5051498} {\bibfield  {journal} {\bibinfo
  {journal} {J. Chem. Phys.}\ }\textbf {\bibinfo {volume} {149}},\ \bibinfo
  {pages} {194114} (\bibinfo {year} {2018})}\BibitemShut {NoStop}%
\bibitem [{\citenamefont {Baiardi}\ and\ \citenamefont
  {Reiher}(2019)}]{baiardi_large-scale_2019}%
  \BibitemOpen
  \bibfield  {author} {\bibinfo {author} {\bibfnamefont {A.}~\bibnamefont
  {Baiardi}}\ and\ \bibinfo {author} {\bibfnamefont {M.}~\bibnamefont
  {Reiher}},\ }\bibfield  {title} {\bibinfo {title} {Large-{Scale} {Quantum}
  {Dynamics} with {Matrix} {Product} {States}},\ }\href
  {https://doi.org/10.1021/acs.jctc.9b00301} {\bibfield  {journal} {\bibinfo
  {journal} {J. Chem. Theory Comput.}\ }\textbf {\bibinfo {volume} {15}},\
  \bibinfo {pages} {3481} (\bibinfo {year} {2019})}\BibitemShut {NoStop}%
\bibitem [{\citenamefont {Gelin}\ \emph {et~al.}(2019)\citenamefont {Gelin},
  \citenamefont {Borrelli},\ and\ \citenamefont {Domcke}}]{gelin_origin_2019}%
  \BibitemOpen
  \bibfield  {author} {\bibinfo {author} {\bibfnamefont {M.~F.}\ \bibnamefont
  {Gelin}}, \bibinfo {author} {\bibfnamefont {R.}~\bibnamefont {Borrelli}},\
  and\ \bibinfo {author} {\bibfnamefont {W.}~\bibnamefont {Domcke}},\
  }\bibfield  {title} {\bibinfo {title} {Origin of {Unexpectedly} {Simple}
  {Oscillatory} {Responses} in the {Excited}-{State} {Dynamics} of {Disordered}
  {Molecular} {Aggregates}},\ }\href
  {https://doi.org/10.1021/acs.jpclett.9b00840} {\bibfield  {journal} {\bibinfo
   {journal} {J. Phys. Chem. Lett.}\ }\textbf {\bibinfo {volume} {10}},\
  \bibinfo {pages} {2806} (\bibinfo {year} {2019})}\BibitemShut {NoStop}%
\bibitem [{\citenamefont {Kloss}\ \emph {et~al.}(2019)\citenamefont {Kloss},
  \citenamefont {Reichman},\ and\ \citenamefont
  {Tempelaar}}]{kloss_multiset_2019}%
  \BibitemOpen
  \bibfield  {author} {\bibinfo {author} {\bibfnamefont {B.}~\bibnamefont
  {Kloss}}, \bibinfo {author} {\bibfnamefont {D.~R.}\ \bibnamefont
  {Reichman}},\ and\ \bibinfo {author} {\bibfnamefont {R.}~\bibnamefont
  {Tempelaar}},\ }\bibfield  {title} {\bibinfo {title} {Multiset {Matrix}
  {Product} {State} {Calculations} {Reveal} {Mobile} {Franck}-{Condon}
  {Excitations} {Under} {Strong} {Holstein}-{Type} {Coupling}},\ }\href
  {https://doi.org/10.1103/PhysRevLett.123.126601} {\bibfield  {journal}
  {\bibinfo  {journal} {Phys. Rev. Lett.}\ }\textbf {\bibinfo {volume} {123}},\
  \bibinfo {pages} {126601} (\bibinfo {year} {2019})}\BibitemShut {NoStop}%
\bibitem [{\citenamefont {Xie}\ \emph {et~al.}(2019)\citenamefont {Xie},
  \citenamefont {Liu}, \citenamefont {Yao}, \citenamefont {Schollw{\"o}ck},
  \citenamefont {Liu},\ and\ \citenamefont {Ma}}]{xie_time-dependent_2019}%
  \BibitemOpen
  \bibfield  {author} {\bibinfo {author} {\bibfnamefont {X.}~\bibnamefont
  {Xie}}, \bibinfo {author} {\bibfnamefont {Y.}~\bibnamefont {Liu}}, \bibinfo
  {author} {\bibfnamefont {Y.}~\bibnamefont {Yao}}, \bibinfo {author}
  {\bibfnamefont {U.}~\bibnamefont {Schollw{\"o}ck}}, \bibinfo {author}
  {\bibfnamefont {C.}~\bibnamefont {Liu}},\ and\ \bibinfo {author}
  {\bibfnamefont {H.}~\bibnamefont {Ma}},\ }\bibfield  {title} {\bibinfo
  {title} {Time-dependent density matrix renormalization group quantum dynamics
  for realistic chemical systems},\ }\href {https://doi.org/10.1063/1.5125945}
  {\bibfield  {journal} {\bibinfo  {journal} {J. Chem. Phys.}\ }\textbf
  {\bibinfo {volume} {151}},\ \bibinfo {pages} {224101} (\bibinfo {year}
  {2019})}\BibitemShut {NoStop}%
\bibitem [{\citenamefont {Li}\ \emph {et~al.}(2019)\citenamefont {Li},
  \citenamefont {Ren},\ and\ \citenamefont {Shuai}}]{li_numerical_2019}%
  \BibitemOpen
  \bibfield  {author} {\bibinfo {author} {\bibfnamefont {W.}~\bibnamefont
  {Li}}, \bibinfo {author} {\bibfnamefont {J.}~\bibnamefont {Ren}},\ and\
  \bibinfo {author} {\bibfnamefont {Z.}~\bibnamefont {Shuai}},\ }\bibfield
  {title} {\bibinfo {title} {Numerical assessment for accuracy and GPU acceleration of TD-DMRG time evolution schemes},\ }\href {https://doi.org/10.1063/1.5135363} {\bibfield  {journal} {\bibinfo
  		{journal} {J. Chem. Phys.}\ }\textbf {\bibinfo {volume} {152}},\ \bibinfo
  	{pages} {024127} (\bibinfo {year} {2020})}\BibitemShut {NoStop}%
\bibitem [{\citenamefont {Hager}\ \emph {et~al.}(2004)\citenamefont {Hager},
  \citenamefont {Jeckelmann}, \citenamefont {Fehske},\ and\ \citenamefont
  {Wellein}}]{hager_parallelization_2004}%
  \BibitemOpen
  \bibfield  {author} {\bibinfo {author} {\bibfnamefont {G.}~\bibnamefont
  {Hager}}, \bibinfo {author} {\bibfnamefont {E.}~\bibnamefont {Jeckelmann}},
  \bibinfo {author} {\bibfnamefont {H.}~\bibnamefont {Fehske}},\ and\ \bibinfo
  {author} {\bibfnamefont {G.}~\bibnamefont {Wellein}},\ }\bibfield  {title}
  {\bibinfo {title} {Parallelization strategies for density matrix
  renormalization group algorithms on shared-memory systems},\ }\href
  {https://doi.org/10.1016/j.jcp.2003.09.018} {\bibfield  {journal} {\bibinfo
  {journal} {J. Comput. Phys.}\ }\textbf {\bibinfo {volume} {194}},\ \bibinfo
  {pages} {795} (\bibinfo {year} {2004})}\BibitemShut {NoStop}%
\bibitem [{\citenamefont {Nemes}\ \emph {et~al.}(2014)\citenamefont {Nemes},
  \citenamefont {Barcza}, \citenamefont {Nagy}, \citenamefont {Legeza},\ and\
  \citenamefont {Szolgay}}]{nemes_density_2014}%
  \BibitemOpen
  \bibfield  {author} {\bibinfo {author} {\bibfnamefont {C.}~\bibnamefont
  {Nemes}}, \bibinfo {author} {\bibfnamefont {G.}~\bibnamefont {Barcza}},
  \bibinfo {author} {\bibfnamefont {Z.}~\bibnamefont {Nagy}}, \bibinfo {author}
  {\bibfnamefont {{\"O}.}~\bibnamefont {Legeza}},\ and\ \bibinfo {author}
  {\bibfnamefont {P.}~\bibnamefont {\hbox{Szolgay}}},\ }\bibfield  {title} {\bibinfo
  {title} {The density matrix renormalization group algorithm on kilo-processor
  architectures: {Implementation} and trade-offs},\ }\href
  {https://doi.org/10.1016/j.cpc.2014.02.021} {\bibfield  {journal} {\bibinfo
  {journal} {Comput. Phys. Commun.}\ }\textbf {\bibinfo {volume} {185}},\
  \bibinfo {pages} {1570} (\bibinfo {year} {2014})}\BibitemShut {NoStop}%
\bibitem [{\citenamefont {Kantian}\ \emph {et~al.}(2019)\citenamefont
  {Kantian}, \citenamefont {Dolfi}, \citenamefont {Troyer},\ and\ \citenamefont
  {Giamarchi}}]{kantian_understanding_2019}%
  \BibitemOpen
  \bibfield  {author} {\bibinfo {author} {\bibfnamefont {A.}~\bibnamefont
  {Kantian}}, \bibinfo {author} {\bibfnamefont {M.}~\bibnamefont {Dolfi}},
  \bibinfo {author} {\bibfnamefont {M.}~\bibnamefont {Troyer}},\ and\ \bibinfo
  {author} {\bibfnamefont {T.}~\bibnamefont {Giamarchi}},\ }\bibfield  {title}
  {\bibinfo {title} {Understanding repulsively mediated superconductivity of
  correlated electrons via massively parallel density matrix renormalization
  group},\ }\href {https://doi.org/10.1103/PhysRevB.100.075138} {\bibfield
  {journal} {\bibinfo  {journal} {Phys. Rev. B}\ }\textbf {\bibinfo {volume}
  {100}},\ \bibinfo {pages} {075138} (\bibinfo {year} {2019})}\BibitemShut
  {NoStop}%
\bibitem [{\citenamefont {Kurashige}\ and\ \citenamefont
  {Yanai}(2009)}]{kurashige_high-performance_2009}%
  \BibitemOpen
  \bibfield  {author} {\bibinfo {author} {\bibfnamefont {Y.}~\bibnamefont
  {Kurashige}}\ and\ \bibinfo {author} {\bibfnamefont {T.}~\bibnamefont
  {Yanai}},\ }\bibfield  {title} {\bibinfo {title} {High-performance ab initio
  density matrix renormalization group method: {Applicability} to large-scale
  multireference problems for metal compounds},\ }\href
  {https://doi.org/10.1063/1.3152576} {\bibfield  {journal} {\bibinfo
  {journal} {J. Chem. Phys.}\ }\textbf {\bibinfo {volume} {130}},\ \bibinfo
  {pages} {234114} (\bibinfo {year} {2009})}\BibitemShut {NoStop}%
\bibitem [{\citenamefont {Chan}(2004)}]{chan_algorithm_2004}%
  \BibitemOpen
  \bibfield  {author} {\bibinfo {author} {\bibfnamefont {G.~K.-L.}\
  \bibnamefont {Chan}},\ }\bibfield  {title} {\bibinfo {title} {An algorithm
  for large scale density matrix renormalization group calculations},\ }\href
  {https://doi.org/10.1063/1.1638734} {\bibfield  {journal} {\bibinfo
  {journal} {J. Chem. Phys.}\ }\textbf {\bibinfo {volume} {120}},\ \bibinfo
  {pages} {3172} (\bibinfo {year} {2004})}\BibitemShut {NoStop}%
\bibitem [{\citenamefont {Skaugen}(2013)}]{skaugen_time_2013}%
  \BibitemOpen
  \bibfield  {author} {\bibinfo {author} {\bibfnamefont {A.}~\bibnamefont
  {Skaugen}},\ }\emph {\bibinfo {title} {Time evolution of magnetic {Bloch}
  oscillations using {TEBD}}},\ {Master thesis},\ \bibinfo
  {school} {University of Oslo},\ \bibinfo {year} {2013},\ \url
  {https://www.duo.uio.no/handle/10852/38773}\BibitemShut {NoStop}%
\bibitem [{\citenamefont {Wall}(2015)}]{wall_quantum_2015}%
  \BibitemOpen
  \bibfield  {author} {\bibinfo {author} {\bibfnamefont {M.~L.}\ \bibnamefont
  {Wall}},\ }\href {https://doi.org/10.1007/978-3-319-14252-4} {\emph {\bibinfo
  {title} {Quantum {Many}-{Body} {Physics} of {Ultracold} {Molecules} in
  {Optical} {Lattices}}}},\ Springer {Theses}\ (\bibinfo  {publisher}
  {Springer, Cham},\ \bibinfo {year} {2015})\BibitemShut {NoStop}%
\bibitem [{\citenamefont {Urbanek}\ and\ \citenamefont
  {Sold{\'a}n}(2016)}]{urbanek_parallel_2016}%
  \BibitemOpen
  \bibfield  {author} {\bibinfo {author} {\bibfnamefont {M.}~\bibnamefont
  {Urbanek}}\ and\ \bibinfo {author} {\bibfnamefont {P.}~\bibnamefont
  {Sold{\'a}n}},\ }\bibfield  {title} {\bibinfo {title} {Parallel
  implementation of the time-evolving block decimation algorithm for the
  {Bose}{\textendash}{Hubbard} model},\ }\href
  {https://doi.org/https://doi.org/10.1016/j.cpc.2015.10.016} {\bibfield
  {journal} {\bibinfo  {journal} {Comput. Phys. Commun.}\ }\textbf {\bibinfo {volume} {199}},\ \bibinfo {pages} {170} (\bibinfo {year}
  {2016})}\BibitemShut{NoStop}%
\bibitem [{vol(2019)}]{volokitin_propagating_2019}%
  \BibitemOpen \bibfield {author} {\bibinfo {author} {\bibfnamefont {V.}~\bibnamefont {Volokitin}}, \bibinfo {author} {\bibfnamefont {I.}~\bibnamefont {Vakulchyk}}, \bibinfo {author} {\bibfnamefont {E.}~\bibnamefont {Kozinov}},
  \bibinfo {author} {\bibfnamefont {A.}~\bibnamefont {\hbox{Liniov}}}, \bibinfo {author} {\bibfnamefont {I.}~\bibnamefont {Meyerov}}, \bibinfo {author} {\bibfnamefont {M.}~\bibnamefont {Ivanchenko}}, {\bibinfo {author} {\bibfnamefont {T.}~\bibnamefont {Laptyeva}}}, \ and\ \bibinfo {author} {\bibfnamefont {S.}~\bibnamefont {Denisov}},} \bibfield  {title} {\bibinfo {title} {Propagating large open quantum systems towards their asymptotic states: cluster implementation of the time-evolving block decimation scheme},\ }\href {https://doi.org/10.1088/1742-6596/1392/1/012061}{\bibfield {journal} {\bibinfo {journal}
  {J.Phys.: Conf. Ser.}\ } \textbf {\bibinfo {volume} {1392}} \bibinfo {pages} {012061}\ (\bibinfo {year} {2019})}\BibitemShut
  {NoStop}%
\bibitem [{\citenamefont {Stoudenmire}\ and\ \citenamefont
  {White}(2013)}]{stoudenmire_real-space_2013}%
  \BibitemOpen
  \bibfield  {author} {\bibinfo {author} {\bibfnamefont {E.~M.}\ \bibnamefont
  {Stoudenmire}}\ and\ \bibinfo {author} {\bibfnamefont {S.~R.}\ \bibnamefont
  {White}},\ }\bibfield  {title} {\bibinfo {title} {Real-space parallel density
  matrix renormalization group},\ }\href
  {https://doi.org/10.1103/PhysRevB.87.155137} {\bibfield  {journal} {\bibinfo
  {journal} {Phys. Rev. B}\ }\textbf {\bibinfo {volume} {87}},\ \bibinfo
  {pages} {155137} (\bibinfo {year} {2013})}\BibitemShut {NoStop}%
\bibitem [{\citenamefont {Depenbrock}(2013)}]{depenbrock_tensor_2013}%
  \BibitemOpen
  \bibfield  {author} {\bibinfo {author} {\bibfnamefont {S.}~\bibnamefont
  {Depenbrock}},\ }\emph {\bibinfo {title} {Tensor {Networks} for the
  {Simulation} of {Strongly} {Correlated} {Systems}}},\ {\bibinfo {type} {{PhD} {Thesis}}},\
  \bibinfo  {school} {Ludwig-Maximilians-Universit{\"a}t M{\"u}nchen},\ \bibinfo
  {year} {2013},\ \url
  {https://edoc.ub.uni-muenchen.de/15963/}\BibitemShut {NoStop}%
\bibitem [{\citenamefont {Ueda}(2018)}]{ueda_infinite-size_2018}%
  \BibitemOpen
  \bibfield  {author} {\bibinfo {author} {\bibfnamefont {H.}~\bibnamefont
  {Ueda}},\ }\bibfield  {title} {\bibinfo {title} {Infinite-{Size} {Density}
  {Matrix} {Renormalization} {Group} with {Parallel} {Hida}{\textquoteright}s
  {Algorithm}},\ }\href {https://doi.org/10.7566/JPSJ.87.074005} {\bibfield
  {journal} {\bibinfo  {journal} {J. Phys. Soc. Jpn.}\ }\textbf {\bibinfo
  {volume} {87}},\ \bibinfo {pages} {074005} (\bibinfo {year}
  {2018})}\BibitemShut {NoStop}%
\bibitem [{\citenamefont {Lubasch}\ \emph
  {et~al.}(2014{\natexlab{a}})\citenamefont {Lubasch}, \citenamefont {Cirac},\
  and\ \citenamefont {Ba{\~n}uls}}]{lubasch_unifying_2014}%
  \BibitemOpen
  \bibfield  {author} {\bibinfo {author} {\bibfnamefont {M.}~\bibnamefont
  {Lubasch}}, \bibinfo {author} {\bibfnamefont {J.~I.}\ \bibnamefont {Cirac}},\
  and\ \bibinfo {author} {\bibfnamefont {M.-C.}\ \bibnamefont {Ba{\~n}uls}},\
  }\bibfield  {title} {\bibinfo {title} {Unifying projected entangled pair
  state contractions},\ }\href {https://doi.org/10.1088/1367-2630/16/3/033014}
  {\bibfield  {journal} {\bibinfo  {journal} {New J. Phys.}\ }\textbf {\bibinfo
  {volume} {16}},\ \bibinfo {pages} {033014} (\bibinfo {year}
  {2014}{\natexlab{a}})}\BibitemShut {NoStop}%
\bibitem [{\citenamefont {Lubasch}\ \emph
  {et~al.}(2014{\natexlab{b}})\citenamefont {Lubasch}, \citenamefont {Cirac},\
  and\ \citenamefont {Ba{\~n}uls}}]{lubasch_algorithms_2014}%
  \BibitemOpen
  \bibfield  {author} {\bibinfo {author} {\bibfnamefont {M.}~\bibnamefont
  {Lubasch}}, \bibinfo {author} {\bibfnamefont {J.~I.}\ \bibnamefont {Cirac}},\
  and\ \bibinfo {author} {\bibfnamefont {M.-C.}\ \bibnamefont {Ba{\~n}uls}},\
  }\bibfield  {title} {\bibinfo {title} {Algorithms for finite projected
  entangled pair states},\ }\href {https://doi.org/10.1103/PhysRevB.90.064425}
  {\bibfield  {journal} {\bibinfo  {journal} {Phys. Rev. B}\ }\textbf {\bibinfo
  {volume} {90}},\ \bibinfo {pages} {064425} (\bibinfo {year}
  {2014}{\natexlab{b}})}\BibitemShut {NoStop}%
\bibitem [{\citenamefont {Oseledets}(2011)}]{oseledets_tensor-train_2011}%
  \BibitemOpen
  \bibfield  {author} {\bibinfo {author} {\bibfnamefont {I.~V.}\ \bibnamefont
  {Oseledets}},\ }\bibfield  {title} {\bibinfo {title} {Tensor-{Train}
  {Decomposition}},\ }\href {https://doi.org/10.1137/090752286} {\bibfield
  {journal} {\bibinfo  {journal} {SIAM J. Sci. Comput.}\ }\textbf {\bibinfo
  {volume} {33}},\ \bibinfo {pages} {2295} (\bibinfo {year}
  {2011})}\BibitemShut {NoStop}%
\bibitem [{\citenamefont {Perez-Garcia}\ \emph {et~al.}(2007)\citenamefont
  {Perez-Garcia}, \citenamefont {Verstraete}, \citenamefont {Wolf},\ and\
  \citenamefont {Cirac}}]{perez-garcia_matrix_2007}%
  \BibitemOpen
  \bibfield  {author} {\bibinfo {author} {\bibfnamefont {D.}~\bibnamefont
  {Perez-Garcia}}, \bibinfo {author} {\bibfnamefont {F.}~\bibnamefont
  {Verstraete}}, \bibinfo {author} {\bibfnamefont {M.~M.}\ \bibnamefont
  {Wolf}},\ and\ \bibinfo {author} {\bibfnamefont {J.~I.}\ \bibnamefont
  {Cirac}},\ }\bibfield  {title} {\bibinfo {title} {Matrix {Product} {State}
  {Representations}},\ }\href
{http://www.rintonpress.com/journals/qiconline.html#v7n56}{\bibfield
  {journal} {\bibinfo {journal} {Quantum Inf. Comput.}\ }\textbf {\bibinfo
  {volume} {7}},\ \bibinfo {pages} {401} (\bibinfo {year} {2007})}\BibitemShut
  {NoStop}%
\bibitem [{\citenamefont {Penrose}(1971)}]{penrose_applications_1971}%
  \BibitemOpen
  \bibfield  {author} {\bibinfo {author} {\bibfnamefont {R.}~\bibnamefont
  {Penrose}},\ }\bibfield  {title} {\bibinfo {title} {Applications of negative
  dimensional tensors},\ }in\ \href@noop {} {\emph {\bibinfo {booktitle}
  {Combinatorial {Mathematics} and its {Applications}}}},\ \bibinfo {editor}
  {edited by\ \bibinfo {editor} {\bibfnamefont {D.~J.~A.}\ \bibnamefont
  {Welsh}}}\ (\bibinfo  {publisher} {Academic Press},\ \bibinfo {address}
  {London},\ \bibinfo {year} {1971})\BibitemShut {NoStop}%
\bibitem [{\citenamefont {Bridgeman}\ and\ \citenamefont
  {Chubb}(2017)}]{bridgeman_hand-waving_2017}%
  \BibitemOpen
  \bibfield  {author} {\bibinfo {author} {\bibfnamefont {J.~C.}\ \bibnamefont
  {Bridgeman}}\ and\ \bibinfo {author} {\bibfnamefont {C.~T.}\ \bibnamefont
  {Chubb}},\ }\bibfield  {title} {\bibinfo {title} {Hand-waving and
  interpretive dance: an introductory course on tensor networks},\ }\href
  {https://doi.org/10.1088/1751-8121/aa6dc3} {\bibfield  {journal} {\bibinfo
  {journal} {J. Phys. A: Math. Theor.}\ }\textbf {\bibinfo {volume} {50}},\
  \bibinfo {pages} {223001} (\bibinfo {year} {2017})}\BibitemShut {NoStop}%
\bibitem [{\citenamefont {Szalay}\ \emph {et~al.}(2015)\citenamefont {Szalay},
  \citenamefont {Pfeffer}, \citenamefont {Murg}, \citenamefont {Barcza},
  \citenamefont {Verstraete}, \citenamefont {Schneider},\ and\ \citenamefont
  {Legeza}}]{szalay_tensor_2015}%
  \BibitemOpen
  \bibfield  {author} {\bibinfo {author} {\bibfnamefont {S.}~\bibnamefont
  {Szalay}}, \bibinfo {author} {\bibfnamefont {M.}~\bibnamefont {Pfeffer}},
  \bibinfo {author} {\bibfnamefont {V.}~\bibnamefont {Murg}}, \bibinfo {author}
  {\bibfnamefont {G.}~\bibnamefont {Barcza}}, \bibinfo {author} {\bibfnamefont
  {F.}~\bibnamefont {Verstraete}}, \bibinfo {author} {\bibfnamefont
  {R.}~\bibnamefont {Schneider}},\ and\ \bibinfo {author} {\bibfnamefont
  {{\"O}.}~\bibnamefont {Legeza}},\ }\bibfield  {title} {\bibinfo {title}
  {Tensor product methods and entanglement optimization for ab initio quantum
  chemistry},\ }\href {https://doi.org/10.1002/qua.24898} {\bibfield  {journal}
  {\bibinfo  {journal} {Int. J. Quantum Chem.}\ }\textbf {\bibinfo {volume}
  {115}},\ \bibinfo {pages} {1342} (\bibinfo {year} {2015})}\BibitemShut
  {NoStop}%
\bibitem [{\citenamefont {Dongarra}\ \emph {et~al.}(2018)\citenamefont
  {Dongarra}, \citenamefont {Gates}, \citenamefont {Haidar}, \citenamefont
  {Kurzak}, \citenamefont {Luszczek}, \citenamefont {Tomov},\ and\
  \citenamefont {Yamazaki}}]{dongarra_singular_2018}%
  \BibitemOpen
  \bibfield  {author} {\bibinfo {author} {\bibfnamefont {J.}~\bibnamefont
  {Dongarra}}, \bibinfo {author} {\bibfnamefont {M.}~\bibnamefont {Gates}},
  \bibinfo {author} {\bibfnamefont {A.}~\bibnamefont {Haidar}}, \bibinfo
  {author} {\bibfnamefont {J.}~\bibnamefont {Kurzak}}, \bibinfo {author}
  {\bibfnamefont {P.}~\bibnamefont {Luszczek}}, \bibinfo {author}
  {\bibfnamefont {S.}~\bibnamefont {Tomov}},\ and\ \bibinfo {author}
  {\bibfnamefont {I.}~\bibnamefont {Yamazaki}},\ }\bibfield  {title} {\bibinfo
  {title} {The {Singular} {Value} {Decomposition}: {Anatomy} of {Optimizing} an
  {Algorithm} for {Extreme} {Scale}},\ }\href
  {https://doi.org/10.1137/17M1117732} {\bibfield  {journal} {\bibinfo
  {journal} {SIAM Rev.}\ }\textbf {\bibinfo {volume} {60}},\ \bibinfo {pages}
  {808} (\bibinfo {year} {2018})}\BibitemShut {NoStop}%
\bibitem [{\citenamefont {McCulloch}(2007)}]{mcculloch_density-matrix_2007}%
  \BibitemOpen
  \bibfield  {author} {\bibinfo {author} {\bibfnamefont {I.~P.}\ \bibnamefont
  {McCulloch}},\ }\bibfield  {title} {\bibinfo {title} {From density-matrix
  renormalization group to matrix product states},\ }\href
  {https://doi.org/10.1088/1742-5468/2007/10/P10014} {\bibfield  {journal}
  {\bibinfo  {journal} {J. Stat. Mech.}\ }\textbf {\bibinfo {volume} {2007}},\
  \bibinfo {pages} {P10014} (\bibinfo {year} {2007})}\BibitemShut {NoStop}%
\bibitem [{\citenamefont {Crosswhite}\ and\ \citenamefont
  {Bacon}(2008)}]{crosswhite_finite_2008}%
  \BibitemOpen
  \bibfield  {author} {\bibinfo {author} {\bibfnamefont {G.~M.}\ \bibnamefont
  {Crosswhite}}\ and\ \bibinfo {author} {\bibfnamefont {D.}~\bibnamefont
  {Bacon}},\ }\bibfield  {title} {\bibinfo {title} {Finite automata for caching
  in matrix product algorithms},\ }\href
  {https://doi.org/10.1103/PhysRevA.78.012356} {\bibfield  {journal} {\bibinfo
  {journal} {Phys. Rev. A}\ }\textbf {\bibinfo {volume} {78}},\ \bibinfo
  {pages} {012356} (\bibinfo {year} {2008})}\BibitemShut {NoStop}%
\bibitem [{\citenamefont {Crosswhite}\ \emph {et~al.}(2008)\citenamefont
  {Crosswhite}, \citenamefont {Doherty},\ and\ \citenamefont
  {Vidal}}]{crosswhite_applying_2008}%
  \BibitemOpen
  \bibfield  {author} {\bibinfo {author} {\bibfnamefont {G.~M.}\ \bibnamefont
  {Crosswhite}}, \bibinfo {author} {\bibfnamefont {A.~C.}\ \bibnamefont
  {Doherty}},\ and\ \bibinfo {author} {\bibfnamefont {G.}~\bibnamefont
  {Vidal}},\ }\bibfield  {title} {\bibinfo {title} {Applying matrix product
  operators to model systems with long-range interactions},\ }\href
  {https://doi.org/10.1103/PhysRevB.78.035116} {\bibfield  {journal} {\bibinfo
  {journal} {Phys. Rev. B}\ }\textbf {\bibinfo {volume} {78}},\ \bibinfo
  {pages} {035116} (\bibinfo {year} {2008})}\BibitemShut {NoStop}%
\bibitem [{\citenamefont {Fr{\"o}wis}\ \emph {et~al.}(2010)\citenamefont
  {Fr{\"o}wis}, \citenamefont {Nebendahl},\ and\ \citenamefont
  {D{\"u}r}}]{frowis_tensor_2010}%
  \BibitemOpen
  \bibfield  {author} {\bibinfo {author} {\bibfnamefont {F.}~\bibnamefont
  {Fr{\"o}wis}}, \bibinfo {author} {\bibfnamefont {V.}~\bibnamefont
  {Nebendahl}},\ and\ \bibinfo {author} {\bibfnamefont {W.}~\bibnamefont
  {D{\"u}r}},\ }\bibfield  {title} {\bibinfo {title} {Tensor operators:
  {Constructions} and applications for long-range interaction systems},\ }\href
  {https://doi.org/10.1103/PhysRevA.81.062337} {\bibfield  {journal} {\bibinfo
  {journal} {Phys. Rev. A}\ }\textbf {\bibinfo {volume} {81}},\ \bibinfo
  {pages} {062337} (\bibinfo {year} {2010})}\BibitemShut {NoStop}%
\bibitem [{\citenamefont {Levenberg}(1944)}]{levenberg_method_1944}%
  \BibitemOpen
  \bibfield  {author} {\bibinfo {author} {\bibfnamefont {K.}~\bibnamefont
  {Levenberg}},\ }\bibfield  {title} {\bibinfo {title} {A method for the
  solution of certain non-linear problems in least squares},\ }\href
  {https://doi.org/10.1090/qam/10666} {\bibfield  {journal} {\bibinfo
  {journal} {Quart. Appl. Math.}\ }\textbf {\bibinfo {volume} {2}},\ \bibinfo
  {pages} {164} (\bibinfo {year} {1944})}\BibitemShut {NoStop}%
\bibitem [{\citenamefont {Marquardt}(1963)}]{marquardt_algorithm_1963}%
  \BibitemOpen
  \bibfield  {author} {\bibinfo {author} {\bibfnamefont {D.~W.}\ \bibnamefont
  {Marquardt}},\ }\bibfield  {title} {\bibinfo {title} {An {Algorithm} for
  {Least}-{Squares} {Estimation} of {Nonlinear} {Parameters}},\ }\href
  {https://doi.org/10.1137/0111030} {\bibfield  {journal} {\bibinfo  {journal}
  {J. Soc. Ind. Appl. Math.}\ }\textbf {\bibinfo {volume} {11}},\ \bibinfo
  {pages} {431} (\bibinfo {year} {1963})}\BibitemShut {NoStop}%
\bibitem [{\citenamefont {{{The MathWorks,
  Inc.}}}()}]{noauthor_least-squares_nodate}%
  \BibitemOpen
  \bibfield  {author} {\bibinfo {author} {\bibnamefont {{{The MathWorks,
  Inc.}}}},\ }\href@noop {} {\bibinfo {title} {Least-{Squares} ({Model}
  {Fitting}) {Algorithms} - \textsc{matlab} \& {Simulink}}},\ \bibinfo
  {howpublished}
  {\url{https://www.mathworks.com/help/optim/ug/least-squares-model-fitting-algorithms.html}
  (accessed 29 {October} 2019)}\BibitemShut {NoStop}%
\bibitem{supp_material}%
\BibitemOpen
\bibfield {note} {\bibinfo
	{note} {See Supplemental Material below for a comparison of different methods of approximating power laws by sums of exponentials, as well as details of our numerical experiments}}
\BibitemShut {NoStop}%
\bibitem [{\citenamefont {McLachlan}(1964)}]{mclachlan_variational_1964}%
  \BibitemOpen
  \bibfield  {author} {\bibinfo {author} {\bibfnamefont {A.~D.}\ \bibnamefont
  {McLachlan}},\ }\bibfield  {title} {\bibinfo {title} {A variational solution
  of the time-dependent {Schrodinger} equation},\ }\href
  {https://doi.org/10.1080/00268976400100041} {\bibfield  {journal} {\bibinfo
  {journal} {Mol. Phys.}\ }\textbf {\bibinfo {volume} {8}},\ \bibinfo {pages}
  {39} (\bibinfo {year} {1964})}\BibitemShut {NoStop}%
\bibitem [{\citenamefont {Broeckhove}\ \emph {et~al.}(1988)\citenamefont
  {Broeckhove}, \citenamefont {Lathouwers}, \citenamefont {Kesteloot},\ and\
  \citenamefont {Van~Leuven}}]{broeckhove_equivalence_1988}%
  \BibitemOpen
  \bibfield  {author} {\bibinfo {author} {\bibfnamefont {J.}~\bibnamefont
  {Broeckhove}}, \bibinfo {author} {\bibfnamefont {L.}~\bibnamefont
  {Lathouwers}}, \bibinfo {author} {\bibfnamefont {E.}~\bibnamefont
  {Kesteloot}},\ and\ \bibinfo {author} {\bibfnamefont {P.}~\bibnamefont
  {Van~Leuven}},\ }\bibfield  {title} {\bibinfo {title} {On the equivalence of
  time-dependent variational principles},\ }\href
  {https://doi.org/10.1016/0009-2614(88)80380-4} {\bibfield  {journal}
  {\bibinfo  {journal} {Chem. Phys. Lett.}\ }\textbf {\bibinfo {volume}
  {149}},\ \bibinfo {pages} {547} (\bibinfo {year} {1988})}\BibitemShut
  {NoStop}%
\bibitem [{\citenamefont {Holtz}\ \emph {et~al.}(2012)\citenamefont {Holtz},
  \citenamefont {Rohwedder},\ and\ \citenamefont
  {Schneider}}]{holtz_manifolds_2012}%
  \BibitemOpen
  \bibfield  {author} {\bibinfo {author} {\bibfnamefont {S.}~\bibnamefont
  {Holtz}}, \bibinfo {author} {\bibfnamefont {T.}~\bibnamefont {Rohwedder}},\
  and\ \bibinfo {author} {\bibfnamefont {R.}~\bibnamefont {Schneider}},\
  }\bibfield  {title} {\bibinfo {title} {On manifolds of tensors of fixed
  {TT}-rank},\ }\href {https://doi.org/10.1007/s00211-011-0419-7} {\bibfield
  {journal} {\bibinfo  {journal} {Numer. Math.}\ }\textbf {\bibinfo {volume}
  {120}},\ \bibinfo {pages} {701} (\bibinfo {year} {2012})}\BibitemShut
  {NoStop}%
\bibitem [{\citenamefont {Uschmajew}\ and\ \citenamefont
  {Vandereycken}(2013)}]{uschmajew_geometry_2013}%
  \BibitemOpen
  \bibfield  {author} {\bibinfo {author} {\bibfnamefont {A.}~\bibnamefont
  {Uschmajew}}\ and\ \bibinfo {author} {\bibfnamefont {B.}~\bibnamefont
  {Vandereycken}},\ }\bibfield  {title} {\bibinfo {title} {The geometry of
  algorithms using hierarchical tensors},\ }\href
  {https://doi.org/10.1016/j.laa.2013.03.016} {\bibfield  {journal} {\bibinfo
  {journal} {Linear Algebra Its Appl.}\ }\textbf {\bibinfo {volume} {439}},\
  \bibinfo {pages} {133} (\bibinfo {year} {2013})}\BibitemShut {NoStop}%
\bibitem [{\citenamefont {Haegeman}\ \emph {et~al.}(2014)\citenamefont
  {Haegeman}, \citenamefont {Mari{\"e}n}, \citenamefont {Osborne},\ and\
  \citenamefont {Verstraete}}]{haegeman_geometry_2014}%
  \BibitemOpen
  \bibfield  {author} {\bibinfo {author} {\bibfnamefont {J.}~\bibnamefont
  {Haegeman}}, \bibinfo {author} {\bibfnamefont {M.}~\bibnamefont
  {Mari{\"e}n}}, \bibinfo {author} {\bibfnamefont {T.~J.}\ \bibnamefont
  {Osborne}},\ and\ \bibinfo {author} {\bibfnamefont {F.}~\bibnamefont
  {\hbox{Verstraete}}},\ }\bibfield  {title} {\bibinfo {title} {Geometry of matrix
  product states: {Metric}, parallel transport, and curvature},\ }\href
  {https://doi.org/10.1063/1.4862851} {\bibfield  {journal} {\bibinfo
  {journal} {J. Math. Phys.}\ }\textbf {\bibinfo {volume} {55}},\ \bibinfo
  {pages} {021902} (\bibinfo {year} {2014})}\BibitemShut {NoStop}%
\bibitem [{\citenamefont {Hatano}\ and\ \citenamefont
  {Suzuki}(2005)}]{hatano_finding_2005}%
  \BibitemOpen
  \bibfield  {author} {\bibinfo {author} {\bibfnamefont {N.}~\bibnamefont
  {Hatano}}\ and\ \bibinfo {author} {\bibfnamefont {M.}~\bibnamefont
  {Suzuki}},\ }\bibfield  {title} {\bibinfo {title} {Finding {Exponential}
  {Product} {Formulas} of {Higher} {Orders}},\ }in\ \href
  {https://doi.org/10.1007/11526216_2} {\emph {\bibinfo {booktitle} {Quantum
  {Annealing} and {Other} {Optimization} {Methods}}}},\ \bibinfo {series and
  number} {Lecture {Notes} in {Physics}},\ \bibinfo {editor} {edited by\
  \bibinfo {editor} {\bibfnamefont {A.}~\bibnamefont {Das}}\ and\ \bibinfo
  {editor} {\bibfnamefont {B.}~\bibnamefont {K.~Chakrabarti}}}\ (\bibinfo
  {publisher} {Springer Berlin Heidelberg},\ \bibinfo {address} {Berlin,
  Heidelberg},\ \bibinfo {year} {2005})\ pp.\ \bibinfo {pages}
  {37--68}\BibitemShut {NoStop}%
\bibitem [{\citenamefont {White}(2005)}]{white_density_2005}%
  \BibitemOpen
  \bibfield  {author} {\bibinfo {author} {\bibfnamefont {S.~R.}\ \bibnamefont
  {White}},\ }\bibfield  {title} {\bibinfo {title} {Density matrix
  renormalization group algorithms with a single center site},\ }\href
  {https://doi.org/10.1103/PhysRevB.72.180403} {\bibfield  {journal} {\bibinfo
  {journal} {Phys. Rev. B}\ }\textbf {\bibinfo {volume} {72}},\ \bibinfo
  {pages} {180403} (\bibinfo {year} {2005})}\BibitemShut {NoStop}%
\bibitem [{\citenamefont {Dolgov}\ and\ \citenamefont
  {Savostyanov}(2014)}]{dolgov_alternating_2014}%
  \BibitemOpen
  \bibfield  {author} {\bibinfo {author} {\bibfnamefont {S.}~\bibnamefont
  {Dolgov}}\ and\ \bibinfo {author} {\bibfnamefont {D.}~\bibnamefont
  {Savostyanov}},\ }\bibfield  {title} {\bibinfo {title} {Alternating {Minimal}
  {Energy} {Methods} for {Linear} {Systems} in {Higher} {Dimensions}},\ }\href
  {https://doi.org/10.1137/140953289} {\bibfield  {journal} {\bibinfo
  {journal} {SIAM J. Sci. Comput.}\ }\textbf {\bibinfo {volume} {36}},\
  \bibinfo {pages} {A2248} (\bibinfo {year} {2014})}\BibitemShut {NoStop}%
\bibitem [{\citenamefont {Hubig}\ \emph {et~al.}(2015)\citenamefont {Hubig},
  \citenamefont {McCulloch}, \citenamefont {Schollw{\"o}ck},\ and\
  \citenamefont {Wolf}}]{hubig_strictly_2015}%
  \BibitemOpen
  \bibfield  {author} {\bibinfo {author} {\bibfnamefont {C.}~\bibnamefont
  {Hubig}}, \bibinfo {author} {\bibfnamefont {I.~P.}\ \bibnamefont
  {McCulloch}}, \bibinfo {author} {\bibfnamefont {U.}~\bibnamefont
  {Schollw{\"o}ck}},\ and\ \bibinfo {author} {\bibfnamefont {F.~A.}\
  \bibnamefont {Wolf}},\ }\bibfield  {title} {\bibinfo {title} {Strictly
  single-site {DMRG} algorithm with subspace expansion},\ }\href
  {https://doi.org/10.1103/PhysRevB.91.155115} {\bibfield  {journal} {\bibinfo
  {journal} {Phys. Rev. B}\ }\textbf {\bibinfo {volume} {91}},\ \bibinfo
  {pages} {155115} (\bibinfo {year} {2015})}\BibitemShut {NoStop}%
\bibitem [{\citenamefont {Jaschke}\ \emph {et~al.}(2018)\citenamefont
  {Jaschke}, \citenamefont {Wall},\ and\ \citenamefont
  {Carr}}]{jaschke_open_2018}%
  \BibitemOpen
  \bibfield  {author} {\bibinfo {author} {\bibfnamefont {D.}~\bibnamefont
  {Jaschke}}, \bibinfo {author} {\bibfnamefont {M.~L.}\ \bibnamefont {Wall}},\
  and\ \bibinfo {author} {\bibfnamefont {L.~D.}\ \bibnamefont {Carr}},\
  }\bibfield  {title} {\bibinfo {title} {Open source {Matrix} {Product}
  {States}: {Opening} ways to simulate entangled many-body quantum systems in
  one dimension},\ }\href {https://doi.org/10.1016/j.cpc.2017.12.015}
  {\bibfield  {journal} {\bibinfo  {journal} {Comput. Phys. Commun.}\ }\textbf
  {\bibinfo {volume} {225}},\ \bibinfo {pages} {59} (\bibinfo {year}
  {2018})}\BibitemShut {NoStop}%
\bibitem [{\citenamefont {Park}\ and\ \citenamefont
  {Light}(1986)}]{park_unitary_1986}%
  \BibitemOpen
  \bibfield  {author} {\bibinfo {author} {\bibfnamefont {T.~J.}\ \bibnamefont
  {Park}}\ and\ \bibinfo {author} {\bibfnamefont {J.~C.}\ \bibnamefont
  {Light}},\ }\bibfield  {title} {\bibinfo {title} {Unitary quantum time
  evolution by iterative {Lanczos} reduction},\ }\href
  {https://doi.org/10.1063/1.451548} {\bibfield  {journal} {\bibinfo  {journal}
  {J. Chem. Phys.}\ }\textbf {\bibinfo {volume} {85}},\ \bibinfo {pages} {5870}
  (\bibinfo {year} {1986})}\BibitemShut {NoStop}%
\bibitem [{\citenamefont {Al-Mohy}\ and\ \citenamefont
  {Higham}(2011)}]{al-mohy_computing_2011}%
  \BibitemOpen
  \bibfield  {author} {\bibinfo {author} {\bibfnamefont {A.}~\bibnamefont
  {Al-Mohy}}\ and\ \bibinfo {author} {\bibfnamefont {N.}~\bibnamefont
  {Higham}},\ }\bibfield  {title} {\bibinfo {title} {Computing the {Action} of
  the {Matrix} {Exponential}, with an {Application} to {Exponential}
  {Integrators}},\ }\href {https://doi.org/10.1137/100788860} {\bibfield
  {journal} {\bibinfo  {journal} {SIAM J. Sci. Comput.}\ }\textbf {\bibinfo
  {volume} {33}},\ \bibinfo {pages} {488} (\bibinfo {year} {2011})}\BibitemShut
  {NoStop}%
\bibitem [{\citenamefont {Goto}\ and\ \citenamefont
  {Danshita}(2019)}]{goto_performance_2019}%
  \BibitemOpen
  \bibfield  {author} {\bibinfo {author} {\bibfnamefont {S.}~\bibnamefont
  {Goto}}\ and\ \bibinfo {author} {\bibfnamefont {I.}~\bibnamefont
  {Danshita}},\ }\bibfield  {title} {\bibinfo {title} {Performance of the
  time-dependent variational principle for matrix product states in the
  long-time evolution of a pure state},\ }\href
  {https://doi.org/10.1103/PhysRevB.99.054307} {\bibfield  {journal} {\bibinfo
  {journal} {Phys. Rev. B}\ }\textbf {\bibinfo {volume} {99}},\ \bibinfo
  {pages} {054307} (\bibinfo {year} {2019})}\BibitemShut {NoStop}%
\bibitem [{\citenamefont {Hubig}\ \emph {et~al.}(2018)\citenamefont {Hubig},
  \citenamefont {Haegeman},\ and\ \citenamefont
  {Schollw{\"o}ck}}]{hubig_error_2018}%
  \BibitemOpen
  \bibfield  {author} {\bibinfo {author} {\bibfnamefont {C.}~\bibnamefont
  {Hubig}}, \bibinfo {author} {\bibfnamefont {J.}~\bibnamefont {Haegeman}},\
  and\ \bibinfo {author} {\bibfnamefont {U.}~\bibnamefont {Schollw{\"o}ck}},\
  }\bibfield  {title} {\bibinfo {title} {Error estimates for extrapolations
  with matrix-product states},\ }\href
  {https://doi.org/10.1103/PhysRevB.97.045125} {\bibfield  {journal} {\bibinfo
  {journal} {Phys. Rev. B}\ }\textbf {\bibinfo {volume} {97}},\ \bibinfo
  {pages} {045125} (\bibinfo {year} {2018})}\BibitemShut {NoStop}%
\bibitem [{\citenamefont {Schachenmayer}\ \emph {et~al.}(2013)\citenamefont
  {Schachenmayer}, \citenamefont {Lanyon}, \citenamefont {Roos},\ and\
  \citenamefont {Daley}}]{schachenmayer_entanglement_2013}%
  \BibitemOpen
  \bibfield  {author} {\bibinfo {author} {\bibfnamefont {J.}~\bibnamefont
  {Schachenmayer}}, \bibinfo {author} {\bibfnamefont {B.~P.}\ \bibnamefont
  {Lanyon}}, \bibinfo {author} {\bibfnamefont {C.~F.}\ \bibnamefont {Roos}},\
  and\ \bibinfo {author} {\bibfnamefont {A.~J.}\ \bibnamefont {Daley}},\
  }\bibfield  {title} {\bibinfo {title} {Entanglement {Growth} in {Quench}
  {Dynamics} with {Variable} {Range} {Interactions}},\ }\href
  {https://doi.org/10.1103/PhysRevX.3.031015} {\bibfield  {journal} {\bibinfo
  {journal} {Phys. Rev. X}\ }\textbf {\bibinfo {volume} {3}},\ \bibinfo {pages}
  {031015} (\bibinfo {year} {2013})}\BibitemShut {NoStop}%
\bibitem [{\citenamefont {Singh}\ \emph {et~al.}(2017)\citenamefont {Singh},
  \citenamefont {Moessner},\ and\ \citenamefont {Roy}}]{singh_effect_2017}%
  \BibitemOpen
  \bibfield  {author} {\bibinfo {author} {\bibfnamefont {R.}~\bibnamefont
  {Singh}}, \bibinfo {author} {\bibfnamefont {R.}~\bibnamefont {Moessner}},\
  and\ \bibinfo {author} {\bibfnamefont {D.}~\bibnamefont {Roy}},\ }\bibfield
  {title} {\bibinfo {title} {Effect of long-range hopping and interactions on
  entanglement dynamics and many-body localization},\ }\href
  {https://doi.org/10.1103/PhysRevB.95.094205} {\bibfield  {journal} {\bibinfo
  {journal} {Phys. Rev. B}\ }\textbf {\bibinfo {volume} {95}},\ \bibinfo
  {pages} {094205} (\bibinfo {year} {2017})}\BibitemShut {NoStop}%
\bibitem [{\citenamefont {Lerose}\ and\ \citenamefont
  {Pappalardi}(2019)}]{lerose_origin_2018}%
  \BibitemOpen
  \bibfield  {author} {\bibinfo {author} {\bibfnamefont {A.}~\bibnamefont
  {Lerose}}\ and\ \bibinfo {author} {\bibfnamefont {S.}~\bibnamefont
  {Pappalardi}},\ }\bibfield  {title} {\bibinfo {title} {Origin of the slow
  growth of entanglement entropy in long-range interacting spin systems},\ }\href {https://doi.org/10.1103/PhysRevResearch.2.012041}
{\bibfield  {journal} {\bibinfo  {journal} {Phys. Rev. Research}\ }\textbf
	{\bibinfo {volume} {2}},\ \bibinfo {pages} {012041(R)} (\bibinfo {year}
	{2020})}\BibitemShut {NoStop}%
\bibitem [{\citenamefont {Liu}\ \emph {et~al.}(2019)\citenamefont {Liu},
  \citenamefont {Lundgren}, \citenamefont {Titum}, \citenamefont {Pagano},
  \citenamefont {Zhang}, \citenamefont {Monroe},\ and\ \citenamefont
  {Gorshkov}}]{liu_confined_2019}%
  \BibitemOpen
  \bibfield  {author} {\bibinfo {author} {\bibfnamefont {F.}~\bibnamefont
  {Liu}}, \bibinfo {author} {\bibfnamefont {R.}~\bibnamefont {Lundgren}},
  \bibinfo {author} {\bibfnamefont {P.}~\bibnamefont {Titum}}, \bibinfo
  {author} {\bibfnamefont {G.}~\bibnamefont {Pagano}}, \bibinfo {author}
  {\bibfnamefont {J.}~\bibnamefont {Zhang}}, \bibinfo {author} {\bibfnamefont
  {C.}~\bibnamefont {Monroe}},\ and\ \bibinfo {author} {\bibfnamefont {A.~V.}\
  \bibnamefont {Gorshkov}},\ }\bibfield  {title} {\bibinfo {title} {Confined
  {Quasiparticle} {Dynamics} in {Long}-{Range} {Interacting} {Quantum} {Spin}
  {Chains}},\ }\href {https://doi.org/10.1103/PhysRevLett.122.150601}
  {\bibfield  {journal} {\bibinfo  {journal} {Phys. Rev. Lett.}\ }\textbf
  {\bibinfo {volume} {122}},\ \bibinfo {pages} {150601} (\bibinfo {year}
  {2019})}\BibitemShut {NoStop}%
\bibitem [{\citenamefont {{The University of
  Bath}}()}]{noauthor_balena_nodate}%
  \BibitemOpen
  \bibfield  {author} {\bibinfo {author} {\bibnamefont {{The University of
  Bath}}},\ }\href@noop {} {\bibinfo {title} {Balena {HPC} cluster}},\ \bibinfo
  {howpublished}
  {\url{https://www.bath.ac.uk/corporate-information/balena-hpc-cluster/}
  (accessed 29 {October} 2019)}\BibitemShut {NoStop}%
\bibitem [{\citenamefont {Al-Assam}\ \emph {et~al.}(2016)\citenamefont
  {Al-Assam}, \citenamefont {Clark}, \citenamefont {Jaksch},\ and\
  \citenamefont {{TNT Development team}}}]{noauthor_tntlibrary_nodate}%
  \BibitemOpen
  \bibfield  {author} {\bibinfo {author} {\bibfnamefont {S.}~\bibnamefont
  {Al-Assam}}, \bibinfo {author} {\bibfnamefont {S.~R.}\ \bibnamefont {Clark}},
  \bibinfo {author} {\bibfnamefont {D.}~\bibnamefont {Jaksch}},\ and\ \bibinfo
  {author} {\bibnamefont {{TNT \hbox{Development} team}}},\ }\href@noop {} {\bibinfo
  {title} {{Tensor Network Theory Library beta version 1.2.1}}},\ \bibinfo
  {howpublished} {\url{https://gitlab.physics.ox.ac.uk/tntlibrary}} (\bibinfo
  {year} {2016})\BibitemShut {NoStop}%
\bibitem [{\citenamefont {Goodyer}(2013)}]{goodyer_tnt_nodate}%
  \BibitemOpen
  \bibfield  {author} {\bibinfo {author} {\bibfnamefont {C.}~\bibnamefont
  {Goodyer}},\ }\href@noop {} {\bibinfo {title} {{TNT Library: tensor
  manipulation and storage}}},\ \bibinfo {year} {2013},\  \url{http://www.hector.ac.uk/cse/distributedcse/reports/UniTNT/UniTNT/index.html}\BibitemShut {NoStop}%
\bibitem [{\citenamefont {Al-Assam}\ \emph {et~al.}(2017)\citenamefont
  {Al-Assam}, \citenamefont {Clark},\ and\ \citenamefont
  {Jaksch}}]{al-assam_tensor_2017}%
  \BibitemOpen
  \bibfield  {author} {\bibinfo {author} {\bibfnamefont {S.}~\bibnamefont
  {Al-Assam}}, \bibinfo {author} {\bibfnamefont {S.~R.}\ \bibnamefont
  {Clark}},\ and\ \bibinfo {author} {\bibfnamefont {D.}~\bibnamefont
  {Jaksch}},\ }\bibfield  {title} {\bibinfo {title} {The tensor network theory
  library},\ }\href {https://doi.org/10.1088/1742-5468/aa7df3} {\bibfield
  {journal} {\bibinfo  {journal} {J. Stat. Mech.}\ }\textbf {\bibinfo {volume}
  {2017}},\ \bibinfo {pages} {093102} (\bibinfo {year} {2017})}\BibitemShut
  {NoStop}%
\bibitem [{\citenamefont {Coulthard}(2018)}]{coulthard_engineering_2018}%
  \BibitemOpen
  \bibfield  {author} {\bibinfo {author} {\bibfnamefont {J.}~\bibnamefont
  {Coulthard}},\ }\emph {\bibinfo {title} {Engineering quantum states of
  fermionic many-body systems}},\ {\bibinfo {type} {{DPhil} {Thesis}}},\ \bibinfo  {school} {University of
  Oxford},\ \bibinfo {year} {2018},\ \url
{https://ora.ox.ac.uk/objects/uuid:c2ad8834-7202-45cf-8348-4f9b07e942d1}\BibitemShut {NoStop}%
\bibitem [{\citenamefont {{{The Mathworks, Inc.}}}()}]{matlab_r2017_nodate}%
  \BibitemOpen
  \bibfield  {author} {\bibinfo {author} {\bibnamefont {{{The Mathworks,
  Inc.}}}},\ }\href@noop {} {\bibinfo {title} {{\textsc{matlab} version
  9.3.0.713579 (R2017b)}}},\ \bibinfo {howpublished}
  {\url{https://www.mathworks.com/products/matlab.html}}\BibitemShut {NoStop}%
\bibitem [{\citenamefont {Lehoucq}\ \emph {et~al.}()\citenamefont {Lehoucq},
  \citenamefont {Maschhoff}, \citenamefont {Sorensen},\ and\ \citenamefont
  {Yang}}]{lehoucq_arpack_nodate}%
  \BibitemOpen
  \bibfield  {author} {\bibinfo {author} {\bibfnamefont {R.}~\bibnamefont
  {Lehoucq}}, \bibinfo {author} {\bibfnamefont {K.}~\bibnamefont {Maschhoff}},
  \bibinfo {author} {\bibfnamefont {D.}~\bibnamefont {Sorensen}},\ and\
  \bibinfo {author} {\bibfnamefont {C.}~\bibnamefont {Yang}},\ }\href@noop {}
  {\bibinfo {title} {{\textsc{arpack}} - {Arnoldi} {Package}}},\ \bibinfo
  {howpublished} {\url{https://www.caam.rice.edu/software/ARPACK/}}\BibitemShut
  {NoStop}%
\bibitem [{\citenamefont {{opencollab}}()}]{cornet_opencollab/arpack-ng_2019}%
  \BibitemOpen
  \bibfield  {author} {\bibinfo {author} {\bibnamefont {{opencollab}}},\
  }\href@noop {} {\bibinfo {title} {{\textsc{arpack-ng}} version 3.4.0}},\
  \bibinfo {howpublished}
  {\url{https://github.com/opencollab/arpack-ng}}\BibitemShut {NoStop}%
\bibitem [{ajo()}]{ajolleyx_intel_nodate}%
  \BibitemOpen
  \href@noop {} {\bibinfo {title} {Intel{\textregistered} {Math} {Kernel}
  {Library} ({Intel}{\textregistered} {MKL}) version 64/18.0.128}},\ \bibinfo
  {howpublished} {\url{https://software.intel.com/mkl}}\BibitemShut {NoStop}%
\bibitem [{mpi()}]{mpi_intel_nodate}%
  \BibitemOpen
  \href@noop {} {\bibinfo {title} {Intel{\textregistered} {MPI} {Library}
  version 64/18.0.128}},\ \bibinfo {howpublished}
  {\url{https://software.intel.com/mpi-library}}\BibitemShut {NoStop}%
\bibitem [{\citenamefont {Anderson}\ \emph {et~al.}()\citenamefont {Anderson},
  \citenamefont {Bai}, \citenamefont {Bischof}, \citenamefont {Blackford},
  \citenamefont {Demmel}, \citenamefont {Dongarra}, \citenamefont {Du~Croz},
  \citenamefont {Greenbaum}, \citenamefont {Hammarling}, \citenamefont
  {McKenney},\ and\ \citenamefont {Sorensen}}]{lapack}%
  \BibitemOpen
  \bibfield  {author} {\bibinfo {author} {\bibfnamefont {E.}~\bibnamefont
  {Anderson}}, \bibinfo {author} {\bibfnamefont {Z.}~\bibnamefont {Bai}},
  \bibinfo {author} {\bibfnamefont {C.}~\bibnamefont {Bischof}}, \bibinfo
  {author} {\bibfnamefont {S.}~\bibnamefont {Blackford}}, \bibinfo {author}
  {\bibfnamefont {J.}~\bibnamefont {\hbox{Demmel}}}, \bibinfo {author} {\bibfnamefont
  {J.}~\bibnamefont {Dongarra}}, \bibinfo {author} {\bibfnamefont
  {J.}~\bibnamefont {Du~Croz}}, \bibinfo {author} {\bibfnamefont
  {A.}~\bibnamefont {\hbox{Greenbaum}}}, \bibinfo {author} {\bibfnamefont
  {S.}~\bibnamefont {\hbox{Hammarling}}}, \bibinfo {author} {\bibfnamefont
  {A.}~\bibnamefont {McKenney}},\ and\ \bibinfo {author} {\bibfnamefont
  {D.}~\bibnamefont {Sorensen}},\ }\href@noop {} {\emph {\bibinfo {title}
  {{\textsc{\hbox{lapack}}} Users' Guide}}},\ \bibinfo {edition} {3rd}\ ed.\ (\bibinfo
   {publisher} {SIAM},\ \bibinfo {address} {Philadelphia, PA},\ \bibinfo {year} {1987}),\ \url{https://www.netlib.org/lapack/lug/}\BibitemShut
  {NoStop}%
\bibitem [{\citenamefont {Cuppen}(1980)}]{cuppen_divide_1980}%
  \BibitemOpen
  \bibfield  {author} {\bibinfo {author} {\bibfnamefont {J.~J.~M.}\
  \bibnamefont {Cuppen}},\ }\bibfield  {title} {\bibinfo {title} {A divide and
  conquer method for the symmetric tridiagonal eigenproblem},\ }\href
  {https://doi.org/10.1007/BF01396757} {\bibfield  {journal} {\bibinfo
  {journal} {Numer. Math.}\ }\textbf {\bibinfo {volume} {36}},\ \bibinfo
  {pages} {177} (\bibinfo {year} {1980})}\BibitemShut {NoStop}%
\bibitem [{\citenamefont {Gu}\ and\ \citenamefont
  {Eisenstat}(1994)}]{gu_stable_1994}%
  \BibitemOpen
  \bibfield  {author} {\bibinfo {author} {\bibfnamefont {M.}~\bibnamefont
  {Gu}}\ and\ \bibinfo {author} {\bibfnamefont {S.~C.}\ \bibnamefont
  {Eisenstat}},\ }\bibfield  {title} {\bibinfo {title} {A {Stable} and
  {Efficient} {Algorithm} for the {Rank}-{One} {Modification} of the
  {Symmetric} {Eigenproblem}},\ }\href
  {https://doi.org/10.1137/S089547989223924X} {\bibfield  {journal} {\bibinfo
  {journal} {SIAM J. Matrix Anal. Appl.}\ }\textbf {\bibinfo {volume} {15}},\
  \bibinfo {pages} {1266} (\bibinfo {year} {1994})}\BibitemShut {NoStop}%
\bibitem [{\citenamefont {Rutter}(1994)}]{Rutter:CSD-94-799}%
  \BibitemOpen
  \bibfield  {author} {\bibinfo {author} {\bibfnamefont {J.~D.}\ \bibnamefont
  {Rutter}},\ } {\emph {\bibinfo
  {title} {A serial implementation of \hbox{Cuppen's} divide and conquer algorithm for
  the symmetric eigenvalue problem}}},\ \bibinfo {type} {Tech. Rep.}\ \bibinfo
  {number} {UCB/CSD-94-799}\ \bibinfo  {institution} {EECS Department,
  University of California, Berkeley},\ \bibinfo {year} {1994},\ \url
{http://www2.eecs.berkeley.edu/Pubs/TechRpts/1994/6315.html}\ \BibitemShut
  {NoStop}%
\bibitem [{\citenamefont {{{The} {MathWorks,}
  {Inc.}}}()}]{noauthor_numerically_nodate}%
  \BibitemOpen
  \bibfield  {author} {\bibinfo {author} {\bibnamefont {{{The} {MathWorks,}
  {Inc.}}}},\ }\href@noop {} {\bibinfo {title} {Numerically evaluate double
  integral - {\textsc{matlab}} integral2}},\ \bibinfo {howpublished}
  {\url{https://www.mathworks.com/help/matlab/ref/integral2.html} (accessed 29
  {October} 2019)}\BibitemShut {NoStop}%
\bibitem [{\citenamefont {Nauts}\ and\ \citenamefont
  {Wyatt}(1983)}]{nauts_new_1983}%
  \BibitemOpen
  \bibfield  {author} {\bibinfo {author} {\bibfnamefont {A.}~\bibnamefont
  {Nauts}}\ and\ \bibinfo {author} {\bibfnamefont {R.~E.}\ \bibnamefont
  {Wyatt}},\ }\bibfield  {title} {\bibinfo {title} {New {Approach} to
  {Many}-{State} {Quantum} {Dynamics}: {The} {Recursive}-{Residue}-{Generation}
  {Method}},\ }\href {https://doi.org/10.1103/PhysRevLett.51.2238} {\bibfield
  {journal} {\bibinfo  {journal} {Phys. Rev. Lett.}\ }\textbf {\bibinfo
  {volume} {51}},\ \bibinfo {pages} {2238} (\bibinfo {year}
  {1983})}\BibitemShut {NoStop}%
\bibitem [{\citenamefont {Luitz}\ and\ \citenamefont
  {Lev}(2017)}]{luitz_ergodic_2017}%
  \BibitemOpen
  \bibfield  {author} {\bibinfo {author} {\bibfnamefont {D.~J.}\ \bibnamefont
  {Luitz}}\ and\ \bibinfo {author} {\bibfnamefont {Y.~B.}\ \bibnamefont
  {Lev}},\ }\bibfield  {title} {\bibinfo {title} {The ergodic side of the
  many-body localization transition},\ }\href
  {https://doi.org/10.1002/andp.201600350} {\bibfield  {journal} {\bibinfo
  {journal} {Ann. Phys. (Berl.)}\ }\textbf {\bibinfo {volume} {529}},\ \bibinfo
  {pages} {1600350} (\bibinfo {year} {2017})}\BibitemShut {NoStop}%
\bibitem [{\citenamefont {Kormos}\ \emph {et~al.}(2017)\citenamefont {Kormos},
  \citenamefont {Collura}, \citenamefont {Tak{\'a}cs},\ and\ \citenamefont
  {Calabrese}}]{kormos_real-time_2017}%
  \BibitemOpen
  \bibfield  {author} {\bibinfo {author} {\bibfnamefont {M.}~\bibnamefont
  {Kormos}}, \bibinfo {author} {\bibfnamefont {M.}~\bibnamefont {Collura}},
  \bibinfo {author} {\bibfnamefont {G.}~\bibnamefont {Tak{\'a}cs}},\ and\
  \bibinfo {author} {\bibfnamefont {P.}~\bibnamefont {Calabrese}},\ }\bibfield
  {title} {\bibinfo {title} {Real-time confinement following a quantum quench
  to a non-integrable model},\ }\href {https://doi.org/10.1038/nphys3934}
  {\bibfield  {journal} {\bibinfo  {journal} {Nat. Phys.}\ }\textbf {\bibinfo
  {volume} {13}},\ \bibinfo {pages} {246} (\bibinfo {year} {2017})}\BibitemShut
  {NoStop}%
\bibitem [{\citenamefont {Lieb}\ and\ \citenamefont
  {Robinson}(1972)}]{lieb_finite_1972}%
  \BibitemOpen
  \bibfield  {author} {\bibinfo {author} {\bibfnamefont {E.~H.}\ \bibnamefont
  {Lieb}}\ and\ \bibinfo {author} {\bibfnamefont {D.~W.}\ \bibnamefont
  {Robinson}},\ }\bibfield  {title} {\bibinfo {title} {The finite group
  velocity of quantum spin systems},\ }\href
  {https://doi.org/10.1007/BF01645779} {\bibfield  {journal} {\bibinfo
  {journal} {Commun. Math. Phys.}\ }\textbf {\bibinfo {volume} {28}},\ \bibinfo
  {pages} {251} (\bibinfo {year} {1972})}\BibitemShut {NoStop}%
\bibitem [{\citenamefont {Calabrese}\ and\ \citenamefont
  {Cardy}(2006)}]{calabrese_time_2006}%
  \BibitemOpen
  \bibfield  {author} {\bibinfo {author} {\bibfnamefont {P.}~\bibnamefont
  {Calabrese}}\ and\ \bibinfo {author} {\bibfnamefont {J.}~\bibnamefont
  {Cardy}},\ }\bibfield  {title} {\bibinfo {title} {Time {Dependence} of
  {Correlation} {Functions} {Following} a {Quantum} {Quench}},\ }\href
  {https://doi.org/10.1103/PhysRevLett.96.136801} {\bibfield  {journal}
  {\bibinfo  {journal} {Phys. Rev. Lett.}\ }\textbf {\bibinfo {volume} {96}},\
  \bibinfo {pages} {136801} (\bibinfo {year} {2006})}\BibitemShut {NoStop}%
\bibitem [{\citenamefont {Bravyi}\ \emph {et~al.}(2006)\citenamefont {Bravyi},
  \citenamefont {Hastings},\ and\ \citenamefont
  {Verstraete}}]{bravyi_lieb-robinson_2006}%
  \BibitemOpen
  \bibfield  {author} {\bibinfo {author} {\bibfnamefont {S.}~\bibnamefont
  {Bravyi}}, \bibinfo {author} {\bibfnamefont {M.~B.}\ \bibnamefont
  {Hastings}},\ and\ \bibinfo {author} {\bibfnamefont {F.}~\bibnamefont
  {Verstraete}},\ }\bibfield  {title} {\bibinfo {title} {Lieb-{Robinson}
  {Bounds} and the {Generation} of {Correlations} and {Topological} {Quantum}
  {Order}},\ }\href {https://doi.org/10.1103/PhysRevLett.97.050401} {\bibfield
  {journal} {\bibinfo  {journal} {Phys. Rev. Lett.}\ }\textbf {\bibinfo
  {volume} {97}},\ \bibinfo {pages} {050401} (\bibinfo {year}
  {2006})}\BibitemShut {NoStop}%
\bibitem [{\citenamefont {Cevolani}\ \emph {et~al.}(2018)\citenamefont
  {Cevolani}, \citenamefont {Despres}, \citenamefont {Carleo}, \citenamefont
  {Tagliacozzo},\ and\ \citenamefont
  {Sanchez-Palencia}}]{cevolani_universal_2018}%
  \BibitemOpen
  \bibfield  {author} {\bibinfo {author} {\bibfnamefont {L.}~\bibnamefont
  {Cevolani}}, \bibinfo {author} {\bibfnamefont {J.}~\bibnamefont {Despres}},
  \bibinfo {author} {\bibfnamefont {G.}~\bibnamefont {Carleo}}, \bibinfo
  {author} {\bibfnamefont {L.}~\bibnamefont {Tagliacozzo}},\ and\ \bibinfo
  {author} {\bibfnamefont {L.}~\bibnamefont {Sanchez-Palencia}},\ }\bibfield
  {title} {\bibinfo {title} {Universal scaling laws for correlation spreading
  in quantum systems with short- and long-range interactions},\ }\href
  {https://doi.org/10.1103/PhysRevB.98.024302} {\bibfield  {journal} {\bibinfo
  {journal} {Phys. Rev. B}\ }\textbf {\bibinfo {volume} {98}},\ \bibinfo
  {pages} {024302} (\bibinfo {year} {2018})}\BibitemShut {NoStop}%
\bibitem [{\citenamefont {Chen}\ and\ \citenamefont
  {Lucas}(2019)}]{chen_finite_2019}%
  \BibitemOpen
  \bibfield  {author} {\bibinfo {author} {\bibfnamefont {C.-F.}\ \bibnamefont
  {Chen}}\ and\ \bibinfo {author} {\bibfnamefont {A.}~\bibnamefont {Lucas}},\
  }\bibfield  {title} {\bibinfo {title} {Finite {Speed} of {Quantum} {Scrambling} with {Long} {Range} {Interactions}},\ }\href {https://doi.org/10.1103/PhysRevLett.123.250605}
  {\bibfield  {journal}
  	{\bibinfo  {journal} {Phys. Rev. Lett.}\ }\textbf {\bibinfo {volume} {123}},\ \bibinfo {pages} {250605} (\bibinfo {year} {2019})}\BibitemShut
  {NoStop}%
\bibitem [{\citenamefont {Kuwahara}\ and\ \citenamefont
  {Saito}(2019)}]{kuwahara_strictly_2019}%
  \BibitemOpen
  \bibfield  {author} {\bibinfo {author} {\bibfnamefont {T.}~\bibnamefont
  {Kuwahara}}\ and\ \bibinfo {author} {\bibfnamefont {K.}~\bibnamefont
  {Saito}},\ }\bibfield  {title} {\bibinfo {title} {Strictly linear light cones
  in long-range interacting systems of arbitrary dimensions},\ }\Eprint
  {https://arxiv.org/abs/1910.14477v3} {arXiv:1910.14477v3} [quant-ph]
  \BibitemShut {NoStop}%
\bibitem [{\citenamefont {Eisert}\ \emph {et~al.}(2013)\citenamefont {Eisert},
  \citenamefont {van~den Worm}, \citenamefont {Manmana},\ and\ \citenamefont
  {Kastner}}]{eisert_breakdown_2013}%
  \BibitemOpen
  \bibfield  {author} {\bibinfo {author} {\bibfnamefont {J.}~\bibnamefont
  {Eisert}}, \bibinfo {author} {\bibfnamefont {M.}~\bibnamefont {van~den
  Worm}}, \bibinfo {author} {\bibfnamefont {S.~R.}\ \bibnamefont {Manmana}},\
  and\ \bibinfo {author} {\bibfnamefont {M.}~\bibnamefont {Kastner}},\
  }\bibfield  {title} {\bibinfo {title} {Breakdown of {Quasilocality} in
  {Long}-{Range} {Quantum} {Lattice} {Models}},\ }\href
  {https://doi.org/10.1103/PhysRevLett.111.260401} {\bibfield  {journal}
  {\bibinfo  {journal} {Phys. Rev. Lett.}\ }\textbf {\bibinfo {volume} {111}},\
  \bibinfo {pages} {260401} (\bibinfo {year} {2013})}\BibitemShut {NoStop}%
\bibitem [{\citenamefont {Gong}\ \emph {et~al.}(2014)\citenamefont {Gong},
  \citenamefont {Foss-Feig}, \citenamefont {Michalakis},\ and\ \citenamefont
  {Gorshkov}}]{gong_persistence_2014}%
  \BibitemOpen
  \bibfield  {author} {\bibinfo {author} {\bibfnamefont {Z.-X.}\ \bibnamefont
  {Gong}}, \bibinfo {author} {\bibfnamefont {M.}~\bibnamefont {Foss-Feig}},
  \bibinfo {author} {\bibfnamefont {S.}~\bibnamefont {Michalakis}},\ and\
  \bibinfo {author} {\bibfnamefont {A.~V.}\ \bibnamefont {\hbox{Gorshkov}}},\
  }\bibfield  {title} {\bibinfo {title} {Persistence of {Locality} in {Systems}
  with {Power}-{Law} {Interactions}},\ }\href
  {https://doi.org/10.1103/PhysRevLett.113.030602} {\bibfield  {journal}
  {\bibinfo  {journal} {Phys. Rev. Lett.}\ }\textbf {\bibinfo {volume} {113}},\
  \bibinfo {pages} {030602} (\bibinfo {year} {2014})}\BibitemShut {NoStop}%
\bibitem [{\citenamefont {Storch}\ \emph {et~al.}(2015)\citenamefont {Storch},
  \citenamefont {van~den Worm},\ and\ \citenamefont
  {Kastner}}]{storch_interplay_2015}%
  \BibitemOpen
  \bibfield  {author} {\bibinfo {author} {\bibfnamefont {D.-M.}\ \bibnamefont
  {Storch}}, \bibinfo {author} {\bibfnamefont {M.}~\bibnamefont {van~den
  Worm}},\ and\ \bibinfo {author} {\bibfnamefont {M.}~\bibnamefont {Kastner}},\
  }\bibfield  {title} {\bibinfo {title} {Interplay of soundcone and supersonic
  propagation in lattice models with power law interactions},\ }\href
  {https://doi.org/10.1088/1367-2630/17/6/063021} {\bibfield  {journal}
  {\bibinfo  {journal} {New J. Phys.}\ }\textbf {\bibinfo {volume} {17}},\
  \bibinfo {pages} {063021} (\bibinfo {year} {2015})}\BibitemShut {NoStop}%
\bibitem [{\citenamefont {Luitz}\ and\ \citenamefont
  {Bar~Lev}(2019)}]{luitz_emergent_2019}%
  \BibitemOpen
  \bibfield  {author} {\bibinfo {author} {\bibfnamefont {D.~J.}\ \bibnamefont
  {Luitz}}\ and\ \bibinfo {author} {\bibfnamefont {Y.}~\bibnamefont
  {Bar~Lev}},\ }\bibfield  {title} {\bibinfo {title} {Emergent locality in
  systems with power-law interactions},\ }\href
  {https://doi.org/10.1103/PhysRevA.99.010105} {\bibfield  {journal} {\bibinfo
  {journal} {Phys. Rev. A}\ }\textbf {\bibinfo {volume} {99}},\ \bibinfo
  {pages} {010105} (\bibinfo {year} {2019})}\BibitemShut {NoStop}%
\bibitem [{\citenamefont {Haegeman}()}]{haegeman_notitle_nodate}%
  \BibitemOpen
  \bibfield  {author} {\bibinfo {author} {\bibfnamefont {J.}~\bibnamefont
  {Haegeman}},\ }\bibinfo {note} {(private communication)}\BibitemShut
  {NoStop}%
\bibitem [{\citenamefont {Singh}\ \emph {et~al.}(2011)\citenamefont {Singh},
  \citenamefont {Pfeifer},\ and\ \citenamefont {Vidal}}]{singh_tensor_2011}%
  \BibitemOpen
  \bibfield  {author} {\bibinfo {author} {\bibfnamefont {S.}~\bibnamefont
  {Singh}}, \bibinfo {author} {\bibfnamefont {R.~N.~C.}\ \bibnamefont
  {Pfeifer}},\ and\ \bibinfo {author} {\bibfnamefont {G.}~\bibnamefont
  {Vidal}},\ }\bibfield  {title} {\bibinfo {title} {Tensor network states and
  algorithms in the presence of a global {U}(1) symmetry},\ }\href
  {https://doi.org/10.1103/PhysRevB.83.115125} {\bibfield  {journal} {\bibinfo
  {journal} {Phys. Rev. B}\ }\textbf {\bibinfo {volume} {83}},\ \bibinfo
  {pages} {115125} (\bibinfo {year} {2011})}\BibitemShut {NoStop}%
\bibitem [{\citenamefont {Haldane}(1988)}]{haldane_exact_1988}%
  \BibitemOpen
  \bibfield  {author} {\bibinfo {author} {\bibfnamefont {F.~D.~M.}\
  \bibnamefont {Haldane}},\ }\bibfield  {title} {\bibinfo {title} {Exact
  {Jastrow}-{Gutzwiller} resonating-valence-bond ground state of the
  spin-{$\frac{1}{2}$} antiferromagnetic {Heisenberg} chain with
  {$1/\mathrm{r}^2$} exchange},\ }\href
  {https://doi.org/10.1103/PhysRevLett.60.635} {\bibfield  {journal} {\bibinfo
  {journal} {Phys. Rev. Lett.}\ }\textbf {\bibinfo {volume} {60}},\ \bibinfo
  {pages} {635} (\bibinfo {year} {1988})}\BibitemShut {NoStop}%
\bibitem [{\citenamefont {Shastry}(1988)}]{shastry_exact_1988}%
  \BibitemOpen
  \bibfield  {author} {\bibinfo {author} {\bibfnamefont {B.~S.}\ \bibnamefont
  {Shastry}},\ }\bibfield  {title} {\bibinfo {title} {Exact solution of an
  {S}=1/2 {Heisenberg} antiferromagnetic chain with long-ranged interactions},\
  }\href {https://doi.org/10.1103/PhysRevLett.60.639} {\bibfield  {journal}
  {\bibinfo  {journal} {Phys. Rev. Lett.}\ }\textbf {\bibinfo {volume} {60}},\
  \bibinfo {pages} {639} (\bibinfo {year} {1988})}\BibitemShut {NoStop}%
\bibitem [{\citenamefont {Haldane}\ and\ \citenamefont
  {Zirnbauer}(1993)}]{haldane_exact_1993}%
  \BibitemOpen
  \bibfield  {author} {\bibinfo {author} {\bibfnamefont {F.~D.~M.}\
  \bibnamefont {Haldane}}\ and\ \bibinfo {author} {\bibfnamefont {M.~R.}\
  \bibnamefont {Zirnbauer}},\ }\bibfield  {title} {\bibinfo {title} {Exact
  calculation of the ground-state dynamical spin correlation function of a
  {S}=1/2 antiferromagnetic {Heisenberg} chain with free spinons},\ }\href
  {https://doi.org/10.1103/PhysRevLett.71.4055} {\bibfield  {journal} {\bibinfo
   {journal} {Phys. Rev. Lett.}\ }\textbf {\bibinfo {volume} {71}},\ \bibinfo
  {pages} {4055} (\bibinfo {year} {1993})}\BibitemShut {NoStop}%
\bibitem [{\citenamefont {LeBlanc}\ \emph {et~al.}(2015)\citenamefont
  {LeBlanc}, \citenamefont {\textit{et al.}},\ and\ \citenamefont {{Simons
  Collaboration on the Many-Electron
  Problem}}}]{simons_collaboration_on_the_many-electron_problem_solutions_2015}%
  \BibitemOpen
  \bibfield  {author} {\bibinfo {author} {\bibfnamefont {J.~P.~F.}\
  \bibnamefont {LeBlanc}}, \bibinfo {author} {\bibnamefont {\textit{et al.}}},\
  and\ \bibinfo {author} {\bibnamefont {{Simons Collaboration on the
  Many-Electron Problem}}},\ }\bibfield  {title} {\bibinfo {title} {Solutions
  of the {Two}-{Dimensional} {Hubbard} {Model}: {Benchmarks} and {Results} from
  a {Wide} {Range} of {Numerical} {Algorithms}},\ }\href
  {https://doi.org/10.1103/PhysRevX.5.041041} {\bibfield  {journal} {\bibinfo
  {journal} {Phys. Rev. X}\ }\textbf {\bibinfo {volume} {5}},\ \bibinfo {pages}
  {041041} (\bibinfo {year} {2015})}\BibitemShut {NoStop}%
\bibitem [{\citenamefont {Schr{\"o}der}\ and\ \citenamefont
  {Chin}(2016)}]{schroder_simulating_2016}%
  \BibitemOpen
  \bibfield  {author} {\bibinfo {author} {\bibfnamefont {F.~A. Y.~N.}\
  \bibnamefont {Schr{\"o}der}}\ and\ \bibinfo {author} {\bibfnamefont {A.~W.}\
  \bibnamefont {Chin}},\ }\bibfield  {title} {\bibinfo {title} {Simulating open
  quantum dynamics with time-dependent variational matrix product states:
  {Towards} microscopic correlation of environment dynamics and reduced system
  evolution},\ }\href {https://doi.org/10.1103/PhysRevB.93.075105} {\bibfield
  {journal} {\bibinfo  {journal} {Phys. Rev. B}\ }\textbf {\bibinfo {volume}
  {93}},\ \bibinfo {pages} {075105} (\bibinfo {year} {2016})}\BibitemShut
  {NoStop}%
\bibitem [{\citenamefont {Abanin}\ \emph {et~al.}(2019)\citenamefont {Abanin},
  \citenamefont {Altman}, \citenamefont {Bloch},\ and\ \citenamefont
  {Serbyn}}]{abanin_colloquium:_2019}%
  \BibitemOpen
  \bibfield  {author} {\bibinfo {author} {\bibfnamefont {D.~A.}\ \bibnamefont
  {Abanin}}, \bibinfo {author} {\bibfnamefont {E.}~\bibnamefont {Altman}},
  \bibinfo {author} {\bibfnamefont {I.}~\bibnamefont {Bloch}},\ and\ \bibinfo
  {author} {\bibfnamefont {M.}~\bibnamefont {Serbyn}},\ }\bibfield  {title}
  {\bibinfo {title} {Colloquium: {Many}-body localization, thermalization, and
  entanglement},\ }\href {https://doi.org/10.1103/RevModPhys.91.021001}
  {\bibfield  {journal} {\bibinfo  {journal} {Rev. Mod. Phys.}\ }\textbf
  {\bibinfo {volume} {91}},\ \bibinfo {pages} {021001} (\bibinfo {year}
  {2019})}\BibitemShut {NoStop}%
\bibitem [{\citenamefont {Safavi-Naini}\ \emph {et~al.}(2019)\citenamefont
  {Safavi-Naini}, \citenamefont {Wall}, \citenamefont {Acevedo}, \citenamefont
  {Rey},\ and\ \citenamefont {Nandkishore}}]{safavi-naini_quantum_2019}%
  \BibitemOpen
  \bibfield  {author} {\bibinfo {author} {\bibfnamefont {A.}~\bibnamefont
  {Safavi-Naini}}, \bibinfo {author} {\bibfnamefont {M.~L.}\ \bibnamefont
  {Wall}}, \bibinfo {author} {\bibfnamefont {O.~L.}\ \bibnamefont {Acevedo}},
  \bibinfo {author} {\bibfnamefont {A.~M.}\ \bibnamefont {Rey}},\ and\ \bibinfo
  {author} {\bibfnamefont {R.~M.}\ \bibnamefont {Nandkishore}},\ }\bibfield
  {title} {\bibinfo {title} {Quantum dynamics of disordered spin chains with
  power-law interactions},\ }\href {https://doi.org/10.1103/PhysRevA.99.033610}
  {\bibfield  {journal} {\bibinfo  {journal} {Phys. Rev. A}\ }\textbf {\bibinfo
  {volume} {99}},\ \bibinfo {pages} {033610} (\bibinfo {year}
  {2019})}\BibitemShut {NoStop}%
\bibitem [{\citenamefont {Chanda}\ \emph {et~al.}(2019)\citenamefont {Chanda},
  \citenamefont {Sierant},\ and\ \citenamefont
  {Zakrzewski}}]{chanda_time_2019}%
  \BibitemOpen
  \bibfield  {author} {\bibinfo {author} {\bibfnamefont {T.}~\bibnamefont
  {Chanda}}, \bibinfo {author} {\bibfnamefont {P.}~\bibnamefont {Sierant}},\
  and\ \bibinfo {author} {\bibfnamefont {J.}~\bibnamefont {Zakrzewski}},\
  }\bibfield  {title} {\bibinfo {title} {Time dynamics with matrix product
  states: {Many}-body localization transition of large systems revisited},\
  }\href {https://doi.org/10.1103/PhysRevB.101.035148} {\bibfield  {journal} {\bibinfo
  		{journal} {Phys. Rev. B}\ }\textbf {\bibinfo {volume} {101}},\ \bibinfo
  	{pages} {035148} (\bibinfo {year} {2020})}\BibitemShut {NoStop}%
\bibitem [{\citenamefont {Brockt}\ \emph {et~al.}(2015)\citenamefont {Brockt},
  \citenamefont {Dorfner}, \citenamefont {Vidmar}, \citenamefont
  {Heidrich-Meisner},\ and\ \citenamefont
  {Jeckelmann}}]{brockt_matrix-product-state_2015}%
  \BibitemOpen
  \bibfield  {author} {\bibinfo {author} {\bibfnamefont {C.}~\bibnamefont
  {Brockt}}, \bibinfo {author} {\bibfnamefont {F.}~\bibnamefont {Dorfner}},
  \bibinfo {author} {\bibfnamefont {L.}~\bibnamefont {Vidmar}}, \bibinfo
  {author} {\bibfnamefont {F.}~\bibnamefont {Heidrich-Meisner}},\ and\ \bibinfo
  {author} {\bibfnamefont {E.}~\bibnamefont {Jeckelmann}},\ }\bibfield  {title}
  {\bibinfo {title} {Matrix-product-state method with a dynamical local basis
  optimization for bosonic systems out of equilibrium},\ }\href
  {https://doi.org/10.1103/PhysRevB.92.241106} {\bibfield  {journal} {\bibinfo
  {journal} {Phys. Rev. B}\ }\textbf {\bibinfo {volume} {92}},\ \bibinfo
  {pages} {241106} (\bibinfo {year} {2015})}\BibitemShut {NoStop}%
\bibitem [{\citenamefont {Motruk}\ \emph {et~al.}(2016)\citenamefont {Motruk},
  \citenamefont {Zaletel}, \citenamefont {Mong},\ and\ \citenamefont
  {Pollmann}}]{motruk_density_2016}%
  \BibitemOpen
  \bibfield  {author} {\bibinfo {author} {\bibfnamefont {J.}~\bibnamefont
  {Motruk}}, \bibinfo {author} {\bibfnamefont {M.~P.}\ \bibnamefont {Zaletel}},
  \bibinfo {author} {\bibfnamefont {R.~S.~K.}\ \bibnamefont {Mong}},\ and\
  \bibinfo {author} {\bibfnamefont {F.}~\bibnamefont {\hbox{Pollmann}}},\ }\bibfield
  {title} {\bibinfo {title} {Density matrix renormalization group on a cylinder
  in mixed real and momentum space},\ }\href
  {https://doi.org/10.1103/PhysRevB.93.155139} {\bibfield  {journal} {\bibinfo
  {journal} {Phys. Rev. B}\ }\textbf {\bibinfo {volume} {93}},\ \bibinfo
  {pages} {155139} (\bibinfo {year} {2016})}\BibitemShut {NoStop}%
\bibitem [{\citenamefont {Pastori}\ \emph {et~al.}(2019)\citenamefont
  {Pastori}, \citenamefont {Heyl},\ and\ \citenamefont
  {Budich}}]{pastori_disentangling_2019}%
  \BibitemOpen
  \bibfield  {author} {\bibinfo {author} {\bibfnamefont {L.}~\bibnamefont
  {Pastori}}, \bibinfo {author} {\bibfnamefont {M.}~\bibnamefont {Heyl}},\ and\
  \bibinfo {author} {\bibfnamefont {J.~C.}\ \bibnamefont {Budich}},\ }\bibfield
   {title} {\bibinfo {title} {Disentangling sources of quantum entanglement in
  quench dynamics},\ }\href {https://doi.org/10.1103/PhysRevResearch.1.012007}
  {\bibfield  {journal} {\bibinfo  {journal} {Phys. Rev. Research}\ }\textbf
  {\bibinfo {volume} {1}},\ \bibinfo {pages} {012007} (\bibinfo {year}
  {2019})}\BibitemShut {NoStop}%
\bibitem [{\citenamefont {Rams}\ and\ \citenamefont
  {Zwolak}(2019)}]{rams_breaking_2019}%
  \BibitemOpen
  \bibfield  {author} {\bibinfo {author} {\bibfnamefont {M.~M.}\ \bibnamefont
  {Rams}}\ and\ \bibinfo {author} {\bibfnamefont {M.}~\bibnamefont {Zwolak}},\
  }\bibfield  {title} {\bibinfo {title} {Breaking the {Entanglement} {Barrier}: {Tensor} {Network} {Simulation} of {Quantum} {Transport}},\ }\href
  {https://doi.org/10.1103/PhysRevLett.124.137701} {\bibfield  {journal} {\bibinfo  {journal} {Phys. Rev. Lett.}\ }\textbf
  	{\bibinfo {volume} {124}},\ \bibinfo {pages} {137701} (\bibinfo {year}
  	{2020})}\BibitemShut {NoStop}%
\bibitem [{\citenamefont {Krumnow}\ \emph {et~al.}(2019)\citenamefont
  {Krumnow}, \citenamefont {Eisert},\ and\ \citenamefont
  {Legeza}}]{krumnow_towards_2019}%
  \BibitemOpen
  \bibfield  {author} {\bibinfo {author} {\bibfnamefont {C.}~\bibnamefont
  {Krumnow}}, \bibinfo {author} {\bibfnamefont {J.}~\bibnamefont {Eisert}},\
  and\ \bibinfo {author} {\bibfnamefont {{\"O}.}~\bibnamefont {Legeza}},\
  }\bibfield  {title} {\bibinfo {title} {Towards overcoming the entanglement
  barrier when simulating long-time evolution},\ }\Eprint
  {https://arxiv.org/abs/1904.11999v1} {arXiv:1904.11999v1}
  [cond-mat.stat-mech] \BibitemShut {NoStop}%
\bibitem [{\citenamefont {Phien}\ \emph {et~al.}(2012)\citenamefont {Phien},
  \citenamefont {Vidal},\ and\ \citenamefont
  {McCulloch}}]{phien_infinite_2012}%
  \BibitemOpen
  \bibfield  {author} {\bibinfo {author} {\bibfnamefont {H.~N.}\ \bibnamefont
  {Phien}}, \bibinfo {author} {\bibfnamefont {G.}~\bibnamefont {Vidal}},\ and\
  \bibinfo {author} {\bibfnamefont {I.~P.}\ \bibnamefont {McCulloch}},\
  }\bibfield  {title} {\bibinfo {title} {Infinite boundary conditions for
  matrix product state calculations},\ }\href
  {https://doi.org/10.1103/PhysRevB.86.245107} {\bibfield  {journal} {\bibinfo
  {journal} {Phys. Rev. B}\ }\textbf {\bibinfo {volume} {86}},\ \bibinfo
  {pages} {245107} (\bibinfo {year} {2012})}\BibitemShut {NoStop}%
\bibitem [{\citenamefont {Bauernfeind}\ and\ \citenamefont
  {Aichhorn}(2019)}]{bauernfeind_time_2019}%
  \BibitemOpen
  \bibfield  {author} {\bibinfo {author} {\bibfnamefont {D.}~\bibnamefont
  {Bauernfeind}}\ and\ \bibinfo {author} {\bibfnamefont {M.}~\bibnamefont
  {Aichhorn}},\ }\bibfield  {title} {\bibinfo {title} {Time dependent variational principle for tree tensor networks},\ }\href
  {https://doi.org/10.21468/SciPostPhys.8.2.024} {\bibfield  {journal} {\bibinfo
  		{journal} {SciPost Phys.}\ }\textbf {\bibinfo {volume} {8}},\ \bibinfo
  	{pages} {024} (\bibinfo {year} {2020})}\BibitemShut {NoStop}%
\bibitem [{\citenamefont {Evenbly}(2018)}]{evenbly_gauge_2018}%
  \BibitemOpen
  \bibfield  {author} {\bibinfo {author} {\bibfnamefont {G.}~\bibnamefont
  {Evenbly}},\ }\bibfield  {title} {\bibinfo {title} {Gauge fixing, canonical
  forms, and optimal truncations in tensor networks with closed loops},\ }\href
  {https://doi.org/10.1103/PhysRevB.98.085155} {\bibfield  {journal} {\bibinfo
  {journal} {Phys. Rev. B}\ }\textbf {\bibinfo {volume} {98}},\ \bibinfo
  {pages} {085155} (\bibinfo {year} {2018})}\BibitemShut {NoStop}%
\bibitem [{\citenamefont {Zaletel}\ and\ \citenamefont
  {Pollmann}(2019)}]{zaletel_isometric_2019}%
  \BibitemOpen
  \bibfield  {author} {\bibinfo {author} {\bibfnamefont {M.~P.}\ \bibnamefont
  {Zaletel}}\ and\ \bibinfo {author} {\bibfnamefont {F.}~\bibnamefont
  {Pollmann}},\ }\bibfield  {title} {\bibinfo {title} {Isometric {Tensor}
  {Network} {States} in {Two} {Dimensions}},\ }\href
  {https://doi.org/10.1103/PhysRevLett.124.037201} {\bibfield  {journal} {\bibinfo
  		{journal} {Phys. Rev. Lett.}\ }\textbf {\bibinfo {volume} {124}},\ \bibinfo
  	{pages} {037201} (\bibinfo {year} {2020})} \BibitemShut {NoStop}%
\bibitem [{\citenamefont {Lubasch}\ \emph {et~al.}(2018)\citenamefont
  {Lubasch}, \citenamefont {Moinier},\ and\ \citenamefont
  {Jaksch}}]{lubasch_multigrid_2018}%
  \BibitemOpen
  \bibfield  {author} {\bibinfo {author} {\bibfnamefont {M.}~\bibnamefont
  {Lubasch}}, \bibinfo {author} {\bibfnamefont {P.}~\bibnamefont {Moinier}},\
  and\ \bibinfo {author} {\bibfnamefont {D.}~\bibnamefont {Jaksch}},\
  }\bibfield  {title} {\bibinfo {title} {Multigrid renormalization},\ }\href
  {https://doi.org/10.1016/j.jcp.2018.06.065} {\bibfield  {journal} {\bibinfo
  {journal} {J. Comput. Phys.}\ }\textbf {\bibinfo {volume} {372}},\ \bibinfo
  {pages} {587} (\bibinfo {year} {2018})}\BibitemShut {NoStop}%
\end{thebibliography}
\end{document}


\title{Parallel time-dependent variational principle algorithm for matrix product states (Supplemental Material)}

\author{Paul Secular}
\email{paul@secular.me.uk}
\homepage{http://secular.me.uk/}
\affiliation{Department of Physics, University of Bath, Claverton Down, Bath BA2 7AY, UK}

\author{Nikita Gourianov}
\affiliation{Clarendon Laboratory, Department of Physics, University of Oxford, Oxford OX1 3PU, UK}

\author{Michael Lubasch}
\affiliation{Clarendon Laboratory, Department of Physics, University of Oxford, Oxford OX1 3PU, UK}

\author{Sergey Dolgov}
\affiliation{Department of Mathematical Sciences, University of Bath, Claverton Down, Bath BA2 7AY, UK}

\author{Stephen R. Clark}
\affiliation{H.H. Wills Physics Laboratory, University of Bristol, Bristol BS8 1TL, UK}
\affiliation{Max Planck Institute for the Structure and Dynamics of Matter, CFEL, 22761 Hamburg, Germany}

\author{Dieter Jaksch}
\affiliation{Clarendon Laboratory, Department of Physics, University of Oxford, Oxford OX1 3PU, UK}
\affiliation{Centre for Quantum Technologies, National University of Singapore, 3 Science Drive 2, Singapore 117543, Singapore}

\date{\rule[11pt]{0pt}{0pt}\today}

\begin{abstract}
In this Supplemental Material we present technical information aimed at practitioners wishing to use the parallel two-site time dependent variational principle algorithm. We compare two different methods of approximating power laws by sums of exponentials, which should prove useful for calculations on larger systems. We also provide details of the software, settings, and parameters used to carry out our numerical experiments, in order to aid reproducibility.
\end{abstract}

\maketitle

\section{Implementation}\label{implementation}

The parallel two-site time-dependent variational principle (p2TDVP) code was built on top of the Tensor \hbox{Network} Theory (TNT) Library \cite{noauthor_tntlibrary_nodate, goodyer_tnt_nodate,al-assam_tensor_2017, coulthard_engineering_2018} -- a C code that implements OpenMP \cite{noauthor_openmp_nodate} shared memory parallelism \cite{goodyer_tnt_nodate}, and which supports multithreaded linear algebra libraries. p2TDVP was implemented using the Message Passing Interface (MPI) standard \cite{noauthor_mpi_nodate}. This allowed us to target distributed memory architectures, hence increasing the total amount of random-access memory (RAM) available for simulations. For our benchmark calculations we employed a hybrid parallel approach in which the MPI network-level parallelism was combined with existing shared memory parallelism.

\section{Approximating power laws}\label{powerlaws}

\begin{figure*}
  \includegraphics[width=8.6cm,keepaspectratio]{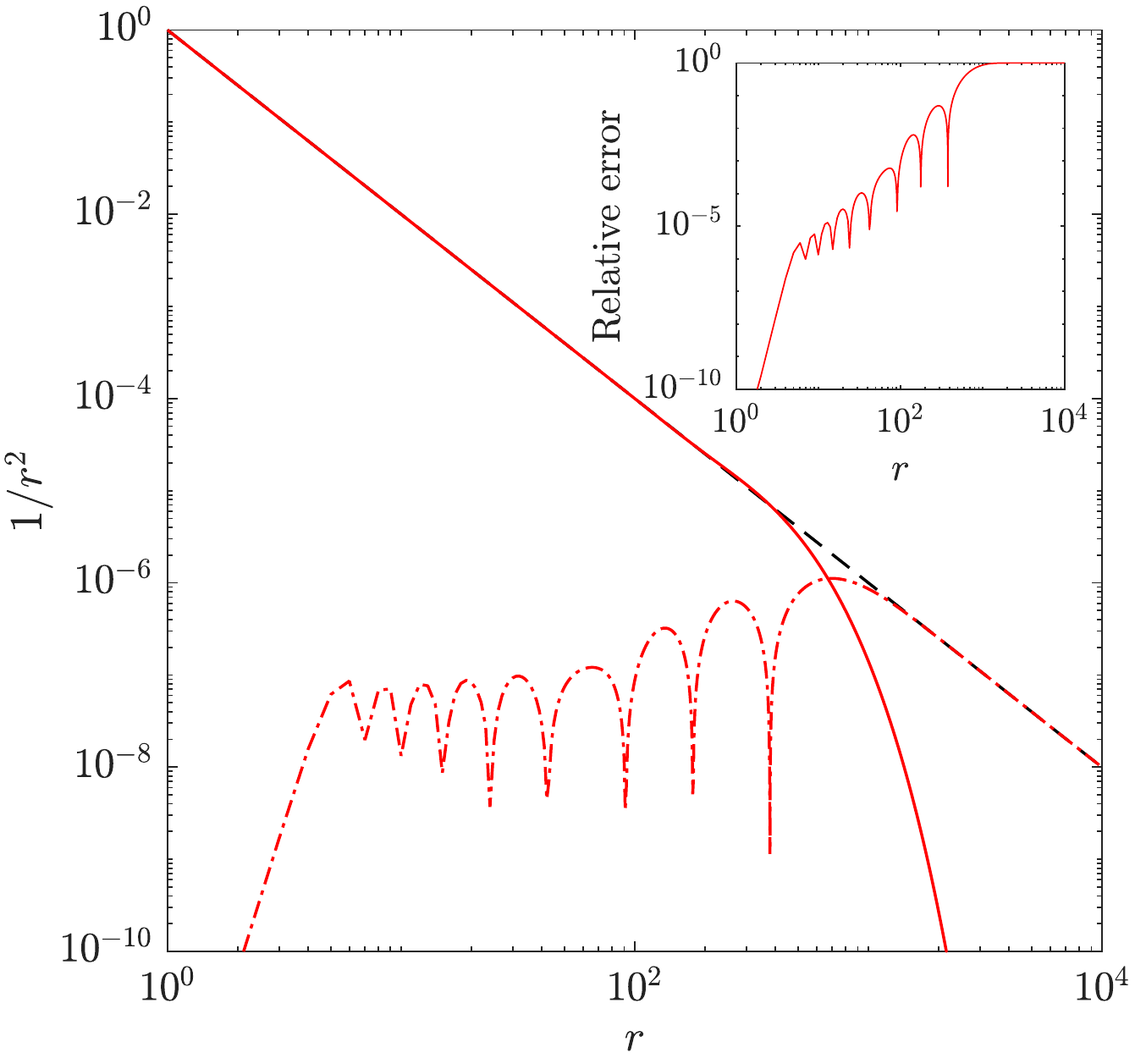}
  \includegraphics[width=8.5cm,keepaspectratio]{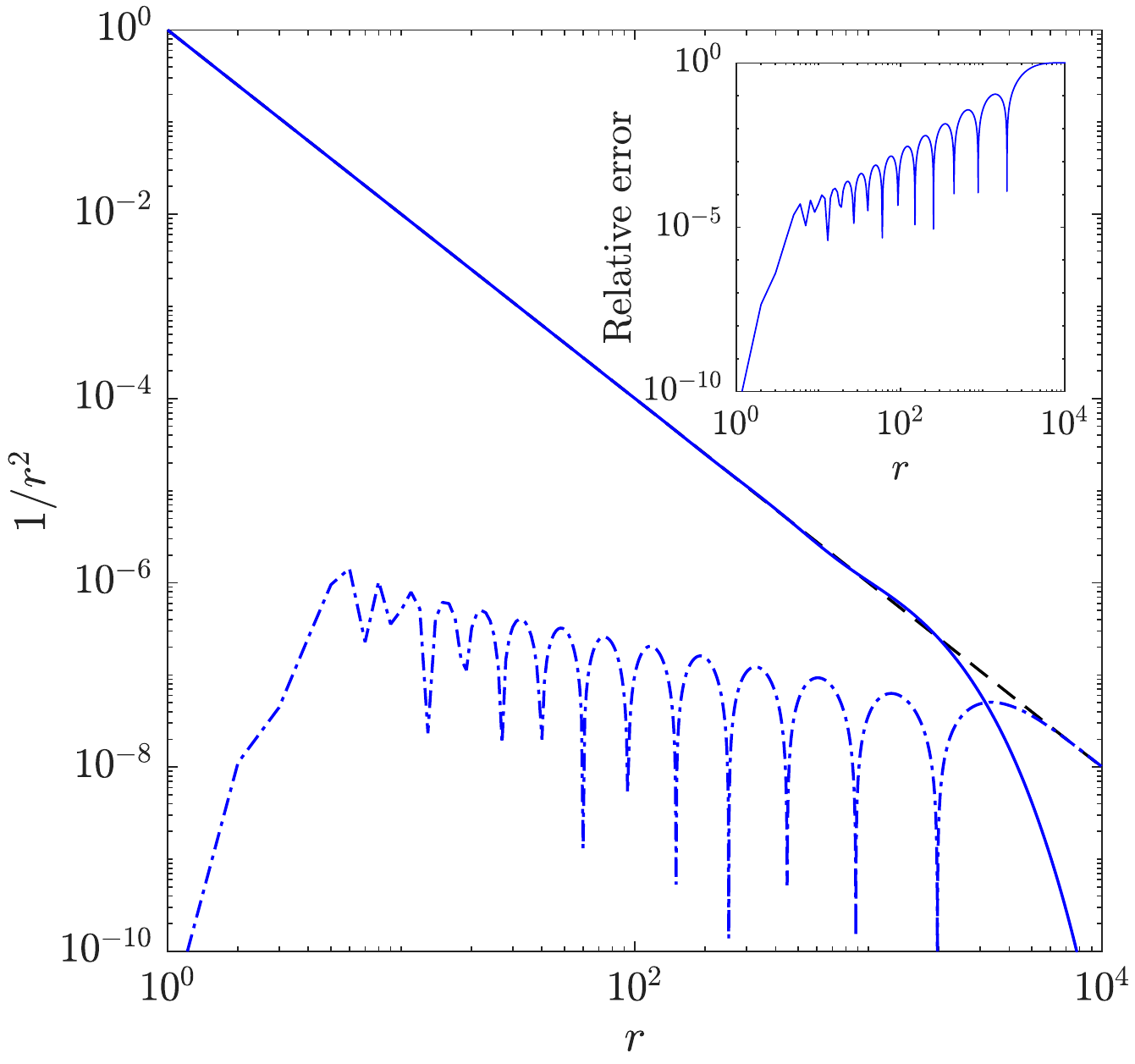}
 \caption{Approximation of $1/r^2$ (dashed lines) by a sum of 9 exponentials (solid lines) using (left panel) the method described by Pirvu \textit{et al.} in Ref. \cite{pirvu_matrix_2010}, and (right panel) the Levenberg-Marquardt nonlinear least-squares algorithm discussed in the text. The broken lines show the absolute error in the approximations, and the insets show the relative error.\label{fig:exps}}
\end{figure*}

The Hamiltonian matrix product operators (MPOs) used in our calculations were defined using a \textsc{\hbox{matlab}} \cite{matlab_r2017_nodate} interface written by Coulthard \cite{coulthard_engineering_2018}. This employs the method described by Pirvu \textit{et al.} in the Appendix of Ref. \cite{pirvu_matrix_2010} to approximate power laws by sums of exponentials. Henceforth, when we refer to the MPO error, we mean the error in this approximation.

Although the approach suggested by Pirvu \textit{et al.} is particularly fast and stable, it does not always give optimal results. In some cases it is preferable to employ a nonlinear least-squares fit. In \fir{fig:exps} we compare the method from Ref. \cite{pirvu_matrix_2010} to the Levenberg-Marquardt nonlinear least-squares algorithm \cite{levenberg_method_1944, marquardt_algorithm_1963}, by approximating $1/r^2$ as a sum of nine exponentials.

The Levenberg-Marquardt fit was calculated in \textsc{\hbox{matlab}} using the \texttt{lsqnonlin()} function \cite{noauthor_least-squares_nodate} with the options shown in Table \ref{tab:matlab-options}. We find that the Levenberg-Marquardt method is slower but gives better results, comparable to those in Ref. \cite{crosswhite_applying_2008}. Using this method should thus allow for a smaller MPO bond dimension, and hence a slight speedup. More importantly, the fit holds over a longer distance, making it valuable for simulations with larger sized systems.

\squeezetable
\begin{table}[t]
\begin{ruledtabular}
\begin{tabular}{ll}
    Option & Value
    \tabularnewline
    \\[-0.255cm]
    \hline
    \\[-0.15cm]
    \texttt{Algorithm} & \texttt{levenberg-marquardt}\\
    \texttt{MaxFunctionEvaluations} & \texttt{10000}\\
    \texttt{MaxIterations} & \texttt{1000}\\
    \texttt{StepTolerance} & \texttt{1E-6}\\
    \texttt{FunctionTolerance} & \texttt{1E-14}
\end{tabular}
\caption{Options used for the \textsc{matlab} \texttt{lsqnonlin()} function when calculating the fit shown in \fir{fig:exps}.}\label{tab:matlab-options}
\end{ruledtabular}
\end{table}

\section{\texorpdfstring{\MakeUppercase{Test Platform}}{Test Platform}\label{appendix-implementation}}

We carried out our benchmarks on the Balena high performance computing (HPC) cluster \cite{noauthor_balena_nodate} at the \hbox{University} of Bath. We had access to a maximum of 32 compute nodes, with a maximum runtime per job of 5 days. All simulations were run on Dell PowerEdge C8220 nodes, which have two Intel E5-2650 v2 CPUs (20 MB Cache, 2.60 GHz base frequency), giving a total of 16 cores per node. Each node has a memory of 64 GB (8 \textrm{GB} $\times$ 8) DDR3 (1866 MHz).

\section{Simulation details}

We linked the TNT Library to \textsc{arpack-ng} \cite{cornet_opencollab/arpack-ng_2019}, and the multithreaded Intel Math Kernel Library (MKL) \cite{ajolleyx_intel_nodate}. We used the same version of the Intel MPI \hbox{Library} \cite{mpi_intel_nodate} and compiler, compiling with the \texttt{-O2} and \texttt{-xHost} optimization flags. We set the OpenMP/MKL environment variables shown in Table \ref{tab:env-vars} to allow dynamic adjustment of the number of threads used (up to a maximum of 16), whilst also disabling nested threading.

\squeezetable
\begin{table}[t]
\begin{ruledtabular}
\begin{tabular}{ll}
    Environment variable & Value
    \tabularnewline
    \\[-0.255cm]
    \hline
    \\[-0.15cm]
    \texttt{OMP\_NUM\_THREADS} & \texttt{16}\\
    \texttt{MKL\_NUM\_THREADS} & \texttt{16}\\
    \texttt{OMP\_NESTED} & \texttt{FALSE}\\
    \texttt{OMP\_DYNAMIC} & \texttt{TRUE}\\
    \texttt{MKL\_DYNAMIC} & \texttt{TRUE}\\
    \texttt{KMP\_AFFINITY} & \texttt{compact,1,0,granularity=fine}
\end{tabular}
\end{ruledtabular}
\caption{Environment variables used to control OpenMP threading in the TNT Library and Intel MKL.}\label{tab:env-vars}
\end{table}

The linear algebra settings used for all calculations were as follows. We used the TNT Library default zero tolerance of $10^{-14}$ for the automatic blocking of matrices \cite{goodyer_tnt_nodate}. We set a relative truncation tolerance of $\varepsilon = 10^{-12}$ for singular value decompositions (SVDs), and used the \textsc{lapack} \cite{lapack} dense matrix divide-and-conquer \cite{cuppen_divide_1980, gu_stable_1994, Rutter:CSD-94-799} routine (as implemented in Intel MKL). For the \hbox{Lanczos} exponentiation in p2TDVP, we created the Krylov subspace using \textsc{arpack-ng} with a maximum of 8 basis vectors, and a convergence tolerance of $10^{-6}$. Density matrix renormalization group (DMRG) calculations used the \textsc{arpack-ng} sparse eigenvalue solver \cite{lehoucq_arpack_nodate, cornet_opencollab/arpack-ng_2019} with these same settings.

\subsection{Long-range Ising model}

\squeezetable
\begin{table}[t]
\begin{ruledtabular}
\begin{tabular}{@{}m{0.4cm}@{}m{2cm}@{}m{2cm}@{}m{2cm}}
 $p$ & Number of sites owned by first process & Number of sites owned by central processes & Number of sites owned by last process
 \tabularnewline
 \\[-0.255cm]
 \hline
 \\[-0.15cm]
 8 & 17 & 16 & 16 \tabularnewline
 16 & 9 & 8 & 8 \tabularnewline
 24 & 10 & 5 & 9 \tabularnewline
 32 & 5 & 4 & 4
\end{tabular}
\end{ruledtabular}
\caption{Partitioning of the 129-site MPS in the long-range Ising model simulations for $p$ parallel processes.}\label{ising-partitioning}
\end{table}

All Hamiltonian MPOs had a maximum absolute error $ \lesssim 10^{-8}$. The other parameters are as described in the main text. The ground state matrix product state (MPS) was calculated using two-site DMRG, and was found to have a maximum bond dimension of $\chi = 22$. For the p2TDVP calculations, the 129-site MPS was partitioned as described in Table \ref{ising-partitioning}.

\subsection{Long-range XY model}

The ground state of the antiferromagnetic XY Hamiltonian is twofold degenerate when there are an odd number of lattice sites. To break this degeneracy we added a small perturbation to the Hamiltonian, so that we actually considered the ground state of
\begin{equation}\label{xyHamiltonianPerturbed}
    H = \frac{1}{2} \sum_{i<j}^L \frac{1}{\lvert i-j \rvert ^\alpha} \left(\sigma_i^x \sigma_j^x + \sigma_i^y \sigma_j^y\right) + \frac{\delta B}{2} \sum_{k}^L \sigma_k^x,
\end{equation}
with $\delta B=10^{-6}$. We calculated this ground state using parallel two-site DMRG \cite{stoudenmire_real-space_2013} with 16 processes. As in Ref. \cite{haegeman_unifying_2016}, we used a maximum MPS bond dimension of $\chi=128$ \cite{haegeman_notitle_nodate}, and a Hamiltonian MPO with maximum absolute error $< 10^{-8}$.

\squeezetable
\begin{table}[t]
\begin{ruledtabular}
\begin{tabular}{@{}m{0.4cm}@{}m{2cm}@{}m{2cm}@{}m{2cm}}
 $p$ & Number of sites owned by first process & Number of sites owned by central processes & Number of sites owned by last process
 \tabularnewline
 \\[-0.255cm]
 \hline
 \\[-0.15cm]
 8 & 15 & 12 & 14 \tabularnewline
 16 & 9 & 6 & 8 \tabularnewline
 24 & 7 & 4 & 6 \tabularnewline
 32 & 6 & 3 & 5
\end{tabular}
\end{ruledtabular}
\caption{Partitioning of the 101-site MPS in the long-range XY model simulations.}\label{xy-partitioning}
\end{table}

For the time evolution we used $\chi_\text{max} = 256$ \cite{haegeman_notitle_nodate} as in Ref. \cite{haegeman_unifying_2016}. No truncation error tolerance was set. The 101-site MPS was partitioned as described in Table \ref{xy-partitioning}.

\subsection{Long-range XXX model}

\squeezetable
\begin{table}[t]
\begin{ruledtabular}
\begin{tabular}{@{}m{0.4cm}@{}m{2cm}@{}m{2cm}@{}m{2cm}}
 $p$ & Number of sites owned by first process & Number of sites owned by central processes & Number of sites owned by last process
 \tabularnewline
 \\[-0.255cm]
 \hline
 \\[-0.15cm]
 32 & 11 & 6 & 10
\end{tabular}
\end{ruledtabular}
\caption{Partitioning of the 201-site MPS in the long-range XXX model simulation.}\label{xxx-partitioning}
\end{table}

To calculate the thermodynamic limit results, we numerically integrated the exact expression for $C_\infty(x,t)$ (given in the main text) using the \texttt{integral2()} function in \textsc{matlab} with the \texttt{iterated} option \cite{noauthor_numerically_nodate}.

For the 201-site p2TDVP calculation we approximated the $1/r^2$ power law by a sum of 12 exponentials, giving an MPO bond dimension of 38 with a maximum absolute error of $1.5 \times 10^{-7}$ and a maximum relative error of $5.3 \times 10^{-3}$. As shown in \fir{fig:exps}, it is possible to use the Levenberg-Marquardt nonlinear least-squares algorithm to approximate the power law more efficiently using just 9 exponentials, albeit with a slightly larger error. This would give an MPO bond dimension of 29, and hence a potential speedup of $\approx 1.3$.

\begin{figure}[t]
 \includegraphics[width=8.4cm]{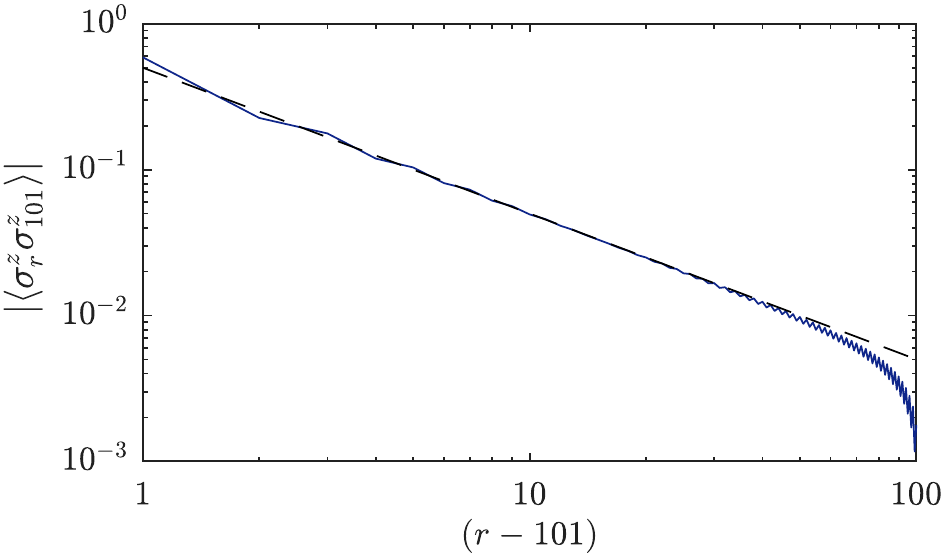}
 \caption{Ground state correlations in the 201-site long-range ($\alpha=2$) XXX model. The dashed line is proportional to $(r-101)^{-1}$.\label{fig:xxx-gs}}
\end{figure}

The ground state of the model $\ket{\psi_0}$ was found using two-site DMRG with a U(1) symmetric MPS of maximum bond dimension $\chi = 512$. The energy per site converged to
\begin{equation}
    E_0/N  = -0.410611165931,
\end{equation}
with a total discarded weight of $2.1\times10^{-9}$. In \fir{fig:xxx-gs} we show the magnitude of the ground state correlation function $\Braket{\sigma_r^z \sigma_k^z}$ for $k=101$. This appears to follow a power law with exponent equal to 1 (dashed line), except towards the edges where the correlations decay exponentially due to the open boundaries.

The initial state for our p2TDVP calculation was $\ket{\psi} = \sigma^z_{101} \ket{\psi_0}$. We time evolved $\ket{\psi}$ on 32 processes, using a truncation error tolerance of $w_\text{max} = 10^{-16}$, and a maximum bond dimension of $\chi_\text{max} = 1024$. The MPS was partitioned as shown in Table \ref{xxx-partitioning}. As in Ref. \cite{zaletel_time-evolving_2015}, we used a timestep of $\delta t = 0.025$ and computed the dynamical spin-spin correlation function,
\begin{equation}\label{HaldaneCorrelation}
    C(r-k,t) = e^{-i E_0 t} \Braket{\psi_0| \sigma^z_r |\psi(t)},
\end{equation}
every eight timesteps.

\squeezetable
\begin{table}[b]
\begin{ruledtabular}
\begin{tabular}{@{}m{3cm}@{}m{0.75cm}@{}m{0.8cm}@{}m{1cm}@{}m{1cm}@{}m{1.18cm}@{}m{0.85cm}}
 Process ID(s) & 0 & 1
 \tabularnewline
 No. of sites owned & 101 & 100
 \tabularnewline
 \\[-0.255cm]
 \hline
 \\[-0.15cm]
 Process ID(s) & 0 & 1--2 & 3
 \tabularnewline
 No. of sites owned & 85 & 16 & 84
 \tabularnewline
 \\[-0.255cm]
 \hline
 \\[-0.15cm]
 Process ID(s) & 0 & 1--6 & 7
 \tabularnewline
 No. of sites owned & 77 & 8 & 76
 \tabularnewline
 \\[-0.255cm]
 \hline
 \\[-0.15cm]
 Process ID(s) & 0 & 1 & 2--13 & 14 & 15
 \tabularnewline
 No. of sites owned & 65 & 12 & 4 & 12 & 64
 \tabularnewline
 \\[-0.255cm]
 \hline
 \\[-0.15cm]
 Process ID(s) & 0 & 1 & 2--21 & 22 & 23
 \tabularnewline
 No. of sites owned & 60 & 11 & 3 & 11 & 59
 \tabularnewline
 \\[-0.255cm]
 \hline
 \\[-0.15cm]
 Process ID(s) & 0 & 1 & 2--3 & 4--27 & 28--29 & 30--31
 \tabularnewline
 No. of sites owned & 33 & 32 & 6 & 2 & 6 & 32
\end{tabular}
\end{ruledtabular}
\caption{(top to botttom) MPS partitions for 2, 4, 8, 16, 24, and 32 processes, corresponding to the single timestep scaling shown in \fir{fig:xxx-scaling}.}\label{xxx-new-partitioning}
\end{table}

Although a full scaling analysis was impractical for this simulation, we used different numbers of processes to time evolve the final state at the end of the simulation for an additional timestep. With one process, this took 71.6 minutes; with 32 processes, it took 7.8 minutes -- a speedup of 9.1 (in comparison, the 65-site simulation gave a speedup of 11.0 on 32 processes). This suggests that our partitioning of the MPS was not optimal. By looking at the final bond dimensions we were able to devise a better partitioning scheme, which gave a speedup of 15.4 with 32 processes. The scaling results are shown in \fir{fig:xxx-scaling}, with the corresponding partitions described in Table \ref{xxx-new-partitioning}. We see close to ideal scaling up to 4 processes, with reasonable scaling up to 16 processes. A speedup of 15.1 was achieved with 24 processes, after which it tails off due to load imbalance. The reason for this is that the dynamics of the system are fairly localized, so the relatively large bond dimension is only saturated by the central tensors.

\begin{figure}
 \includegraphics[width=8.1cm,keepaspectratio]{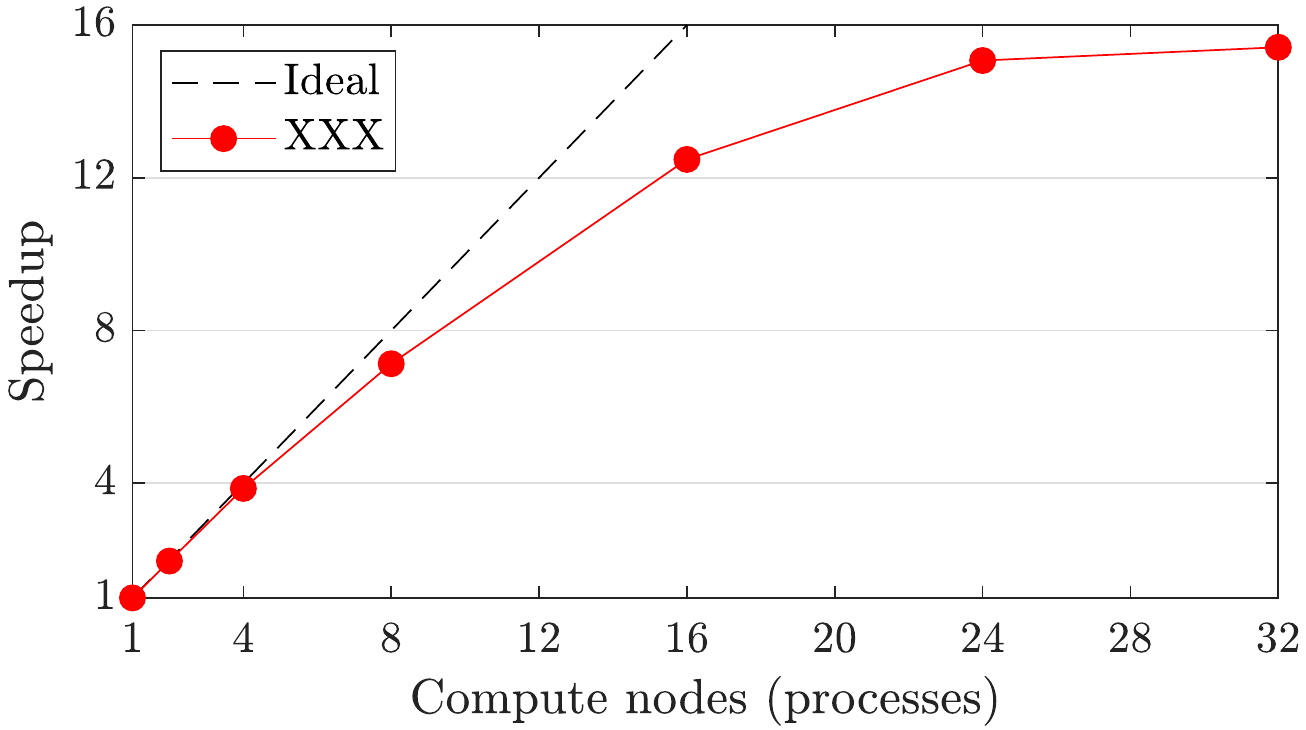}
 \caption{Scaling plot for the extra timestep at the end of the 201-site XXX model p2TDVP simulation.\label{fig:xxx-scaling}}
 \vspace{-0.85em}
 \end{figure}

This example highlights the fact that the choice of partitioning scheme and number of processes is nontrivial for simulations in which the MPS bond dimensions $\chi_j$ grow inhomogeneously. Unless necessitated by memory requirements, using ``too many'' compute nodes is a waste of resources. On the other hand, sub-optimal partitioning is a performance issue. It should be possible to address this using a dynamic load balancer, as has previously been done for the parallel time-evolving block decimation algorithm \cite{urbanek_parallel_2016}.
\vfill

\interlinepenalty=10000